# Assessment of Future Changes in Intensity-Duration-Frequency Curves for Southern Ontario using North American (NA)-CORDEX Models with Nonstationary Methods

Poulomi Ganguli[1], Paulin Coulibaly[1]

[1] McMaster Water Resources and Hydrologic Modeling Group, Department of Civil Engineering, McMaster University, 1280 Main Street West, Hamilton ON L8S 4L7, Canada

**Abstract**

The evaluation of possible climate change consequence on extreme rainfall has significant implications for the design of engineering structure and socioeconomic resources development. While many studies have assessed the impact of climate change on design rainfall using global and regional climate model (RCM) predictions, to date, there has been no comprehensive comparison or evaluation of intensity-duration-frequency (IDF) statistics at regional scale, considering both stationary versus nonstationary models for the future climate. To understand how extreme precipitation may respond to future IDF curves, we used an ensemble of three RCMs participating in the North-American (NA)-CORDEX domain over eight rainfall stations across Southern Ontario, one of the most densely populated and major economic region in Canada. The IDF relationships are derived from multi-model RCM simulations and compared with the station-based observations. We modeled precipitation extremes, at different durations using extreme value distributions considering parameters that are either stationary or nonstationary, as a linear function of time. Our results showed that extreme precipitation intensity driven by future climate forcing shows a significant increase in intensity for 10-year events in 2050s (2030-2070) relative to 1970-2010 baseline period across most of the locations. However, for longer return periods, an opposite trend is noted. Surprisingly, in term of design storms, no significant differences were found when comparing stationary and nonstationary IDF estimation methods for the future (2050s) for the larger return periods. The findings, which are specific to regional precipitation extremes, suggest no immediate reason for alarm, but the need for progressive updating of the design standards in light of global warming.

## 1. Introduction

Extreme precipitation events have significant societal consequences and important implications for urban and rural development, public infrastructure, agriculture and human health [*Rosenzweig et al.*, 2001; *Mailhot and Duchesne*, 2009; *Rosenberg et al.*, 2010; *Hatfield et al.*, 2011; *Murray* *and Ebi*, 2012]. Historical climate records indicate increasing trend in annual precipitation over Canada since 1950s, often statistically significant [*Kunkel et al.*, 1999; *Shephard et al.*, 2014]. The short-duration (sub-hourly to daily) extreme rainfall is expected to intensify due to global warming [*Lehmann et al.*, 2015; *Fischer and Knutti*, 2016; *Prein et al.*, 2016; *Bao et al.*, 2017; *Pfahl et al.*, 2017]. Furthermore, observed long-term trends (1951-2005) in daily annual maximum (AM) precipitation over Northern Hemisphere land areas are attributed to anthropogenic influences, and the magnitude of increase is consistent with an increase in global mean surface temperature [*Zhang et al.*, 2013]. Overall, the mean precipitation in Canada has increased by about 12% (5 to 35% in Southern Canada) during the period 1950 – 1998, although annual snowfall has been significantly decreased over Southern region since 1950s [*Ligeti et al.*, 2006].

Daily and sub-daily precipitation are extensively used in engineering design and infrastructural planning. The design of hydraulic infrastructure, in general, relies on short-duration extreme precipitation statistics obtained from *T*-year return periods at different durations, which are in turn represented in the form of intensity-duration-frequency (IDF) curves [*Bonnin et al.*, 2006]. Although there is clear evidence that climate is changing, the traditional IDF analyses assume stationarity of the climate, means trend in extremes will not vary significantly over time. The concept of nonstationarity in water resources [*Milly et al.*, 2008, 2015], poses a key question to the scientific community whether it is significant enough at the scale relevant to design, operation, and maintenance of hydraulic infrastructures. Further, it is an open question to stakeholders and decision makers how the changing climate information will be translated into the development of future IDFs.

In Canada, the nationwide historical IDF information is produced based on short-duration ground-based rainfall records and archived at Environment Canada (EC) Engineering database [*EC*, 2014]. At regional and local scales, one of the fundamental challenges is the scale mismatch between climate information provided by the climate models and the impact relevant metrics [*McNie*, 2007; *Bonnin*, 2013]. The global climate models (GCMs), which are typically designed to simulate large-scale atmospheric processes (large scale models; horizontal grid spacing of 150-300 km), do not directly simulate all extreme rainfall processes, such as convection [*Zhang et al.*, 2017]. To capture fine-scale regional information for impact, adaptation and vulnerability (IAV) studies, different downscaling methods, such as statistical [*Hayhoe et al.*, 2004, 2008; Dibike and Coulibaly 2006; *Stoner et al.*, 2013] and dynamical [*Giorgi and Mearns*, 1991; *Giorgi*, 2006] methods have been developed. The statistical method is less computationally demanding and is based on finding statistical relationships between a set of predictors and predictands [*Wilby et al.*, 1998]. However, its performance largely relies on good quality long-term observational records and assumes that the observed statistical relationship would remain unchanged in the future [*Wilby and Wigley*, 1997; *Wilby et al.*, 1998]. Furthermore, statistical downscaling approaches are generally applied to aggregate (seasonal or monthly time step) rather than daily time scales. At a daily time scale, the perfect prognosis assumption of statistical downscaling makes them quite susceptible to GCM biases [*Ines and Hansen*, 2006; *Maraun*, 2016]. On the other hand, the dynamical downscaling approach uses nested modeling in which GCM boundary conditions drive regional climate models (RCMs) and involve explicit solving of process-based physical dynamics of the system. Although dynamical downscaling is computationally expensive, it allows simulation of topographic complexity and fine-scale atmospheric dynamics and is not limited to either locations or length of records. Hence dynamically downscaled climate variables, in general, respond more physically consistent ways to external forcing [*Giorgi and Mearns*, 1991; *Wang and Kotamarthi*, 2015] as compared to their statistically downscaled counterparts. The added value [*Feser et al.*, 2011; *Gutmann et al.*, 2012] of dynamical downscaling method in improving spatial resolution typically by the factor of five - ten times as compared to the host GCMs, allows modelers to widely use the downscaled climate information in impact assessment studies [*Gao et al.*, 2012; *Mailhot et al.*, 2012; *Glotter et al.*, 2014; *Ganguli and Ganguly*, 2016a; *Yang et al.*, 2016; *Yoo and Galewsky*, 2016; *Li et al.*, 2017]. However, a common problem associated with the use of RCM outputs directly in hydrologic impact assessment is that the simulated precipitation exhibits significant biases [*Teutschbein and Seibert*, 2010, 2012]. RCM biases stem from the intrinsic RCM physics or from the errors in the lateral boundary conditions provided by the reanalysis data or by a GCM [*Schoetter et al.*, 2012]. Therefore, bias correction is often required to match RCM simulated rainfall to observations for impact assessment studies.

The wide use of dynamically downscaled climate information for impact assessment has accelerated the growth of computational resources. Currently, many large collaborative projects are generating downscaled climate information, which is archived and publicly available. One such effort includes the North American Regional Climate Change Assessment Program (NARCCAP) since 2006, which generated high resolution (50 km) climate projections for the United States, Canada and Northern Mexico [*Mearns et al.*, 2012]. Two of such programs are available for Europe – the Prediction of Regional Scenarios and Uncertainties for Defining European Climate Change Risks and Effects (PRUDENCE) [*Christensen and Christensen*, 2007] and ENSEMBLES [*Van der Linden and Mitchell*, 2009]. Over North America, NARCCAP offers a suite of climate models consist of six different RCMs driven by four different GCMs (http://www.narccap.ucar.edu/data/model-info.html), which helps in estimating climate uncertainties in multi-model ensembles on a regional scales. However, the NARCCAP simulations are available only at a limited 30-year period [1970 – 2000; historical and 2040 – 2077, the end term future runs; *Mearns et al.*, 2012]. Furthermore, to project future climate scenarios, the host GCMs are forced with the SRES (Special Report on Emission Scenarios) 'A2' emission scenario as applied in the Coupled Model Intercomparison Project Phase III (CMIP3). In this context, the latest generation global climate models that are part of the fifth phase of Climate Model Intercomparison Project (CMIP5; *Taylor et al.*, 2012), involve over 20 modeling groups and more than 40 climate models, offers improved horizontal resolutions and physical processes in many models over its earlier version (CMIP3) [*Sillmann et al.*, 2013a, 2013b; *Kumar et al.*, 2014]. The Coordinated Regional Downscaling Experiment (CORDEX) is a program started by the World Climate Research Program (WCRP) to develop fine-scale regional climate projections within the CMIP5 modeling framework in the IPCC AR5 [Intergovernmental Panel of Climate Change Fifth Assessment Report Phase 5; *Stocker et al.*, 2013] timeline. The CORDEX North American domain involves several universities and partner institutions to simulate high-resolution regional climate information (horizontal resolution of ~ 50 km), driven by several global climate models (https://na-cordex.org/) and archived at Earth System Grid (ESG) data portal for regional impact assessment studies.

So far, most efforts to develop future IDF curves are limited to individual city or regional scale assuming stationarity in precipitation extremes [*Coulibaly and Shi*, 2005; *Prodanovic and Simonovic*, 2007; *Hassanzadeh et al.*, 2013; *Rana et al.*, 2013; *Rodríguez et al.*, 2014; *Srivastav and Simonovic*, 2014; *Srivastav et al.*, 2014; *Chandra et al.*, 2015; *Elshorbagy et al.*, 2015; *Lima et al.*, 2016; *Li et al.*, 2017; *Vu et al.*, 2017]. *Mirhosseini et al.*, [2013] analyzed the impact of climate change in the IDF curves at the city of Auburn, Alabama using NARCCAP RCMs. Although their analysis suggests, future precipitation pattern tends towards less intense rainfall for short duration events, for longer durations (> 4-hour) the results are not consistent across the models. *Wang et al.* [2014] developed projected IDF curves for the city of Toronto using dynamically downscaled precipitation data from the regional climate modeling system (PRECIS) driven by a Hadley Center Coupled Model version 3 forced with SRES A1B emission scenario. *Kuo et al.* [2015] assessed climate change impact on future IDFs over the city of Edmonton in central Alberta using MM5 (the Pennsylvania State University/National Center for Atmospheric Research numerical model) regional climate models with boundary conditions from four different global climate models (CGCM3, ECHAM5, CCSM3, and MIROC3.2) forced with A2, A1B, and B1 emission scenarios from CMIP3. Their analysis suggests overall; design storm intensities would gradually increase from 2020 to 2080s for short duration AM intensities of less than 1-hour and return period less than 25-year. *Simonovic et al.* [2016] assessed the impact of climate change on future IDFs across 567 meteorological stations across Canada using a web-based IDF_cc tool [*Simonovic et al.*, 2016]. The results for the end term period (the year 2100) was based on the second generation Canadian Earth System Model (CanESM2) regional climate model and the multi-model ensemble median of 24 global climate models. The CanESM2-based simulations indicated a reduction in AM intensities in central regions of Canada and increased in other regions, whereas ensemble median GCMs simulated lower value of design storm intensity than the RCM. At the regional scale, using third generation Canadian Regional Climate Model (CRCM) *Mailhot et al.*, [2007] performed regional frequency analysis for grid boxes over Southern Quebec region. Comparison of regional frequency estimates of control (1961-1990) versus future (2041-2070) climatic condition reveals return periods corresponding to 2- and 6- hour events would be approximately halved in the projected scenario, whereas they will decrease by a third for 12- and 24-hour events. In an another study, *Mailhot et al.*, [2012] compared return periods of sub-daily, daily, 3-, and 5-day AM precipitation intensity for the future (2041-2070) versus historical (1968-2000) periods using NARCCAP simulations over fourteen climatic regions of Canada and part of Northern United States. Their analysis suggested for 2- to 25-year return periods, inland regions, especially Southern Ontario, the Prairies, and Southern Quebec would experience the largest increase in AM intensities while the coastal regions, the least. *Switzman et al.*, [2017] characterized variability among an ensemble of future IDF curves generated using a combination of 5- different regional and global climate models, 3- climate change scenarios, and 2- downscaling methods. For 2050s, the authors found statistically significant variability in the direction of change and magnitude in the projected IDFs at the Greater Toronto Area (Oshawa WPCP, Toronto, Toronto Pearson International Airport), with some ensemble members showing increase and decrease in rainfall intensity within the ensemble members with no specific trend. On the other hand, over Windsor regions (Windsor Airport, Chatham WPCP, and Harrow), a statistically significant increase in storm intensity is noted for the future. However, there is an evidence of variability in the magnitude of change among ensemble members. Recently, [*Agilan and Umamahesh*, 2016, 2017], compared nonstationary IDFs for the future (2015-2056 and 2057- 2098) versus historical (1961 – 1990) period using an ensemble of 24 global climate model simulations over the city of Hyderabad, India. The results of the study indicate that the estimated return periods using nonstationary IDFs with the trend as a covariate can reasonably cope the effect of climate change for at least next fifty years.

From existing literature on future IDF development, it may be conjectured that to date more scientific efforts are being towards on characterizing the uncertainty in climate change projections than on developing robust adaptation strategies from a range of plausible climate outcomes [*Wilby and Dessai*, 2010]. Secondly, till date, no study has been carried out comparing stationary and nonstationary approaches to design storms in a future climatic condition considering climate uncertainty at a regional scale. Third, a robust assessment of post-processing of climate risk information is required for translating climate change scenarios into decision-relevant metrics through appropriate bias correction schemes taking into account locations, durations and ensemble of climate model runs. Effective management of climate-induced risks requires robust characterization of the probability of extremes accounting both historical nonstationarity and the likelihood of future changes [*Mote et al.*, 2016; *Diffenbaugh et al.*, 2017]. The cities across Southern Ontario, the prominent economic hubs [*Bourne and Simmons*, 2003; *Partridge et al.*, 2007], are especially vulnerable to extreme climatic events given their larger population concentration, property, and aging infrastructure than any other part of Canada [*Kling et al.*, 2003; *Ligeti et al.*, 2006; *Hayhoe et al.*, 2010; *Henstra and Thistlethwaite*, 2017]. Based on regional climate model experiments, two of the recent literature has indicated an increase in frequency and magnitude of rainfall extremes in Southern Ontario by the 2050s [*Mailhot et al.*, 2012; *Deng et al.*, 2016]. Given these challenges, this study has three preliminary objectives:

- To evaluate the credibility of the suite of regional climate models participating in the NA-CORDEX program in attributing extreme rainfall statistics at different durations and return periods with reference to ground-based historical observations.
- To select appropriate bias correction methods based on various skill measures for translating RCM information to hydrometeorological impact assessment.
- To evaluate the impact of climate change on IDF curves by comparing design storms of historical versus projected climate assuming both stationary and nonstationary conditions. Thus we performed a three-way comparison between historical and projected IDFs for impact assessment: *stationary* (*future*) versus *stationary (historical), nonstationary (future)* versus *stationary (historical)* and *nonstationary (future)* versus *nonstationary (historical)*.

To address these research objectives, we analyze the trend and frequency of AM precipitation and temperature data of Southern Ontario region from the station-based observational record. We characterize extreme precipitation frequency at different durations using the concept of "*T*-year return period" (i.e., likelihood of the particular event to occur is once in every *T*-year) and develop stationary and nonstationary IDF relations for the historical and projected time periods. The typical planning horizon of 30 ~ 40 years is used by climatologist to filter out natural variability as well as a tradeoff between data length required to perform nonstationary analysis [*Palutikof et al.*, 1999; *WMO*, 2009]. In this study, estimates of 50-year return period of AM hourly and daily intensity are considered as the representative of rare and extreme events. This return period length and the durations are in agreement with previous studies [*Palutikof et al.*, 1999; *Kharin et al.*, 2007; *Mailhot et al.*, 2012; *Wehner*, 2013]. Since RCM simulations are available at a daily time scales, the RCM model output is temporally downscaled using a multiplicative cascade-based disaggregation technique [*Olsson*, 1995, 1998; *Güntner et al.*, 2001; *Rana et al.*, 2013]. The downscaled climate information is corrected by employing the quantile-based parametric and non-parametric bias correction methodologies [*Panofsky and Brier*, 1965]. First, the statistical characteristics of ground-based extreme rainfall events for the historical period, 1970 – 2010 (centered around 1990), are compared with extreme rainfall statistics from an ensemble of regional climate models. Then, comparative analyses of the future climate, 2030 – 2070 (centered around 2050), relative to the past are performed to estimate expected changes in extreme precipitation statistics for the resilient design of water resources, reliability and resources planning [*Mills*, 2005; *Milly et al.*, 2008]. Since the 2050s represent a reasonable planning horizon for multisector stakeholders' perspective [*Hall et al.*, 2014; *Ganguly et al.*, 2015; *Watts et al.*, 2015], we consider RCM simulations from 2030 – 2070 as projected planning period. We use RCM outputs to project future precipitation for the 2050s considering the high emission scenario, Representative Concentration Pathways 8.5 (RCP8.5 hereafter; *Meinshausen et al.*, [2011]) to maximize adaptation and mitigation strategies. We also explore the effect of bias-correction taking into account the duration and location of observation on model performance for historical simulation and projected scenario.

A flowchart of this study is shown in Figure S1. The organization of the paper is as follows: Section 2 describes study area, observational data, regional climate models, the bias correction method and evaluation statistics; Section 3 presents historical evaluation at sub-daily and daily scales at different return periods; Section 4 discusses relative changes in precipitation intensities under stationary and nonstationary conditions. The summary and discussion follow in Section 5.

## 2. Data and Methods
### 2.1 Study Region
The study is conducted across eight rain gauge locations over Southern Ontario, specifically the Southwest-northeast transect, in the southernmost region of Canada. The Southern Ontario is surrounded by three Great Lakes, Erie, Huron, and Ontario (*Jien and Gough*, 2013). The study sites include eight rain gauge locations of the Windsor-Kingston corridor (*in the order from southwest to northeast*): Windsor Airport, London International Airport, Stratford wastewater treatment plant (WWTP), Shand Dam in Fergus on the Grand River, Hamilton Airport, Toronto International Airport, Oshawa Water Pollution Control Plant (WPCP), Trenton Airport (Figure S2). The Digital Elevation Model (DEM) of the study area indicates a shallow slope with a maximum altitude of 670 m above mean sea level (MSL). The proximity to Great Lakes and topographic effect, especially in areas to the lee of Lakes Erie, Lake Ontario, and the Georgian Bay significantly modifies the climate in the region. Convective storms and thunderstorms primarily modulate the summer rainfall, but fall rainfall is dominated by reduced convective activity and increased lake effect precipitation [*Lapen and Hayhoe*, 2003; *Baldwin et al.*, 2011].

## 2.2 Station-based Observational Data

Since a recent observation-based study has indicated a steady increase in the global warming trend from the 1970s onwards [*Rahmstorf et al.,* 2017], we selected the baseline period for the current analysis from 1970 to 2010. Station-based historical (1970 – 2010) AM precipitation time series at durations ($d$ = 1-, 2-, 6-, 12-, 24- hour) are obtained from Environment Canada Engineering Database (http://climate.weather.gc.ca/prods_servs/engineering_e.html) for the eight rain gauge locations over Southern Ontario. The available data is thoroughly quality controlled [*Shephard et al.*, 2014] and have been previously analyzed for the assessment of national extreme rainfall trends [*Burn and Taleghani*, 2013; *Shephard et al.*, 2014; *Simonovic et al.*, 2016; *Switzman et al.*, 2017]. The extent of missing values in the sub-daily AM rainfall time series ranges between 2 and 20% (average ~ 13%) with the least being in Toronto (only AM rainfall in the year 2005 is missing) and the highest are in Hamilton (1970, and 2004 – 2010 are missing) and Stratford (1973, 1999, 2005 – 2010) respectively. We obtained daily and hourly rainfall records and daily maximum air temperature data from the EC Historical database (http://climate.weather.gc.ca/historical_data/) and Toronto Region Conservation Authority (TRCA; https://trca.ca/). We infilled missing values and updated the extreme precipitation records till 2010 by successively disaggregating daily rainfall values to hourly and sub-hourly time steps using multiplicative random cascade (MRC)-based disaggregation tool [*Olsson*, 1995, 1998; *Güntner et al.*, 2001]. The details of the disaggregation procedure of historical precipitation time series are in the supplementary section.

## 2.3 Regional Climate Model

We used archived prediction data from three RCMs available at NA-CORDEX domain. The specific models include, fourth generation of the Canadian Regional Climate Model (CanRCM4) driven by the second generation of the Canadian Earth System Model (CanESM2); fifth generation of the Canadian Regional Climate Model (CRCM5) driven by CanESM2; and Regional Climate Model version 4 (RegCM4) nested in Hadley Center Global Environmental Model, version 2 (HadGEM2-ES) global climate model. The choice of RCMs are based on their extensive use of the current and previous versions over North America for high-resolution multi-decadal climate change simulations [*Ashfaq et al.*, 2010; *Šeparović et al.*, 2013; *Singh et al.*, 2013; *Naz et al.*, 2016; *Whan and Zwiers*, 2016; *Jalbert et al.*, 2017]. All models' outputs are available in a grid mesh of 0.44° horizontal resolution at a daily time step for the historical and projected scenario, except CanRCM4-CanESM2, for which historical simulations are available at an hourly time step. Use of common horizontal resolution removes large sources of variability between RCMs and helps us to evaluate how the differences in the configuration of the RCMs influence simulation of extremes [*Whan and Zwiers*, 2016]. Both CORDEX generation Canadian RCMs (CanRCM4; [*von Salzen et al.*, 2013; *Scinocca et al.*, 2016] and CRCM5; [*Martynov et al.,* 2013]) share the common dynamical core, however, differ in nesting strategy employed, and land-surface and physics schemes [*Zadra et al.*, 2003; *Whan and Zwiers*, 2016].

For computational purposes, all climate model outputs are regridded to a common grid point of 0.5° latitude/longitude resolution using bilinear interpolation scheme from the climate data operators [*CDO*, 2017]. Further, we consider multi-ensemble approach (multi-model median and associated bounds, defined by minimum and maximum simulation of AM series) to take into account the internal variability of the climate system, which is particularly suitable for near-term impact assessment [*Hawkins and Sutton*, 2009; *Meehl et al.*, 2009; *Hawkins and Sutton*, 2011]. While the ensemble median (multi-model median, hereafter MM-Med) values represent the most likely case, the ensemble minima (multi-model minimum, hereafter MM-Min) and maxima (multi-model maximum, hereafter MM-Max) are considered as the best and worst case scenarios, which indicate the spread of climate model [*Ganguli and Ganguly*, 2016b; *Vousdoukas et al.*, 2017]. Grid-based RCM simulations are downloaded at the four nearest neighbor values of station-based observation, and a distance weighted average remapping was employed [*CDO*, 2017]. Since RCMs, on average simulates a small amount of precipitation on regular time steps, as a threshold for discrimination between wet and dry days a value of 0.1 mm is chosen [*Wehner*, 2013; *Frei et al.*, 2003].

## 2.4 Bias Correction

For bias correction of RCM output, we employ quantile mapping (QM) for historical (1970 – 2010) RCM simulations. For applying bias correction to climate model simulated rainfall, different options are available. For example, bias correction can be applied to entire time series, which then can be used to extract AM series for the IDF development. On the other hand, since only AM rainfall is required for the IDF development, it is also possible to correct AM precipitation. [*Li et al.*, 2017] found that bias correcting AM rainfall based on empirical distribution followed by frequency analysis yields design storm closest to the observations. Hence, we employ quantile mapping bias correction on RCM simulated historical AM time series.

In quantile mapping, a quantile of the present day simulated distribution is replaced by the same quantile of the historical observed distribution [*Maraun*, 2016]. Given a precipitation time series $x$, the method is formulated as [*Ines and Hansen*, 2006; *Maraun*, 2016; *Vu et al.*, 2017]:

$$\tilde{x}_{m-c} = F_{o-c}^{-1}\left[F_{m-c}\left(x_{m-c}\right)\right] \tag{1}$$

Where, $\tilde{x}$ is the bias-corrected precipitation, $F(\bullet)$ is the CDF and $F^{-1}(\bullet)$ is its inverse of either the observations ('o') or model ('m') for the historical training period or current climate ('c') condition. Since RCMs simulate too many wet days (the 'drizzle effect'), the QM is automatically able to adjust the number of wet days [*Gutowski Jr et al.*, 2003; *Hay and Clark*, 2003]. Based on distributional choices, QM can be both parametric and nonparametric. However, for high quantiles, where sampling noise is high, nonparametric QM may produce noisy results and applies random correction [*Maraun*, 2016]. Hence, we evaluate the performance of both parametric [based on Generalized Extreme Value (GEV)] and nonparametric [based on Kernel Density Estimate (KDE)] distributions for QM of historical AM series. For parametric QM, the GEV distribution with three parameters (location, scale and shape) are first fitted to observed and current climate model AM series. Then CDF of the GEV distribution is fitted to the current model data, which is again mapped to the CDF of the observed data. An inverse CDF transformation is then employed to bias correct AM series of the modelled time series [Eqn. 1]. For KDE-based bias correction, nonparametric kernel density function is fitted to the AM series. Mathematically, KDE $\hat{f}(x)$ is [*McGinnis et al.*, 2015]:

$$\hat{f}(x) = \frac{1}{n}\sum_{i=1}^{n} K_h\left(x - x_i\right) \tag{2}$$

Where $K_h$ is the kernel function at a bandwidth '$h$'. Following a previous study [*McGinnis et al.*, 2015] We employ Gaussian kernel function and Silverman's rule of thumb for '$h$' calculation.

However, the basic assumption of QM is that the future distribution properties (such as variance and skew) remain similar to the reference period, and only mean changes. However, owing to nonstationarity of the climate system, this assumption may not hold true in the future [*Milly et al.*, 2008; *Li et al.*, 2010]. Hence, following previous studies [*Li et al.*, 2010; *Srivastav et al.*, 2014; *Vu et al.*, 2017], we applied Equidistant Quantile Mapping (EQM) to bias correct projected AM, which is presented as follows:

$$\tilde{x}_{m-p} = x_{m-p} + F_{o-c}^{-1}\left[F_{m-p}\left(x_{m-p}\right)\right] - F_{m-c}^{-1}\left[F_{m-p}\left(x_{m-p}\right)\right] \tag{3}$$

However, Eq. (3) cannot guarantee that the corrected AM series will have positive value (for example, Multimodel minimum 24-hour AM series in London and Multimodel maximum 12-hour AM series in Toronto when KDE-based bias correction is applied). Hence, in these cases, following an earlier study [*Wang and Chen*, 2014], we employed equiratio QM to the AM series using following mathematical expression:

$$\tilde{x}_{m-p} = x_{m-p} + \frac{F_{o-c}^{-1}\left[F_{m-p}\left(x_{m-p}\right)\right]}{F_{m-c}^{-1}\left[F_{m-p}\left(x_{m-p}\right)\right]} \tag{4}$$

## 2.5 Trend Test

We used non-parametric Mann-Kendall test with correction for ties and autocorrelation [*Hamed and Rao*, 1998; *Reddy and Ganguli*, 2013] to detect signature of monotonic trends in the observed AM historical time series at different durations ($d$ = 1-, 2-, 6-, 12-, 24- hours). In addition, we estimate the trend using Theil Sen's slope estimator [*Sen*, 1968; *Gilbert*, 1987]. To understand the extent of warming we also present results of trend estimates for AM temperature anomaly records of historical observations. Significance of trends are detected at 10% significance level (*i.e.*, p-value < 0.10).

## 2.6 Extreme Value Analysis: Stationary and Nonstationary GEV Models

The Generalized Extreme Value (GEV) distribution is a combination of three different distributions, *viz.*, Fréchet ($\zeta > 0$), Weibull ($\zeta < 0$) and the Gumbel ($\zeta \to 0$) depending on the sign of the shape ($\zeta$) parameter. The cumulative distribution function (CDF) of stationary GEV model is given as [*Coles et al.*, 2001]

$$G(z) = \exp\left\{-\left[1+\zeta\left(\frac{z-\mu}{\sigma}\right)\right]_+^{-1/\zeta}\right\} \quad \sigma > 0, -\infty < \mu, \zeta < \infty \qquad (5)$$

Where, $y_+ = \max\{y, 0\}$ with '+' sign indicates positive part of the argument, $\mu$ is a location parameter, $\sigma$ is a scale parameter and the shape parameter, $\zeta$ determines the symmetry and heaviness of the tail [*El Adlouni et al.*, 2008]. For nonstationary GEV model, we incorporate time-varying covariates into GEV location (GEV I), to describe trends as a linear function of time (in years), *i.e.*, $\mu(t) = \mu_1 t + \mu_0$. Since we limit our planning horizon to 40 years, and modeling temporal changes in shape and scale parameters requires long-term records, following an earlier study [*Cheng et al.*, 2014a], we assume these two parameters as constant. For estimation of GEV parameters, a Bayesian inference is performed combined with Differential Evaluation Markov Chain (DE-MC) Monte Carlo (MC) simulation as suggested by [*Martins et al.*, 2000; *Cheng et al.*, 2014b; *Cheng and AghaKouchak*, 2014]. For the AM time series, the parameters are derived by computing $50^{th}$ (median), $5^{th}$ and $95^{th}$ (the lower and the upper bounds) of the DE-MC sampled GEV parameters. For stationary GEV model, the design storm intensity, $q_p$ associated with *T*-year return period is obtained using following expression [*Coles and Tawn*, 1996]

$$q_p = \mu + \frac{\sigma}{\zeta}\left[1 - \{-\ln(1-p)\}^\zeta\right] \quad \forall \zeta \neq 0 \qquad (6)$$

The computation of nonstationary design storm intensity is similar to the stationary model except the inclusion of time-varying location parameter [*Cheng et al.*, 2014b; *Cheng and AghaKouchak*, 2014]. We perform the calculations following *Cheng and AghaKouchak*, [2014] using an MATLAB-based software package, Nonstationary Extreme Value Analysis (NEVA), Version 2.0].

## 2.6 Historical Model Agreement and Relative Changes

The evaluation of observation and models during present-day (1970 – 2010) climatic condition is performed using following test statistics:

1. Time-series plot of AM series of observed versus multi-model ensemble climate models before and after bias correction.

2. Taylor diagrams [*Taylor*, 2001] of AM series to assess pattern error over eight locations. To better assess models' performance, besides Taylor diagrams, we employ a skill score based on the model's spread and correlation reference to the observation [*Taylor*, 2001]:

$$S = \frac{4(1+R)}{\left(\hat{\sigma}_f + 1/\hat{\sigma}_f\right)^2 (1+R_0)} \quad (7)$$

Where $R$ is the pattern correlation, $\hat{\sigma}_f$ is the normalized standard deviation, $R_0$ is the maximum attainable correlation calculated from the maximum of inter-ensemble correlation.

3. Density plots of observed versus multi-model ensemble climate models at different durations.
4. Box plot of the shape parameters for observed versus MM-Med ensemble to evaluate the credibility of the models' to simulate the fat-tailed behavior of extremes.
5. Intensity (*I*) versus Duration (*D*) plots of observed versus MM-Med ensemble at return periods, *T* = 2-, 5-, 10-, 25-, and 50-year for stationary and nonstationary methods of analysis for three significant stations, *viz.,* Toronto International Airport, Oshawa WPCP, and Windsor Airport.
6. Vertical boxplots of the relative difference between observed and MM-Med ensemble design storm for all eight locations for the durations 1- and 24-hour and *T* = 10-, and 50-year return periods. The relative changes are obtained through the difference between observed and MM-Med ensemble design storm estimates normalized by the baseline values.
7. Intensity (*I*) versus Frequency (*F*) plots of observed versus multi-model climate ensembles at return periods, *T* = 2-, 5-, 10-, 25-, and 50-year for stationary and nonstationary GEV models for all eight locations.
8. The percentage of change between future (calculated from MM-Med ensemble run) and historical period can be attributed to changes in precipitation extremes. Further, the (statistically) significant differences of future versus baseline design intensity is computed using standard z-statistics, which is given by [*Mikkelsen et al.*, 2005; *Madsen et al.*, 2009]

$$z = \frac{\hat{z}_T^{future} - \hat{z}_T^{baseline}}{\sqrt{0.5\left(var\{\hat{z}_T^{future}\} + var\{\hat{z}_T^{baseline}\}\right)}} \quad (8)$$

Where $\hat{z}_T^{future}$ is the *T*-year intensity obtained from the MM-Med climate model ensemble from a stationary/nonstationary model, $\hat{z}_T^{baseline}$ describes the same but with the present-day condition. The denominator indicates predictive uncertainty; $var\{\hat{z}_T^{future}\}$ and $var\{\hat{z}_T^{baseline}\}$ are the estimated variance obtained from the *T*-year event estimates and associated confidence interval (*i.e.*, 5[th], and 95[th] quantiles). The z-statistic can be interpreted as statistically equivalent to quantiles of standard normal distribution, i.e., z = ± 1.96 and ± 1.64 correspond to 5% and 10% significance levels. The null hypothesis of the test assumes that *T*-year event estimate in the future planning period is significantly different from the baseline value.

## 3. RCM Evaluation of Historical Simulation
### 3.1 Climatological Statistics

Figure 1 compares time series plots of observed versus climate ensembles before bias-correction for 1- and 24-hour AM series at eight observation sites. Although models are in-phase with observations, in few cases (for example, London International Airport and Windsor Airport), multi-model ensemble underestimates observed AM series. Hence, we bias corrected AM series by the procedure described in Section 2.4, before performing frequency analysis. As a first step, we obtained statistics from bias-corrected AM series and compared it with observed AM to comprehend if the model simulations are credible enough to project future climate change. Therefore, we graphically depicted bias corrected models' performance using Taylor diagrams [*Taylor*, 2001]. A Taylor diagram provides comprehensive statistical summary of how well models agree with observation in the form of a radial plot where distance from the origin is a normalized standard deviation of the model output (relative to observation) and the cosine of the azimuthal angle is given by the centered pattern correlation factor of model output with respect to observation [*Wehner*, 2013]. Figures 2 and 3 present Taylor diagrams of maximum hourly and daily precipitation extremes for the eight rain gauge locations across Southern Ontario. The bias-corrected (*i.e.,* both GEV and KDE-based) three RCMs, their multi-model ensembles, and observations are represented by different symbols (as in legend), and the different bias correction schemes are shown with different colours. The statistics of observed AM is shown on the x-axis as a reference point. The perfect simulation would be plotted at a unit distance from the origin at an angle of 0°. The radial distance from model's point in the Taylor diagram is proportional to its root mean square error relative to the observation whereas centered pattern correlation provides information of similarity in pattern between model and observation. In general, an agreement for the daily AM series is slightly better than the agreement in hourly AM series. The heat maps of skill score (Eqn. 7) across all durations are graphically presented in Figure 4 and Table S2. The skill score improves gradually from higher to lower temporal resolution (Table S2 – S6). The variability among models is small, and a moderate skill score (varies from more than 0.3 to 0.6) is noted for MM-Med climate ensemble across different durations (Table S1). Among individual model performance, RegCM4 shows the highest skill score over Oshawa whereas CRCM5 exhibits the lowest skill over Fergus Shand dam for 1-hour AM series. Likewise, for daily AM series, RegCM4 performed the best over London, whereas worst over Trenton. In many cases, we found that the single-model performance is better than or comparable to that of the multi-model climate ensembles. However, as noted in the previous studies [*Kumar et al.*, 2014; *Min et al.*, 2014; *Martre et al.*, 2015; *Ganguli and Ganguly*, 2016], the main advantage of using a multi-model ensemble is not the vast improvement over the best performing single model but the consistently better performance of multimodel considering all aspects of projection. Finally, we chose the method of bias correction based on the skill score.

We further evaluate the fidelity of multi-model climate ensembles (MM-Med and associated bounds based on the method of bias correction) relative to observation using probability density function [PDF; *Perkins et al.*, 2007]. Figures 5 and 6 show PDFs of hourly and daily AM series relative to baseline observations. For sub-daily AM series (2- to 12- hour), the density plots are shown in the supplement (Figures S6 – S8). The PDFs are created using a smoothed empirical distribution with Gaussian kernel density function. We found a reasonably well agreement between climate model ensembles and observed AM series; nonetheless, the multi-model RCMs perform better for the hourly extremes as compared to the daily. For hourly extreme, the best agreement is observed for Toronto, while the RCMs do not adequately capture a few outliers on far right tail of Hamilton. For daily extreme, MM-Med simulates reasonably well for Toronto and Stratford. However, for other sites, although multi-model bounds consistently simulate well, MM-Med ensemble exhibits higher peaks relative to the observations, indicating an overestimation of the frequency of extremes. Figure 7 shows time series plots of observed versus bias corrected AM series for all eight locations. Figure 7 indicates improvement in the performance of RCMs in simulating peaks than that of before bias correction (Figure 1).

## 3.2 Analysis of Trend

Next, we analyze the presence of trends in the AM series at different durations to detect the signature of nonstationarity in the time series. For this, first, we analyze the presence of a trend in historical AM series and compare it with CORDEX RCMs. The ability of RCMs in simulating historical trends provides the information regarding the credibility of models in simulating anthropogenic climate change in a future scenario. Tables S7-S11 summarizes results of trend analysis in the historical and MM-Median CORDEX RCMs. In general, we found agreement in the signs of the trend between observation and model, although exceptions exist in many cases. For example, at hourly and sub-hourly rainfall extremes in Toronto and Windsor. In both of these locations, while the observed AM series shows a decreasing trend, the model exhibits an upward trend. Further, in all cases, models over/under-estimate the magnitude of the trend. However, neither in observations nor models, the magnitude of trend estimate is zero. Further, to know the extent of warming we discuss the changes in historical air temperature (Table S12) in the daily AM time series. Except for Windsor, we found an upward trend in historical air temperature at all station locations. During baseline period (1970 – 2010), Stratford shows the highest warming trend (at the rate of 0.46/decade), followed by Toronto (0.37/decade). However, none of these trends are statistically significant. At Windsor, a decrease in temperature extreme is at the rate of -0.035/decade.

## 3.3 Simulation of GEV Shape Parameter

Since shape parameter directly influences the nature of simulated extremes [*Ragulina and Reitan*, 2017], we compare the shape parameters of observed versus RCMs for both stationary and nonstationary GEV models. Figures 8-9 and Table S13 present the shape parameters simulated by MM-Med climate ensembles relative to baseline observation. We found RCMs simulates shape parameters reasonably well, although variability is more for nonstationary GEV models (Figure 8). The shape parameter shows spatial and temporal variations (Figure 8), however, no specific pattern is identified. In most of the cases, the shape parameters vary in the range of -0.30 to +0.30 [Table S13]. For daily precipitation extreme, the spatial average of observed shape parameters across the stations is $\xi = 0.13$ with highest in Toronto ($\xi_{Med} = \sim 0.25$; $\xi_{Med}$ indicates shape parameter obtained from the 50$^{th}$ percentile of DE-MC simulated samples) and least in Fergus

Shand dam (median $\xi_{Med} = 0.05$). On the other hand, the spatial average simulated by MM-Med CORDEX RCM slightly overestimates the observed mean, $\xi = 0.15$ with highest being in Stratford ($\xi_{Med} = \sim 0.29$) and the least in London ($\xi_{Med} = 0.04$). Our results are consistent with a recent study [*Ragulina and Reitan*, 2017], which estimated a global average of shape parameter around 0.14, with a 95% confidence interval between 0.13 and 0.15 using daily gridded precipitation extremes from the Global Historical Climatology Network-Daily database (version 2.60). Further, the authors showed for Southern Ontario the estimates of shape parameter varies from 0.13 – 0.16 [Fig. 5 in *Ragulina and Reitan*, 2017], which is little different in our case, given the fact that we are dealing with extreme precipitation and the ground-based observations are subjected to biases resulting from multiple sources [*New et al.*, 2001; *Adam and Lettenmaier*, 2003].

Next, we investigate the variation of shape parameter in nonstationary models (Figure 9). The shape parameters for daily observed AM series varies between $\xi_{Med} = 0.06$ (for Fergus Shand Dam) and $\xi_{Med} = 0.31$ (for Stratford) with an average estimate of $\xi = 0.19$. On the other hand, for CORDEX RCM, we found models slightly underestimates the spatial average shape parameter; $\xi = 0.16$, which varies between $\xi_{Med} = 0.31$ in Stratford and $\xi_{Med} = 0.05$ in London. Evaluation of shape parameter across different time steps suggests that the most positive shape parameter values occur for 12-hour storm duration in Stratford (in stationary simulation, $\xi_{Med} = 0.34$) and Toronto (in nonstationary simulation, $\xi_{Med} = 0.41$) respectively. On an average, the shape parameter simulated by the nonstationary model is higher than that of the stationary model, indicating the former adequately simulates the presence of heavy upper tail (or higher likelihood of extreme magnitude events) [*Markose and Alentorn*, 2011; *Halmstad et al.*, 2013].

**3.4 IDF Curves: Stationary versus Nonstationary Simulation**

Next, we compare the simulated IDF curves of observed versus CORDEX RCMs. Figures 10 – 12 illustrate a design storm estimates for $T = 2$-, 5-, 10-, 25-, and 50-year return periods at three selected station locations: Toronto International Airport, Oshawa WPCP, and Windsor Airport. Weak trends also have considerable impacts on results of exceedance probability [*Porporato and*

*Ridolfi*, 1998]. Therefore, we modeled nonstationarity considering time as a covariate in the location parameters of the GEV distribution. The uncertainty in design storm is shown using the vertical boxplots. Tables S14 – S16 show parameters of the fitted GEV distributions for stationary and nonstationary models. Table S17 presents a statistical performance of stationary versus nonstationary GEV models for the durations of 1- and 6-hour using Akaike Information Criteria with small sample correction [$AIC_c$; *Akaike*, 1974; *Hurvich and Tsai*, 1990]. The skill of the individual model, as measured by the $AIC_c$ statistics does not exhibit any consistent patterns either based on the locations or the storm durations.

In general, we found CORDEX RCMs able to simulate observed design storm intensity reasonably well for all three locations (Tables S14 – S16, and Figures 10 – 12). This is in contrast with the results of trend estimates in Tables S7 – S11, where we noted the discrepancy in nature (sign) of trends simulated by the CORDEX RCMs. Further, Tables S14 – S16 show the fitted GEV parameters for MM-Med CORDEX ensemble are close to that of the observed AM series. As noted from the plots, the uncertainty in design storm estimates is higher for the lower storm durations ($d$ = 1-, and 2-hour) and higher return periods ($T$ = 25-, and 50-year). For Toronto and Oshawa, at durations, $d$ = 1- to 6-hour, the simulated rainfall intensity with higher return periods tend to underestimate the observed ones. However, for Windsor Airport, we find overestimation of design intensity by the models at $d$ = 1-hour, whereas underestimation at $d$ = 6-hour. Secondly, for Oshawa, we note an increase in observed rainfall intensity in nonstationary models reference to stationary one at short durations of 1-, and 2-hour and return periods, $T$ = 10-year and beyond. This may be attributed to consistent upward trend in AM rainfall intensity across all durations in Oshawa. Further, we find at shorter durations, in nonstationary models, the length of top whisker in the boxplot tend to be longer than the bottom one, indicating the model adequately describes the asymmetric nature of extremes. Our findings are consistent with an earlier study [*Cheng and AghaKouchak*, 2014], in which authors reported the difference between nonstationary and stationary rainfall intensities are more prominent at shorter durations and higher return periods, and the differences gradually diminish at longer durations.

For detailed exploratory analysis, we present relative changes in rainfall intensity for 1 in 10-year (*i.e.,* $T$ = 10-year) and 1 in 50-year events respectively (Figure 13). While the top panel in the figure (Figure 13-a) shows stationary models, the bottom one (Figure 13-b) shows the nonstationary ones, corresponding to design rainfall intensity at $T = $ 10-year and 50-year return periods, which can be themselves are categorized as less and more intense events. At lower return period, $T = $ 10-year, the relative difference between observation and model is close to zero, while the differences tend to increase at higher return period, $T = $ 50-year. Few exceptions exist, for example, London International Airport shows large negative bias for 24-hour rainfall extreme. Next, we present ensemble median and spread of RCMs in simulating extreme rainfall intensity relative to observation (Figures 14-15). For stationary models and 1-hour extreme (*Top panel*; Figure 14), in general, the models simulate observed rainfall intensity reasonably well for $T = $ 2 to 10-year return periods, in which observed intensity is well within the bounds of simulated intensity. However, the variability is larger for $T = $ 25-year and beyond. Exceptions exist for London, in which observed intensity lies outside the simulated bounds, even in the case of shorter return periods, although differences are small (*Top panel;* Figure 14). For nonstationary simulation (*Bottom panel*; Figure 14), the uncertainty or spread of ensemble members are higher than that of the stationary models. However, except London International Airport and Stratford WWTP, in all cases, the observed extremes lie within the bounds of simulated extremes at all return periods. For a 24-hour extreme (Figure 15), except in a few locations and higher return periods ($T = $ 25-year and beyond), the multi-model climate ensembles reasonably simulates the observed intensity. Overall, our analysis suggests CORDEX RCMs show higher confidence in simulating design storms at smaller return periods while lower at return periods of $T = $ 25-year and beyond.

## 4. Projection of Future IDF
### 4.1 Projection of GEV Shape Parameter

In this section, we discuss changes in shape parameter of simulated extreme in future relative to the baseline period. Figures 16 – 17 and Table S18 compares the shape parameters of the GEV distribution in the baseline (1970 – 2010) versus projected (2030 – 2070) scenarios. Barring few exceptions, in general, for both stationary (Figure 16) and nonstationary (Figure 17) models in the near-future, the shape parameter shows a decrease in value, across most of the durations relative to the baseline period. Table S18 shows during 2050s, the shape parameters tend to be negative across all durations irrespective of the nature of simulations. A negative shape parameter indicates the GEV distribution has a bounded upper tail (*i.e.*, an identified upper limit to those extreme events or a Weibull-type distribution), implying a tendency towards heavy lower tail. For stationary GEV models, the values of shape parameter are most negative at 12-hour duration, and ranged from -0.57 (at Trenton) followed by -0.49 (at Toronto) to -0.06 (at Hamilton) with spatial standard deviation of 0.29. At same duration, for nonstationary GEV models, $\xi$ values ranged from -0.66 (at Trenton) followed by -0.65 (at Toronto) to 0.33 (at Stratford) with spatial standard deviation 0.09. These values in shape parameter are more negative than those conventionally considered as physically reasonable in hydrology literature [*Martins et al.*, 2000]. While $\xi$ is less than -0.33, the AM distribution has an infinite third moment, and when $\xi$ is less than -0.5, the distribution has infinite variance as suggested by an earlier study [*Morrison and Smith*, 2002]. A very negative value of the shape parameter suggests that the distribution of AM precipitation tends to have a very heavy tail. However, we found at shorter durations (1 to 6-hour), the $\xi$ values ranged between -0.33 and 0.23, and well within the physically consistent limit.

## 4.2 Future Changes in IDF Statistics

Table S19 lists the performance of stationary and nonstationary GEV models using *AICc* statistics for the selected (*i.e.*, 1- and 6-hour) storm durations. As observed from the Table S19, for 1-hour storm duration, stationary GEV model performs the best in six out of eight sites. However, for longer storm duration (in this case, 6-hour), the spatial pattern is not consistent. Figures 18 and 19 present projected intensity versus duration relations for 1 in 10-year and 1 in 50-year events for nonstationary simulations. The associated IDF curves under stationary assumption are presented in Figures S9 – S10. The uncertainty in rainfall intensities at different durations are shown using boxplots. We find that for 1 in 10-year event (Figures 18 and S9), the MM-Min projected *Intensity versus duration* (*ID*) curves are close to that of the historical curves, and often superimposing over the later for some stations (for example, Toronto, Oshawa, and London). On the other hand, the MM-Max yields the upper bound of *ID* relationships. For 1 in 50-year events (Figures 19 and S10), the nature of *ID* curves follows similar trends, except for the Toronto and Stratford, in which we note widening of uncertainty envelop of RCM ensembles (MM-Min for Stratford and MM-Max for Toronto) in the projected scenario.

To further evaluate how the future intensities are sensitive towards the choice of frequency analysis (*i.e.*, stationary/ nonstationary), the boxplots of difference between future and historical precipitation extremes for 10-year and 50-year events are presented in Figures 20 and 21. The absolute changes in future design storm intensity relative to observations are shown for three different cases: *stationary (future) versus stationary (historical)*; *nonstationary (future) versus stationary (historical)*; and *nonstationary (future) versus nonstationary (historical)*. For 1 in 10-year events, except Stratford, at all locations, the quantile boxes are above zero for 6- and 24-hour durations. For Stratford, at 24-hour duration, the intensity of rainfall remains relatively unchanged over time as indicated by the location of quantile boxes, which is close to zero irrespective of the choice of the return period estimates. For hourly extreme, the increase is observed for Hamilton, Oshawa and Trenton. In contrast, for 1 in 50-year events, except a few stations, for most of the stations, the quantile boxes are below zero for all durations, indicating a decrease in precipitation intensity in the future period. However, we find an increase in rainfall intensity at 2-hour and 1-hour extremes for Windsor and Oshawa respectively. For 24-hour precipitation extreme, the intensity of extreme precipitation remains relatively unchanged for 1 in 50-year events (Figure 19). Further, for both events, although the differences in rainfall intensity are apparent for durations up to 2-hour they follow the same trend irrespective of the choice of frequency analysis. However, the magnitude of differences fade away at durations of 6-hour and beyond.

To quantify the magnitude of changes in projected intensity relative to the baseline period, we present heat maps of the difference in rainfall intensity of the 2050s versus 1990s for 10-, 25- and 50-year return periods (Figures 22-23, and S10). While the left panel of the figures show the *z*-statistics indicating statistical significance of the change, the right panel summarizes the median changes in projected rainfall intensities for the future reference to the baseline period is expressed in terms of percentages. A positive value indicates an increase relative to baseline period, whereas negative ones indicates a decrease. For 10-year return period (Figure 22), under stationarity assumption (Figure 22; *top panel*), the statistically significant increase in design storm is observed for sub-hourly rainfall extremes in Hamilton and Oshawa and hourly precipitation extreme in Toronto International Airport. On the other hand, assuming non-stationarity (Figure 20; *bottom panel*), a significant increase in rainfall intensity is observed for 6-hour extremes in Toronto, Oshawa and Windsor and 24-hour extreme in Toronto and Trenton respectively. A significant decrease in intensity is observed for London. In particular, the projected increase ranges from 1 ~ 40% and 1 ~ 38% respectively, assuming stationarity and non-stationarity methods of frequency analysis. For 25-year return period (Figure S11), the performance is mixed, with a few stations exhibit an increase (for instance, 6- and 24-hour rainfall extreme in Toronto, 6-hour extreme in Oshawa) while the others show a decrease (*i.e.*, 12-hour rainfall extreme in Toronto and Trenton, 1- and 2-hour hour extremes in London and Fergus Shand dam respectively). However, considering stationarity assumption, none of these changes are significant (Figure S11; *top panel*) in contrast to nonstationarity assumption (Figure S11; *bottom panel*), in which magnitude of changes are deemed to be significant. For 50-year return period (Figure 23), assuming non-stationarity, we found a statistically significant decrease in return period for sub-hourly rainfall extremes in Toronto, London, Trenton and Fergus Shand. However, the changes are not significant assuming stationarity in return period estimate. Considering stationarity, the percentage decrease in rainfall intensity ranges between 2 – 37%, while it varies from 2 to 46% across different locations and durations. Taken together, the following broader insights emerge for future precipitation extremes in Southern Ontario: (i) For 1 in 10-year events, an increase in rainfall intensity is observed across most of the stations and durations, irrespective of the choice of frequency analysis. (ii) For more extreme events (*i.e.*, 1 in 25-year and 1 in 50-year), a decrease in rainfall intensity is noted; however, the magnitude of changes in storm intensity are deemed to be statistically significant considering non-stationarity assumption in contrast to the stationary assumptions.

## 5. Discussion and Conclusions

Extreme precipitation poses significant risks for cities across Southern Ontario, Canada due to their dense population, valuable and geographically concentrated property, complex and interdependent infrastructure networks [*Henstra and Thistlethwaite*, 2017]. We characterize extreme rainfall frequency over Southern Ontario using IDF statistics with a suite of high-resolution new generation of coordinated regional climate modeling experiments participating in the NA-CORDEX domain with an elevated global greenhouse forcing. Since the study area comprises urban catchments and most of the municipal infrastructures are designed based on design storm intensity of 50-year or less, the IDF statistics are compared for 2-, 5-, 10-, 25-, and 50-year return periods. The RCM performance in simulating extreme rainfall statistics in the present day climate is validated using ground-based AM observational records. The temporally downscaled NA-CORDEX RCM ensembles are able to simulate the observed IDF statistics reasonably well as demonstrated by various performance metrics. Further, we modeled design rainfall intensity using extreme value distributions under stationarity and nonstationarity assumptions with a time-varying location parameter.

The following conclusions emerge from the present analysis:

- The ground-based short-duration historical (1970 – 2010) precipitation extremes often exhibit a statistically significant increase (or decrease) trends for the selected locations across Southern Ontario. On the other hand, extreme daily temperature anomaly shows an upward trend for most of the locations, although the apparent trend is not statistically significant across the stations.
- A robust bias-correction methodology is employed to correct systematic biases in the RCM outputs for different time scales and across individual station locations. However, the skill of bias correction does not follow any specific trend either based on locations or the durations.
- Statistically significant increase in rainfall intensity is observed for 10-year events over most of the station locations. However, opposite trend is noted for events with longer return periods.

The paper presents a proof of principal results to compare changes in extreme precipitation intensity in the near-future using three regional climate models with the future atmosphere represented by the RCP8.5 scenario. While Canada's national climate change assessment indicates an increase in average surface air temperature and precipitation volume across several regions nationally [*Bush et al.*, 2014]; still most of the ground-based individual station observations do not exhibit any significant trend. Further, typically a limited number of stations show a statistically significant nonstationary trend at a sub-daily and daily time scales [*Shephard et al.*, 2014]. Recently, using station-based AM precipitation [*Shephard et al.*, 2014] have shown that nationwide less than 5% of the stations exhibits significant increasing (or decreasing) trends. Based on their findings the authors recommend that in Canada, the traditional IDF design based on stationarity assumption of precipitation extreme has not been violated. Further, in projected scenario, we find an increase in rainfall intensity at 6- and 24-hour extremes for 10-year events over majority of stations. Our results are consistent with [*Barbero et al.*, 2017], where authors reported except in winter months, there is a significant increase in daily AM precipitation trends as compared to the hourly extremes after analyzing a large number of ground-based observational records across the United States for the present day (1950-2011) condition. Their results indicate that at a station level, the trends in daily precipitation extremes are better detected than that of the hourly extremes in a changing climate. They attributed this to the limited spatial extent of the most extreme events and the inability of sparse station network to accurately measure such events [*Barbero et al.*, 2017].

Some of the caveats of the study include: The heavy computational requirement involving successive spatiotemporal downscaling of extreme precipitation at individual station level has limited us to use only three RCMs. However, uncertainty in climate change projection results from both model responses and internal climate variability [*Hawkins and Sutton*, 2009, 2011]. A series of previous studies [*Deser et al.*, 2012, 2014; *Wettstein and Deser*, 2014; *Kay et al.*, 2015] have examined the effect of intrinsic climate variability and anthropogenic climate change on projected climate using a 40-member climate model ensembles available from National Center for Atmospheric Research (NCAR)'s CESM Large Ensemble Community Project (LENS). These studies explored various aspects of uncertainty resulting from intrinsic climate variability and found substantial climate uncertainty on a global and regional scale. The role of internal climate variability is considerably higher at a regional scale and in a shorter lead time (in a near-term planning horizon) [Figure 4 in *Hawkins and Sutton*, 2009; *Hawkins and Sutton* 2012]. Therefore, using three-member RCMs may not be sufficient to capture the full spectrum of climate variability in a near-term planning period. Next, hourly and sub-daily extreme precipitation, required for engineering design, are often produced by convective events. However, global and regional climate models are not able to simulate such events well because of models' limited spatial and temporal resolution, and convection is not explicitly resolved in these models [*Lenderink and Fowler*, 2017; *Zhang et al.*, 2017]. Although high resolution convective permitting models may provide a reliable representation of local storm dynamics [*Kendon et al.*, 2014], these models still suffer from inherent uncertainties in simulating extremes [*Hagos et al.*, 2014; *Singh and O'Gorman*, 2014].

Secondly, the cascade of uncertainty may also stem from temporal downscaling of RCM output. Third, we assume that the selected bias-correction method at historical period is time independent and same will holds good for correcting bias in the future period. If global warming is solely responsible for changes in extreme precipitation trends, we might expect to see continued increase in extreme precipitation intensity. On the other hand, if large-scale climate variability and/or anthropogenic influences are responsible, the extremes (for instance, July, 2013, extreme storm event caused insured losses of over $850 million [*Canadian Underwriter*, 2013]) may re-occur in the future leading to wide scale damages to life and property. As a final caveat, inferring extreme precipitation frequency solely based on ground-based observation has considerable uncertainty, at least as much as in the estimated projected trends in the future emission scenario. Although we examined sources of uncertainty in model projection by comparing the multi-model spread of future IDF curves (Figures 20 – 21; S9 – S10) in RCM ensembles in the future, there is a need to explore a complete suite of RCM outputs [*Mote et al.*, 2016] archived at the NA-CORDEX domain in developing probability-based IDF curves considering a range of plausible scenarios for risk-based resilient water management [*Hall and Borgomeo*, 2013].

**Acknowledgement**

The annual maximum rainfall data used in this study is downloaded from Environment Canada website: ftp://ftp.tor.ec.gc.ca/Pub/Engineering_Climate_Dataset/IDF/. Hourly and daily rainfall data are obtained from Toronto and Region Conservation Authority (TRCA; https://trca.ca/) and Environment Canada Historical Climate Data Archive (http://climate.weather.gc.ca/). The authors would like to acknowledge financial support from the Natural Science and Engineering Research Council (NSERC) of Canada and the NSERC Canadian FloodNet (Grant Number: NETGP 451456). The first author of the manuscript would like to thank Dr. Jonas Olsson of Swedish Meteorological and Hydrological Institute (SMHI), Norrköping, Sweden for sharing MATLAB-based random cascade disaggregation tools and implementation details through email. The nonstationary GEV analyses were performed using MATLAB-based toolbox NEVA, available at the University of California, Irvine website: http://amir.eng.uci.edu/neva.php (as accessed on May 2016).

**List of Main Figures**

**Figure 1.** Time series of (a) 1- (*top panel*) and (b) 24- (*bottom panel*) hour annual maxima precipitation before bias-correction in 8 selected stations across Southern Ontario. The black lines indicate observed AM precipitation and the blue multi-model median NA-CORDEX simulated precipitation. The shaded bounds in blue indicate the minimum and the maximum RCM simulated precipitation extremes.

**Figure 2.** Taylor diagrams to evaluate the skill of bias-corrected 1-hour AM regional climate model ensembles relative to observation during baseline (1970 – 2010) period. Symbols represent the method of bias-correction (red: Kernel-based and black: GEV-based). The redial distance is the spatial standard deviation ($\sigma_s$) normalized by the observations and is a measure of spatial variation. $\sigma_s$ contours from the origin are shown in black. Contours showing the RMSE difference between models and the observation are shown in green. The cosine of the azimuthal angle is given by the centered pattern correlation (*r*) between the observation and the RCM.

**Figure 3.** Same as Figure 2 but for 24-hour AM precipitation extreme.

**Figure 4.** Heat maps of the skill scores for bias-corrected AM precipitation relative to observation for different durations during baseline (1970 – 2010) period. The top six rows show kernel-based bias correction while the bottom six rows indicate GEV-based bias-correction. The dark shade indicates higher skills, whereas the lighter ones show, the lower skill score.

**Figure 5.** PDF of observed versus multi-model median AM precipitation at 1-hour duration in eight station locations. The minimum and the maximum bounds are shown using dotted blue and red lines respectively.

**Figure 6.** Same as Figure 5 but for 24-hour precipitation extreme.

**Figure 7.** Same as Figure 1 but after bias correction.

**Figure 8.** The distribution of shape parameters (Y-axes) of the simulated GEV distribution at different durations (x-axes) for the present day climate (1970-2010) assuming stationary GEV models. The box-plot in blue indicates the observed distribution and in red represent multi-model median NA-CORDEX RCM simulated distribution. The horizontal segment inside the box shows the median, and the whiskers above and below the box show the locations of the minimum and maximum. The span of the boxes represents the interquartile range or the variability in the distribution.

**Figure 9.** Same as Figure 8 but for nonstationary GEV models.

**Figure 10.** Comprehensive comparison of nonstationary versus stationary IDF distributions of observed versus RCM simulated precipitation extremes for different return periods at Toronto International Airport.

The boxplots in darker shade indicate observed extreme modeled using stationary (red) and nonstationary (blue) GEV models, whereas the one in lighter shade shows NA-CORDEX simulated extreme (stationary: green and orange: nonstationary). The vertical span of the boxes in the boxplot represents uncertainty in the estimated design storm. The legend applies to all figure panels.

**Figure 11.** Same as Figure 10 but for Oshawa WPCP.

**Figure 12.** Same as Figure 11 but for Windsor Airport.

**Figure 13.** The relative difference in design storm estimates of observed versus multi-model median design storm in all station locations at 1- and 24-hour storm durations corresponding to 10- and 50-year return periods during present day climate modelled using (a) stationary (top panel) and (b) nonstationary (bottom panel) GEV models.

**Figure 14.** Return periods versus design storm estimates of present day 1-hour precipitation extreme modelled using (a) stationary and (b) nonstationary GEV models. Colored-boxes express the inter-model variability represented by the multi-model minimum and maximum of RCM simulations (best-worse case). The subplots are shown for the 8 stations across Southern Ontario.

**Figure 15.** Same as Figure 14 but for 24-hour precipitation extreme.

**Figure 16.** The distribution of shape parameters of the simulated GEV distribution at different durations for the future period (2030-2070) assuming stationary GEV models. The box-plot in blue indicates the observed distribution and in red represents multi-model median NA-CORDEX RCM simulated distribution.

**Figure 17.** Same as Figure 16 but for nonstationary GEV models.

**Figure 18.** Nonstationary present-day (in blue) and Projected (in red; 2030-2070) IDF curves for 10-year return period across eight stations in Southern Ontario. The uncertainty in IDF simulation at different duration is expressed using boxplots. The minimum (in green) and maximum (in red) bounds in IDF curves are shown to express the best and worst plausible scenarios.

**Figure 19.** Nonstationary present-day and Projected IDF curves for 50-year return period.

**Figure 20.** Design storm intensity of future scenario versus present day climate for 10-year return period. The variability in the differences in design storm intensity are shown using box-plots representing three different cases: Stationary (2030-2070; or 2050s) versus stationary (1970 – 2010; or 1990s) [in purple]; Nonstationary (2050s) versus stationary (1990s) [in blue]; Nonstationary (2050s) versus nonstationary (1990s) [in orange].

**Figure 21.** Same as Figure 20 but for 50-year return period.

**Figure 22.** Projected changes in rainfall intensity for 10-year return period in three different cases: stationary (2050s) versus stationary (1990s) [*top panel*]; Nonstationary (2050s) versus stationary (1990s) [*middle panel*]; Nonstationary (2050s) versus nonstationary (1990s) [*bottom panel*]. The comparative assessment is performed between projected storm intensity modeled using multi-model median NA-CORDEX RCM ensemble and observed baseline intensity. The shades of the changes express high and low end, with a dark red indicating increase in storm intensity while dark blue show decrease in the intensity. Very small changes [*i.e.,* in and around zero values] are marked with white.

**Figure 23.** Same as in Figure 21 but for 50-year return period.

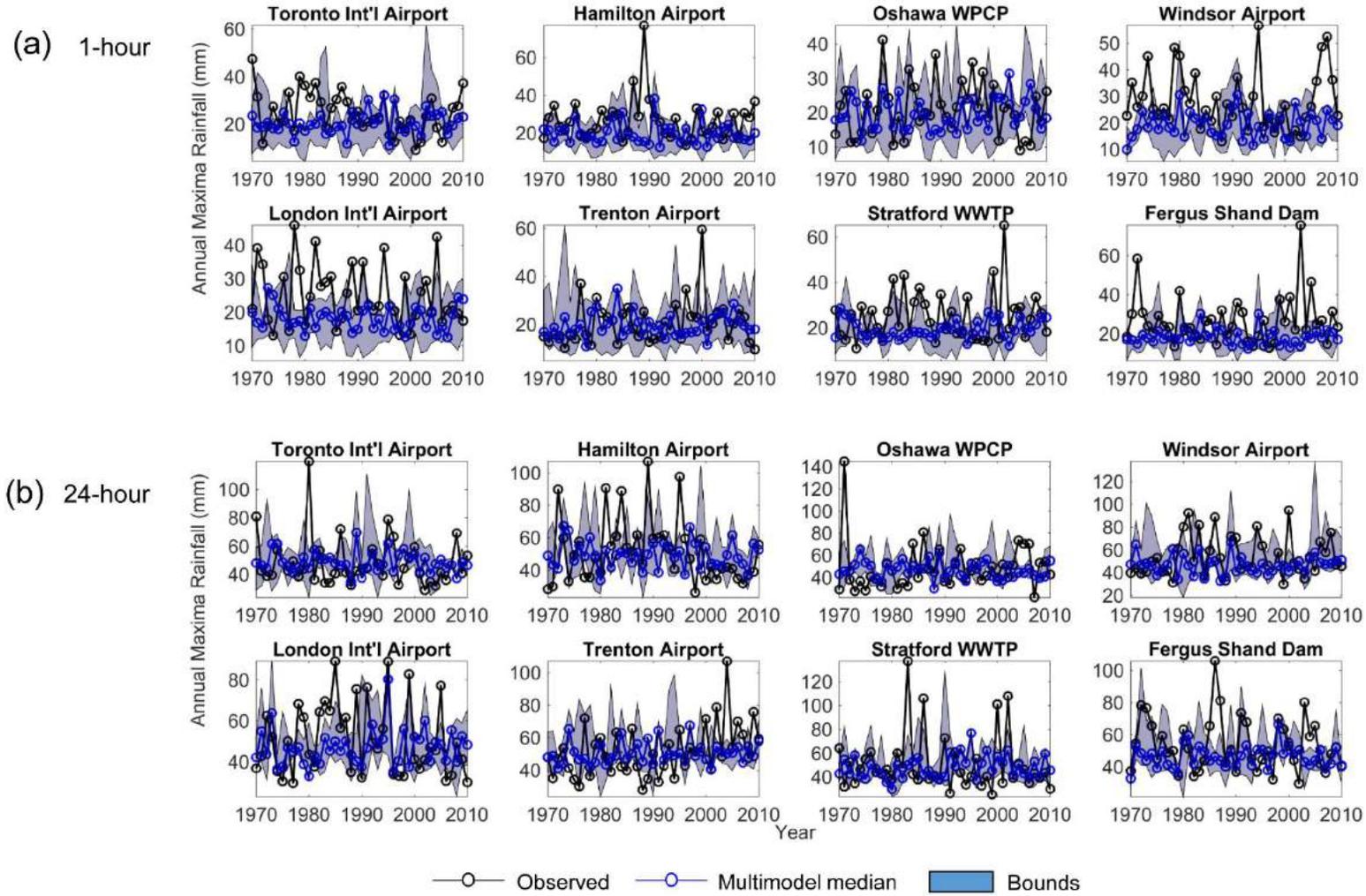

**Figure 1.** Time series of (a) 1- (*top panel*) and (b) 24- (*bottom panel*) hour annual maxima precipitation before bias-correction in 8 selected stations across Southern Ontario. The black lines indicate observed AM precipitation and the blue multi-model median NA-CORDEX simulated precipitation. The shaded bounds in blue indicate the minimum and the maximum RCM simulated precipitation extremes.

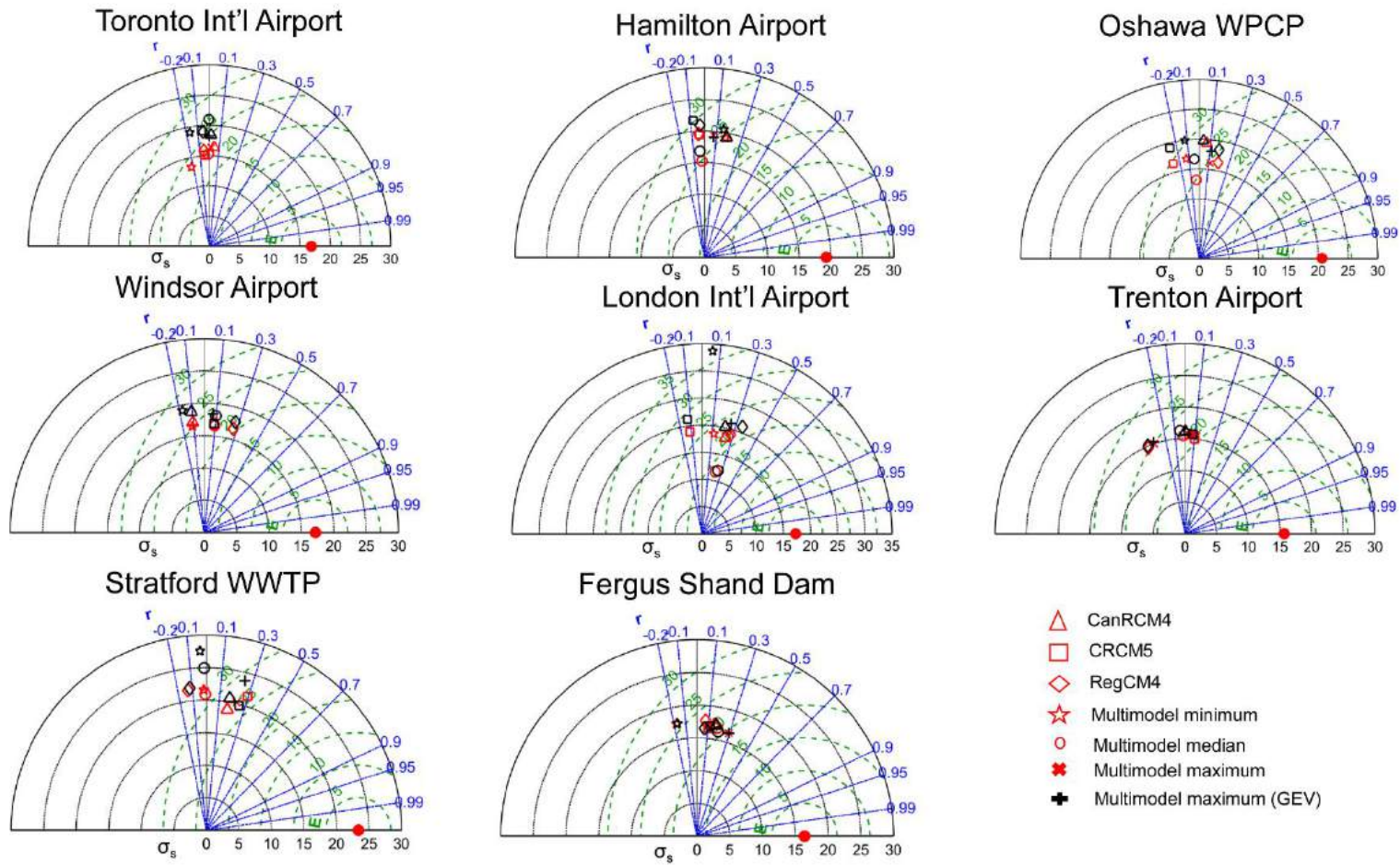

**Figure 2.** Taylor diagrams to evaluate the skill of bias-corrected 1-hour AM regional climate model ensembles relative to observation during baseline (1970 – 2010) period. Symbols represent the method of bias-correction (red: Kernel-based and black: GEV-based). The redial distance is the spatial standard deviation ($\sigma_s$) normalized by the observations and is a measure of spatial variation. $\sigma_s$ contours from the origin are shown in black. Contours showing the RMSE difference between models and the observation are shown in green. The cosine of the azimuthal angle is given by the centered pattern correlation (*r*) between the observation and the RCM.

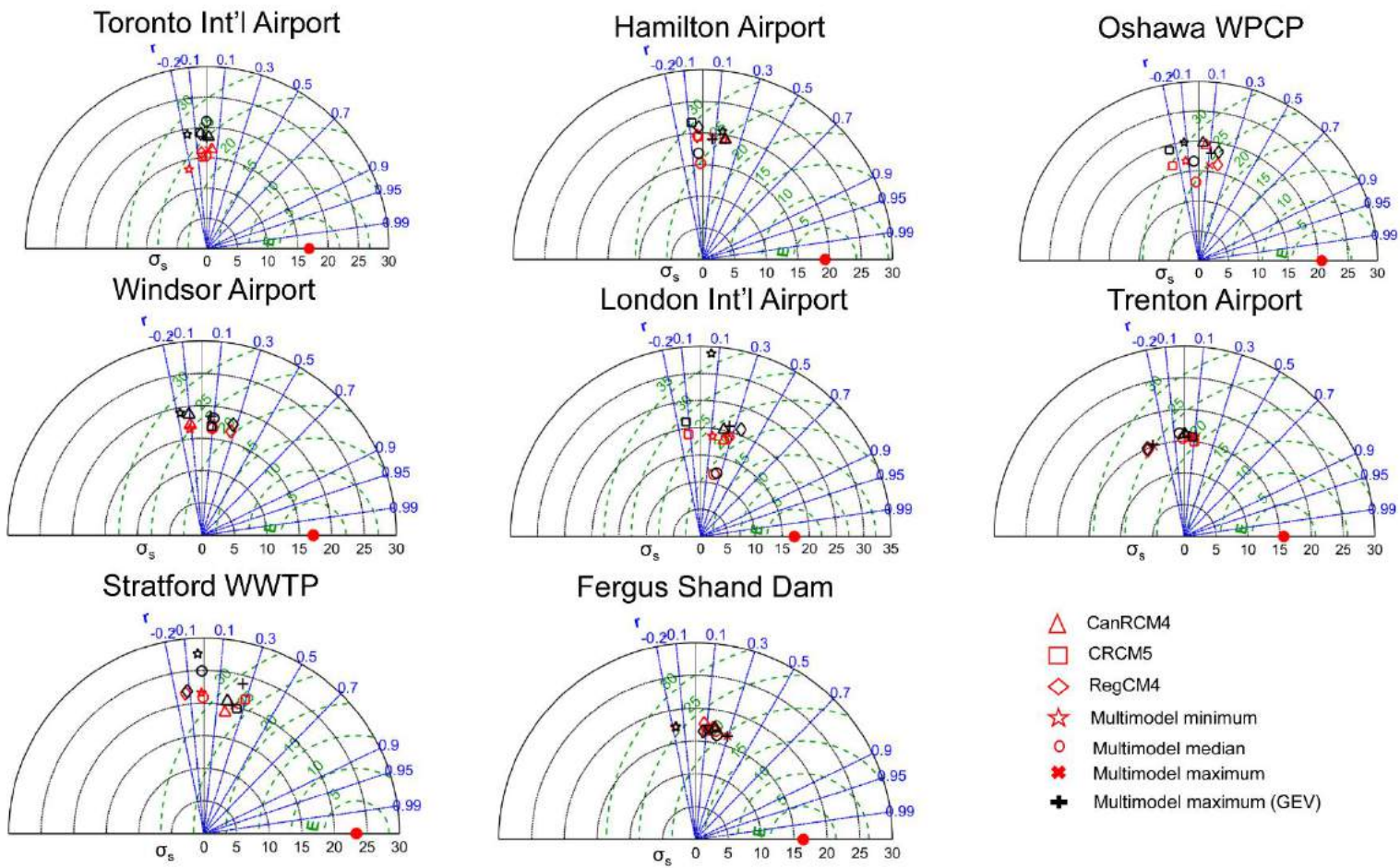

**Figure 3.** Same as Figure 2 but for 24-hour AM precipitation extreme.

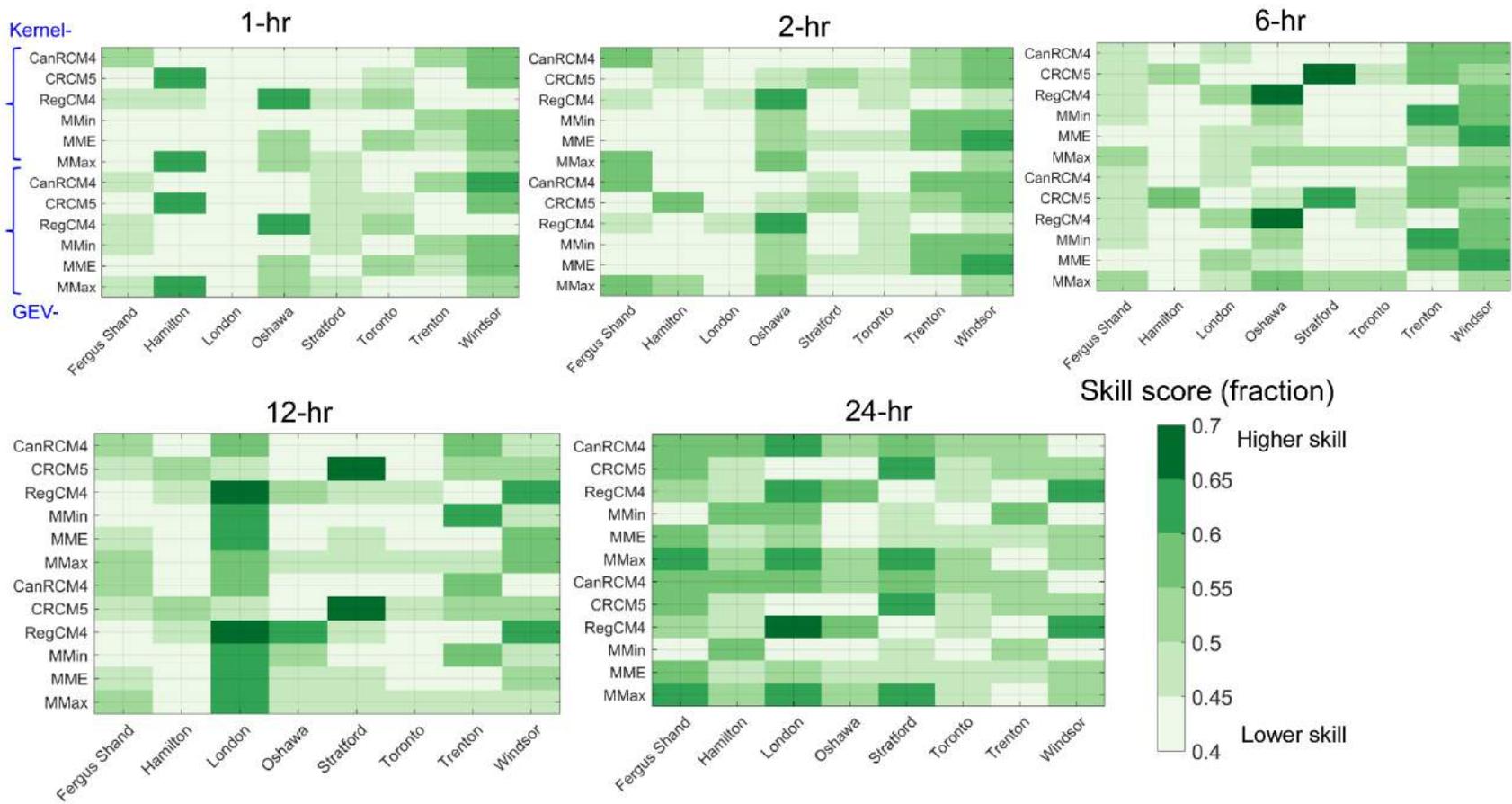

**Figure 4.** Heat maps of the skill scores for bias-corrected AM precipitation relative to observation for different durations during baseline (1970 – 2010) period. The top six rows show kernel-based bias correction while the bottom six rows indicate GEV-based bias-correction. The dark shade indicates higher skills, whereas the lighter ones show, the lower skill score.

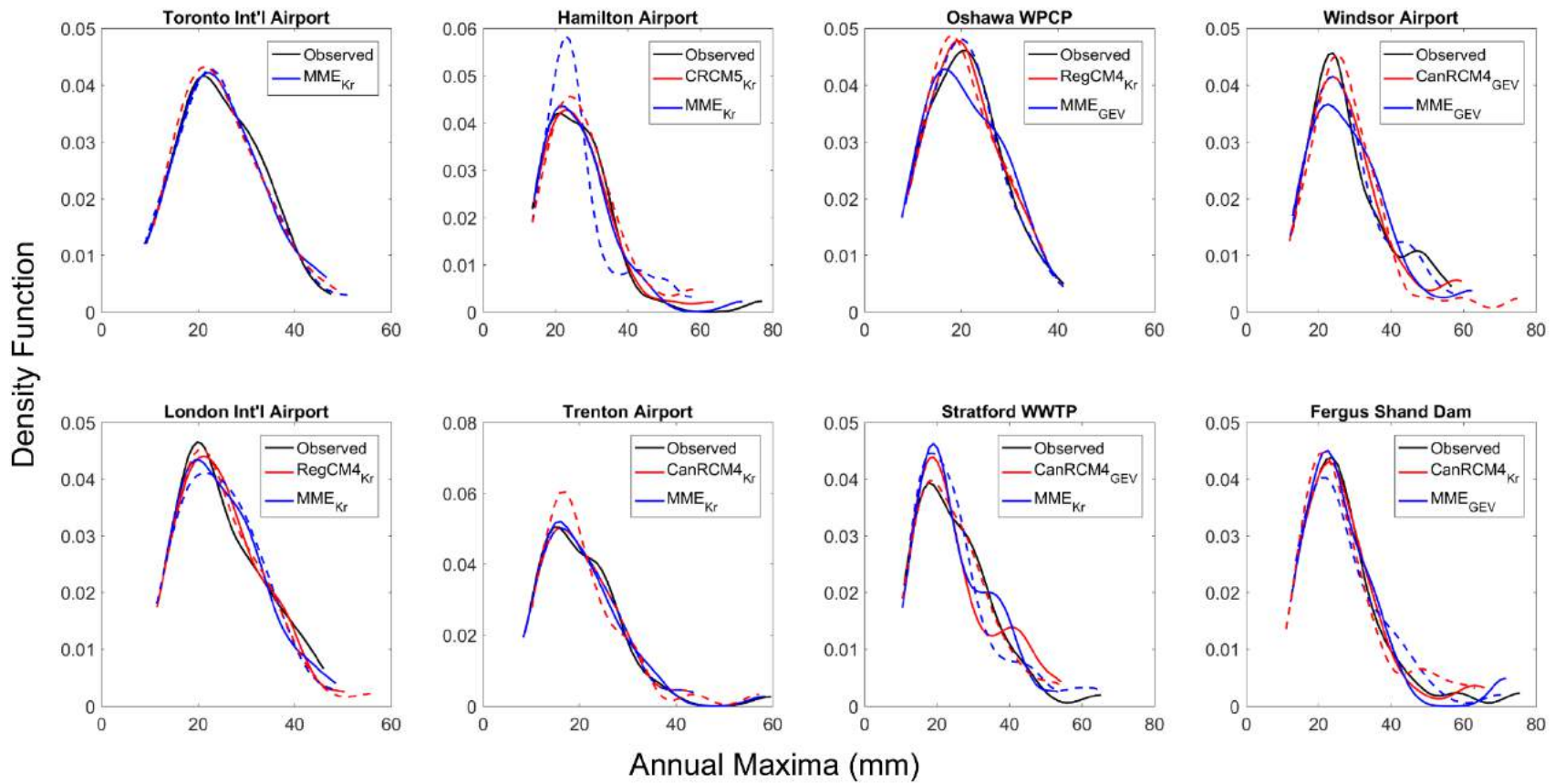

**Figure 5.** PDF of observed versus multi-model median AM precipitation at 1-hour duration in eight station locations. The minimum and the maximum bounds are shown using dotted blue and red lines respectively.

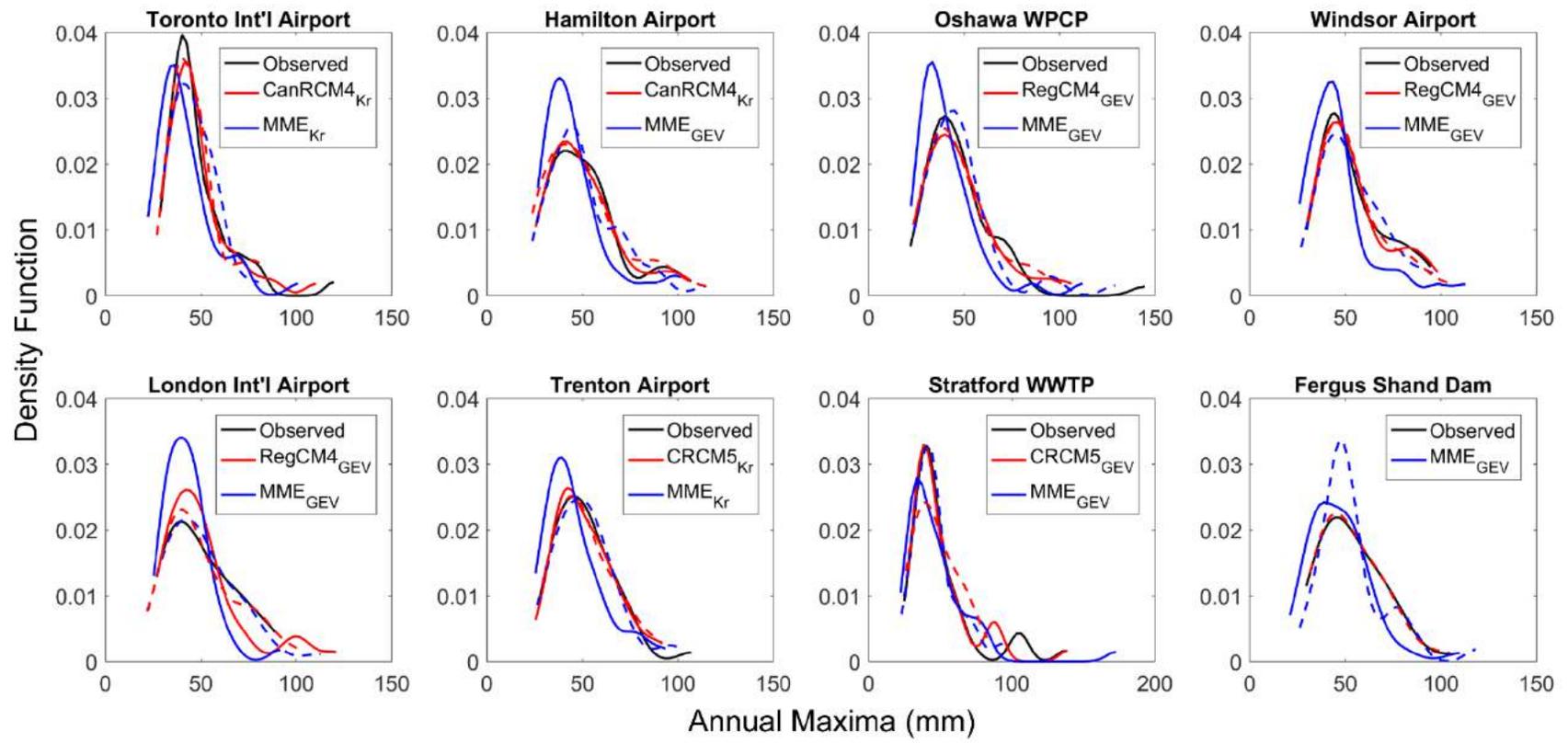

**Figure 6.** Same as Figure 5 but for 24-hour precipitation extreme.

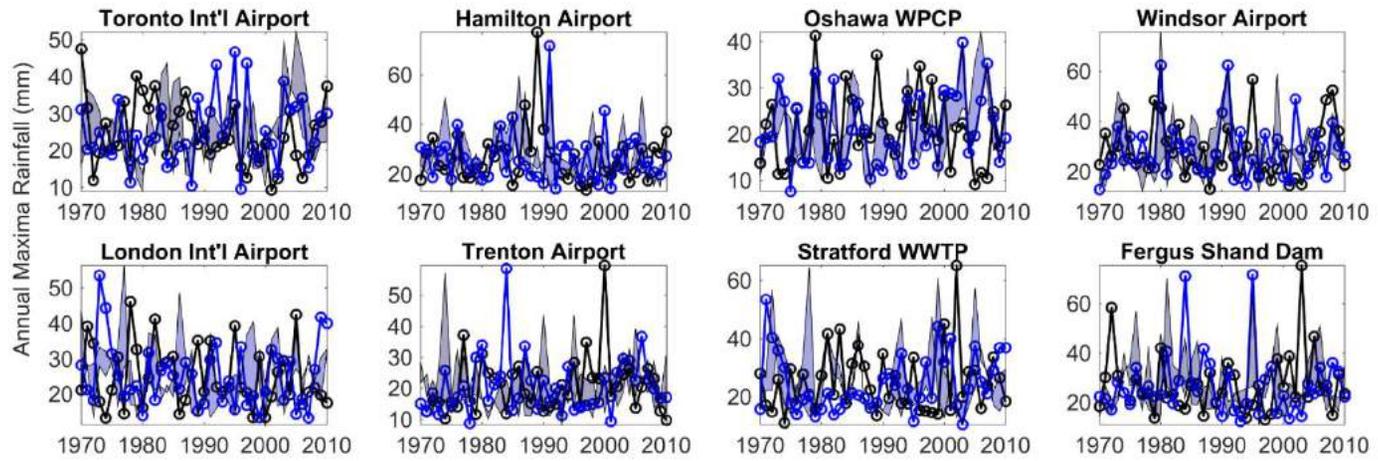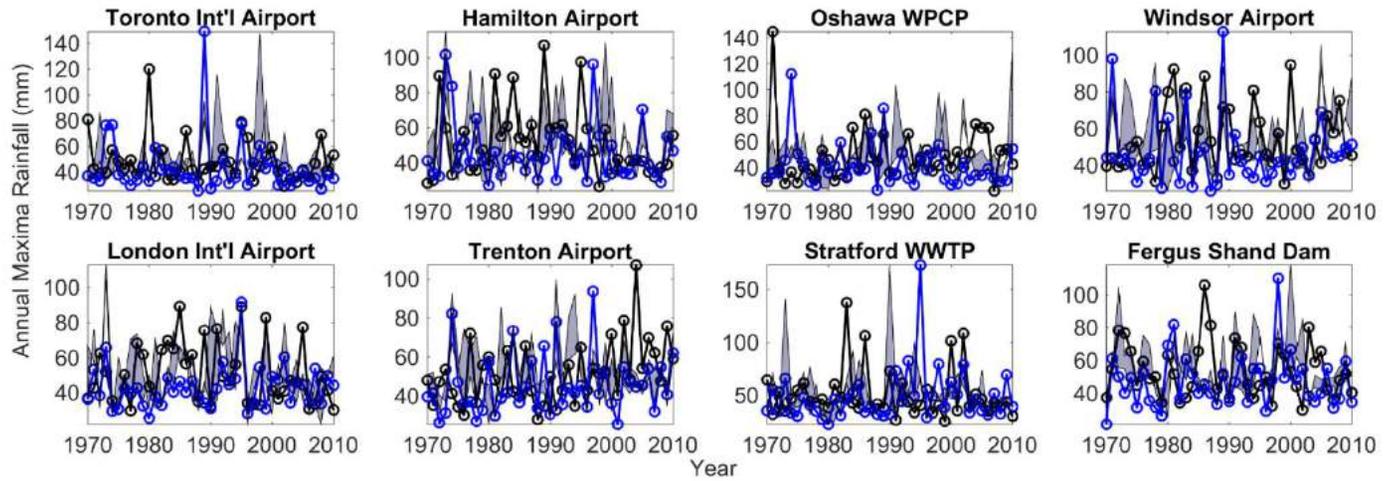

**Figure 7.** Same as Figure 1 but after bias correction.

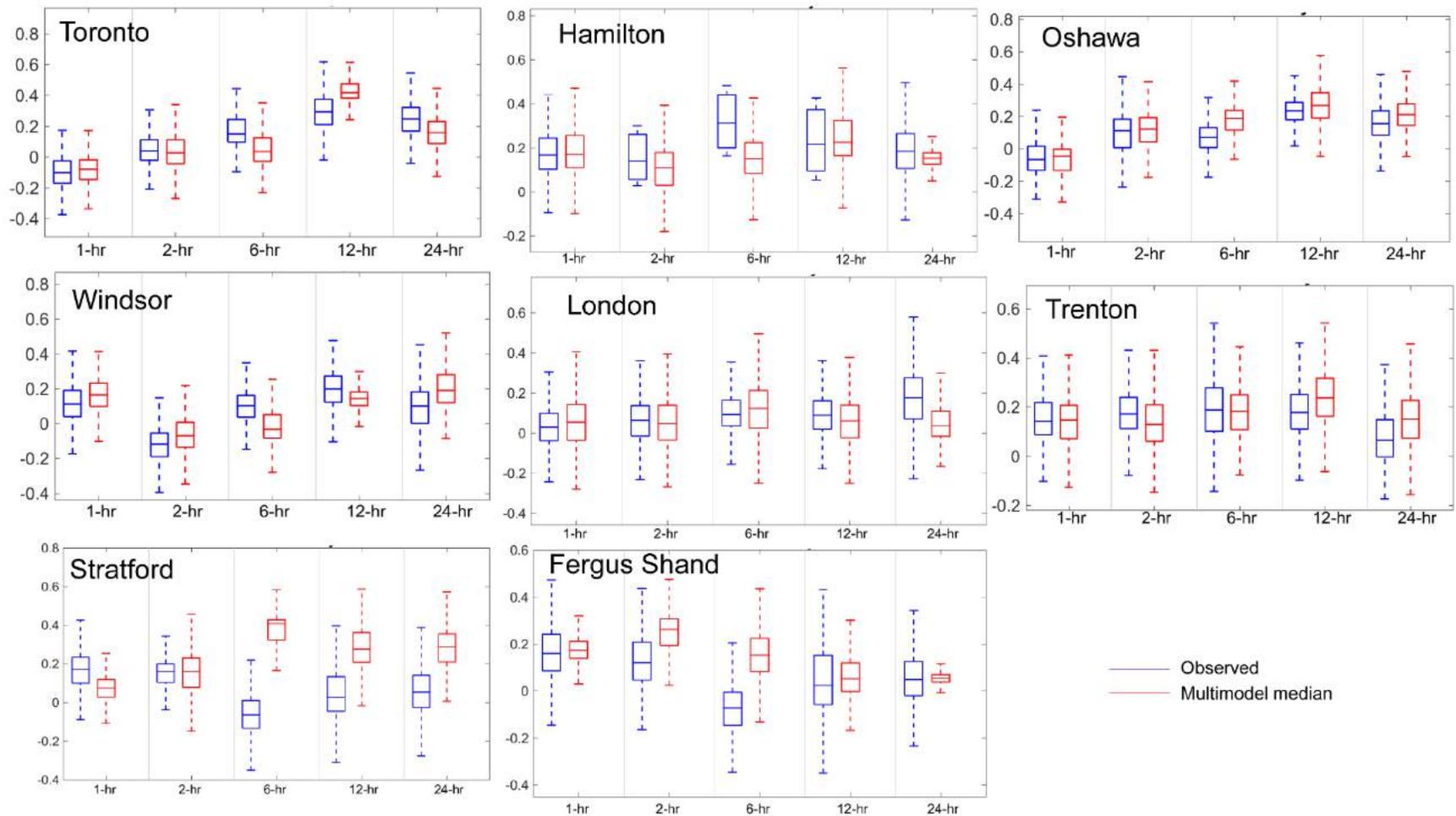

**Figure 8.** The distribution of shape parameters (Y-axes) of the simulated GEV distribution at different durations (x-axes) for the present day climate (1970-2010) assuming stationary GEV models. The box-plot in blue indicates the observed distribution and in red represent multi-model median NA-CORDEX RCM simulated distribution. The horizontal segment inside the box shows the median, and the whiskers above and below the box show the locations of the minimum and maximum. The span of the boxes represents the interquartile range or the variability in the distribution.

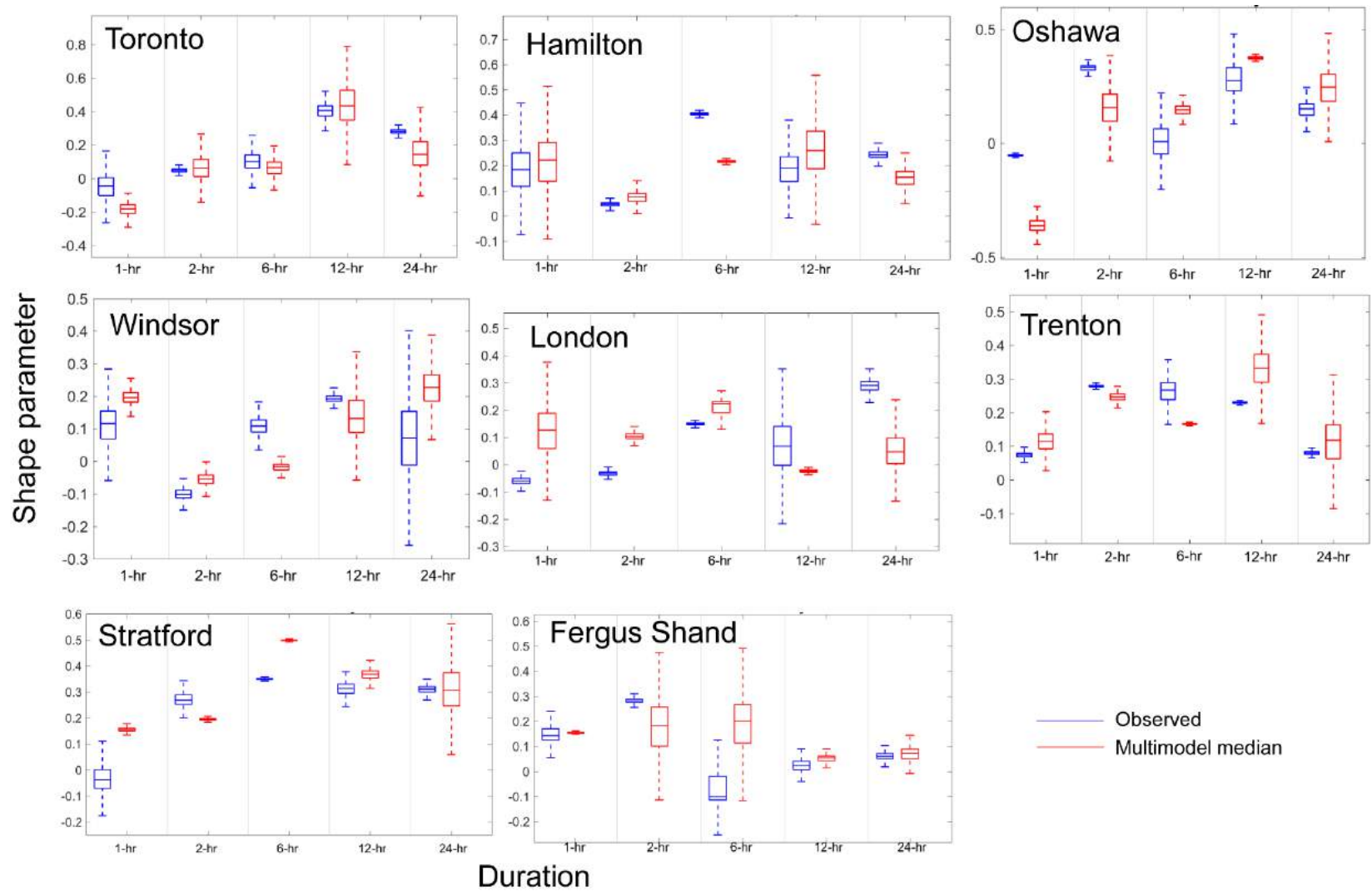

**Figure 9.** Same as Figure 8 but for nonstationary GEV models

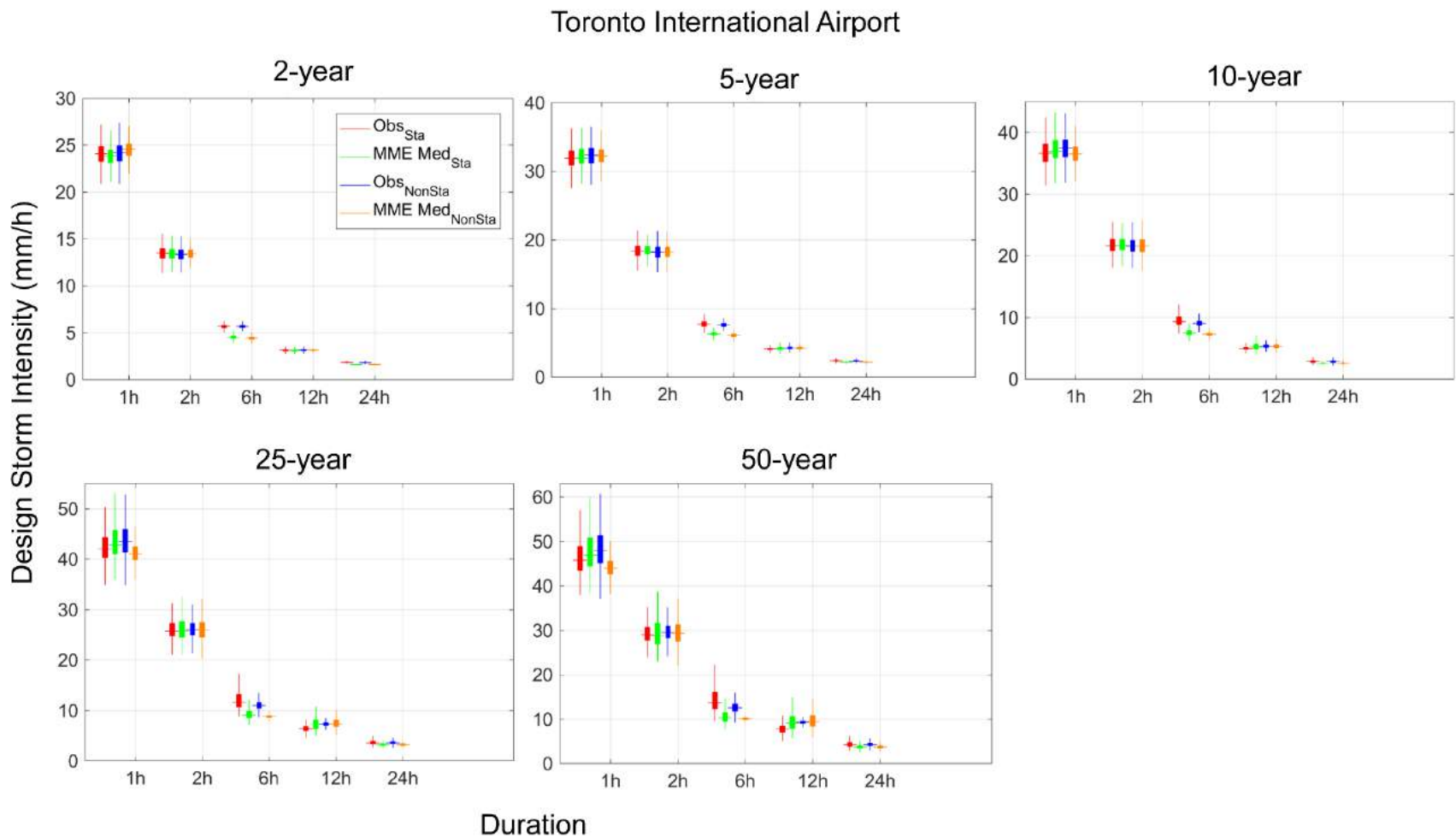

**Figure 10.** Comprehensive comparison of nonstationary versus stationary IDF distributions of observed versus RCM simulated precipitation extremes for different return periods at Toronto International Airport. The boxplots in darker shade indicate observed extreme modeled using stationary (red) and nonstationary (blue) GEV models, whereas the one in lighter shade shows NA-CORDEX simulated extreme (stationary: green and orange: nonstationary). The vertical span of the boxes in the boxplot represents uncertainty in the estimated design storm. The legend applies to all figure panels.

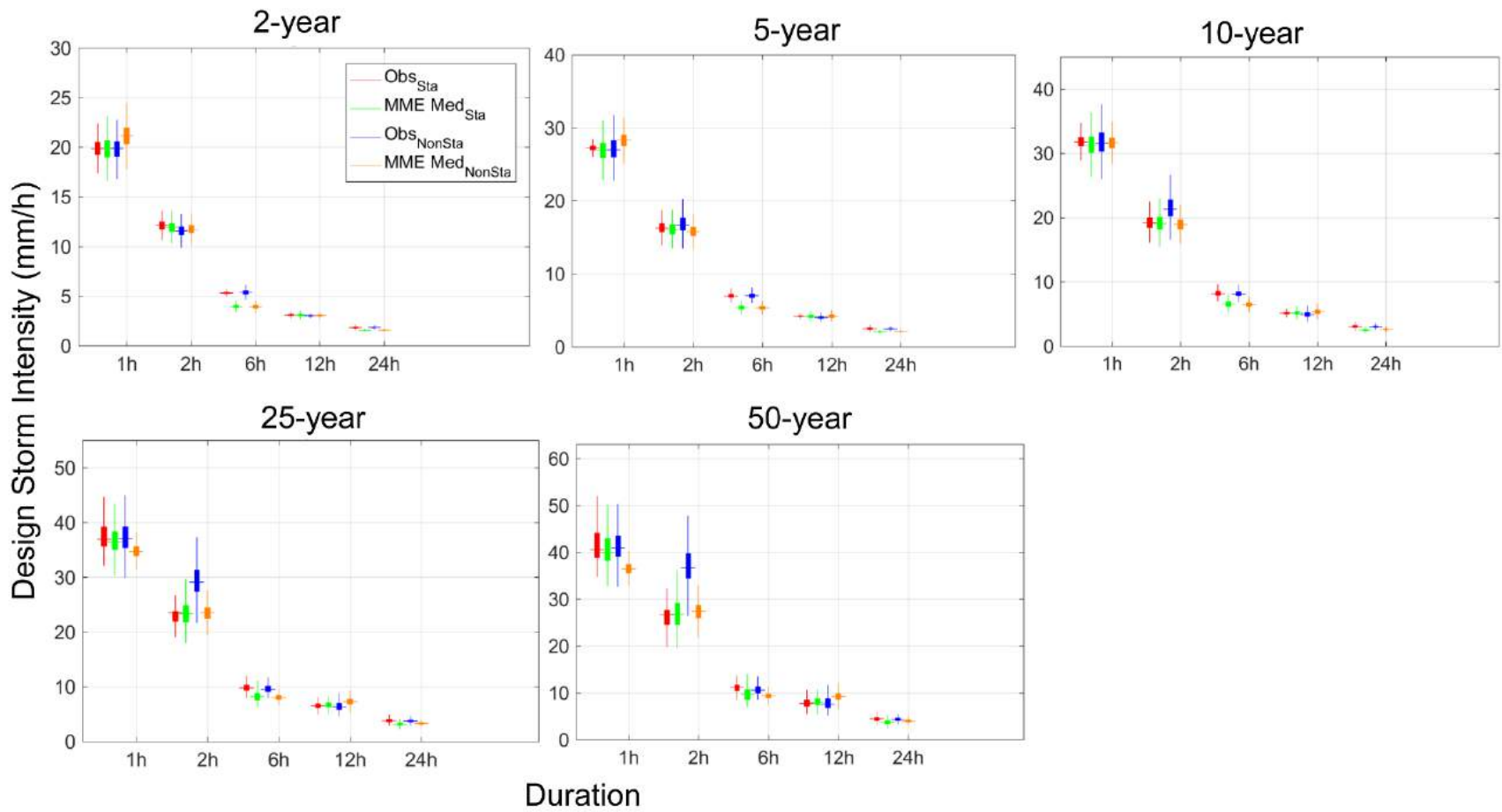

**Figure 11.** Same as Figure 10 but for Oshawa WPCP.

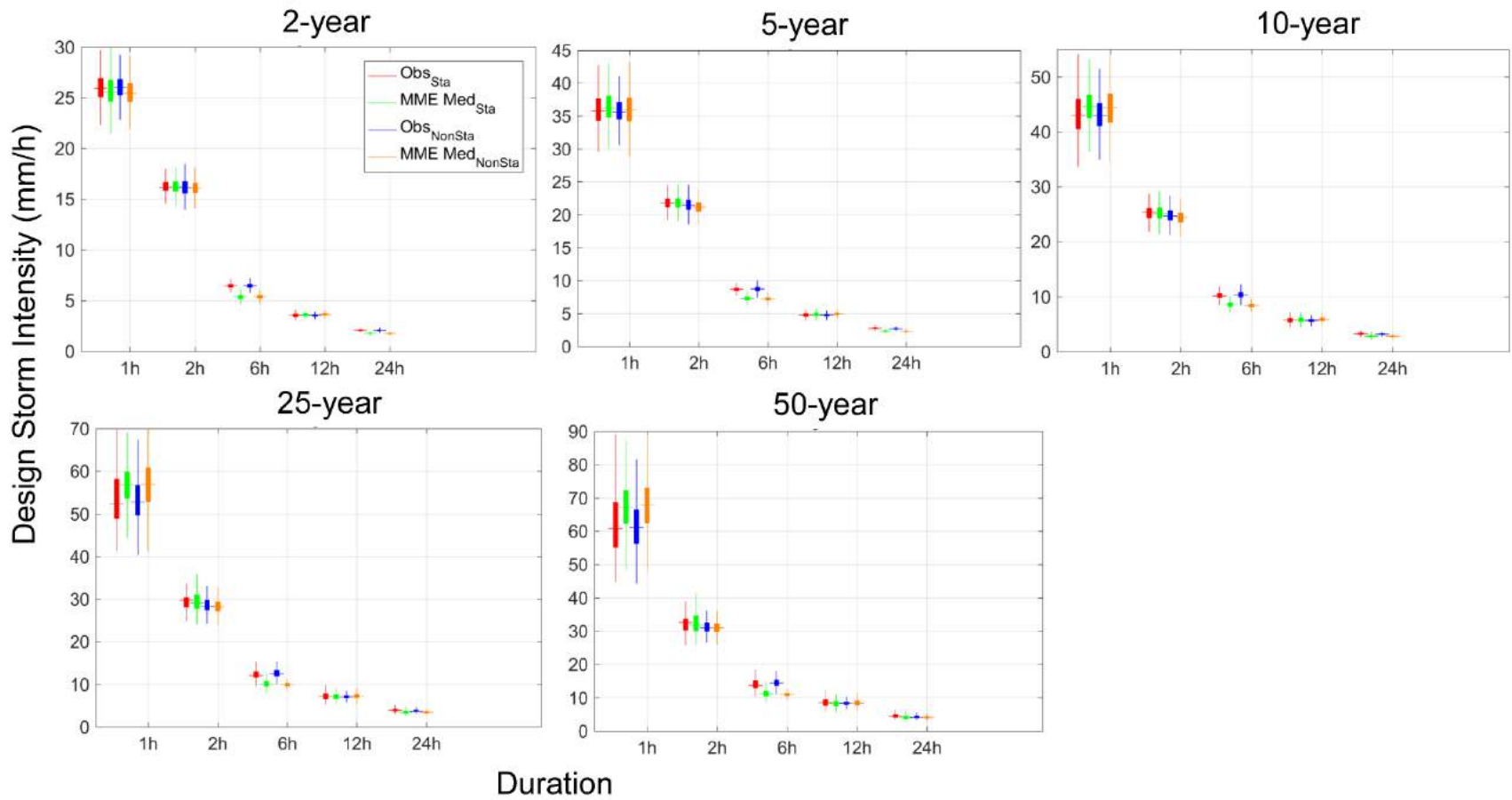

**Figure 12.** Same as Figure 11 but for Windsor Airport.

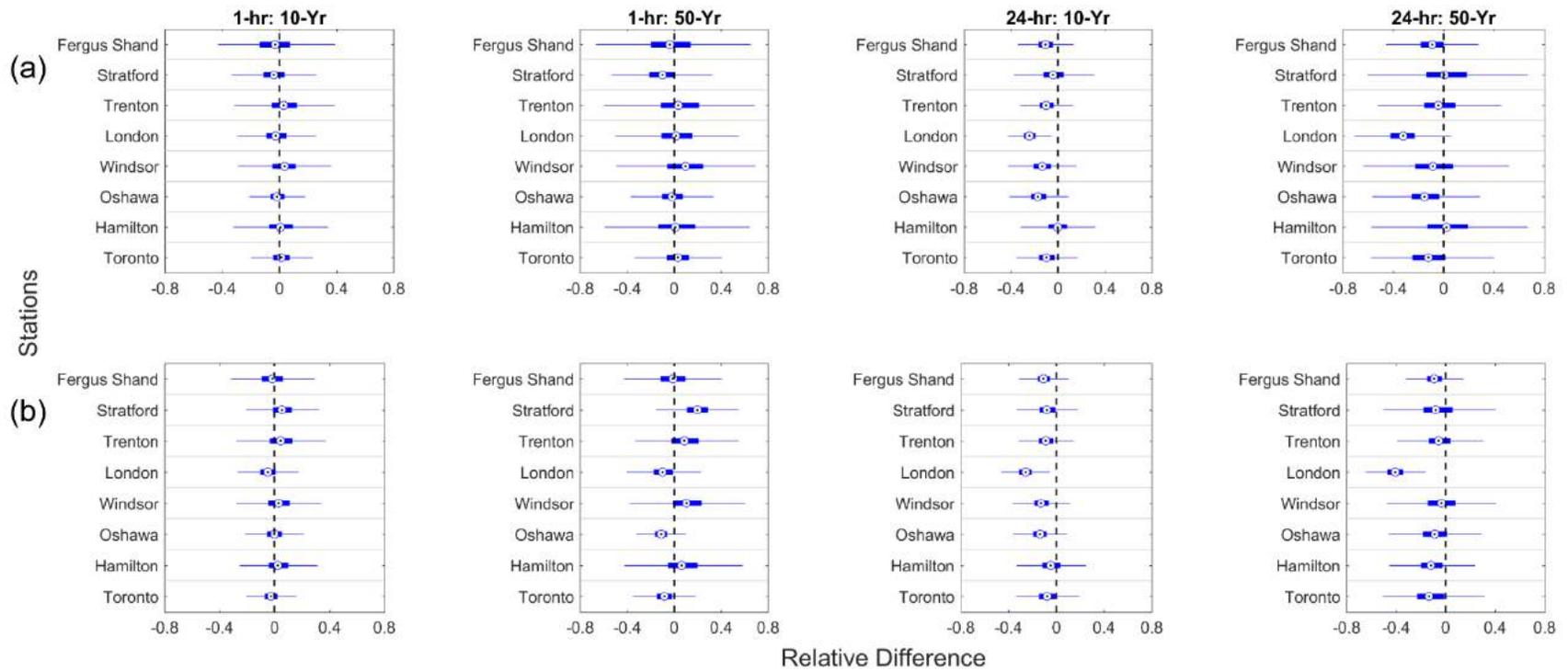

**Figure 13.** The relative difference in design storm estimates of observed versus multi-model median design storm in all station locations at 1- and 24-hour storm durations corresponding to 10- and 50-year return periods during present day climate modelled using (a) stationary (top panel) and (b) nonstationary (bottom panel) GEV models.

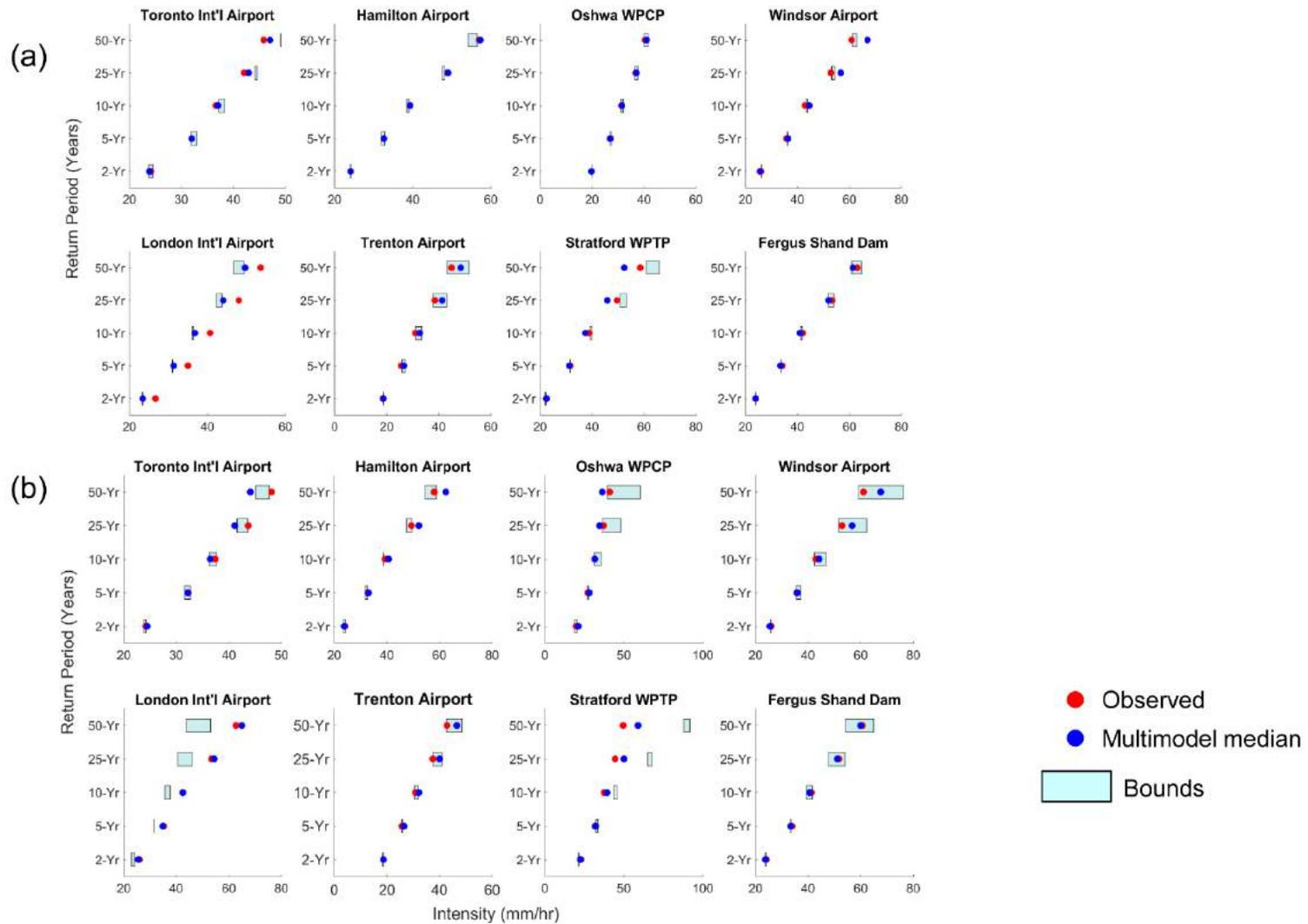

**Figure 14.** Return periods versus design storm estimates of present day 1-hour precipitation extreme modelled using (a) stationary and (b) nonstationary GEV models. Colored-boxes express the inter-model variability represented by the multi-model minimum and maximum of RCM simulations (best-worse case). The subplots are shown for the 8 stations across Southern Ontario.

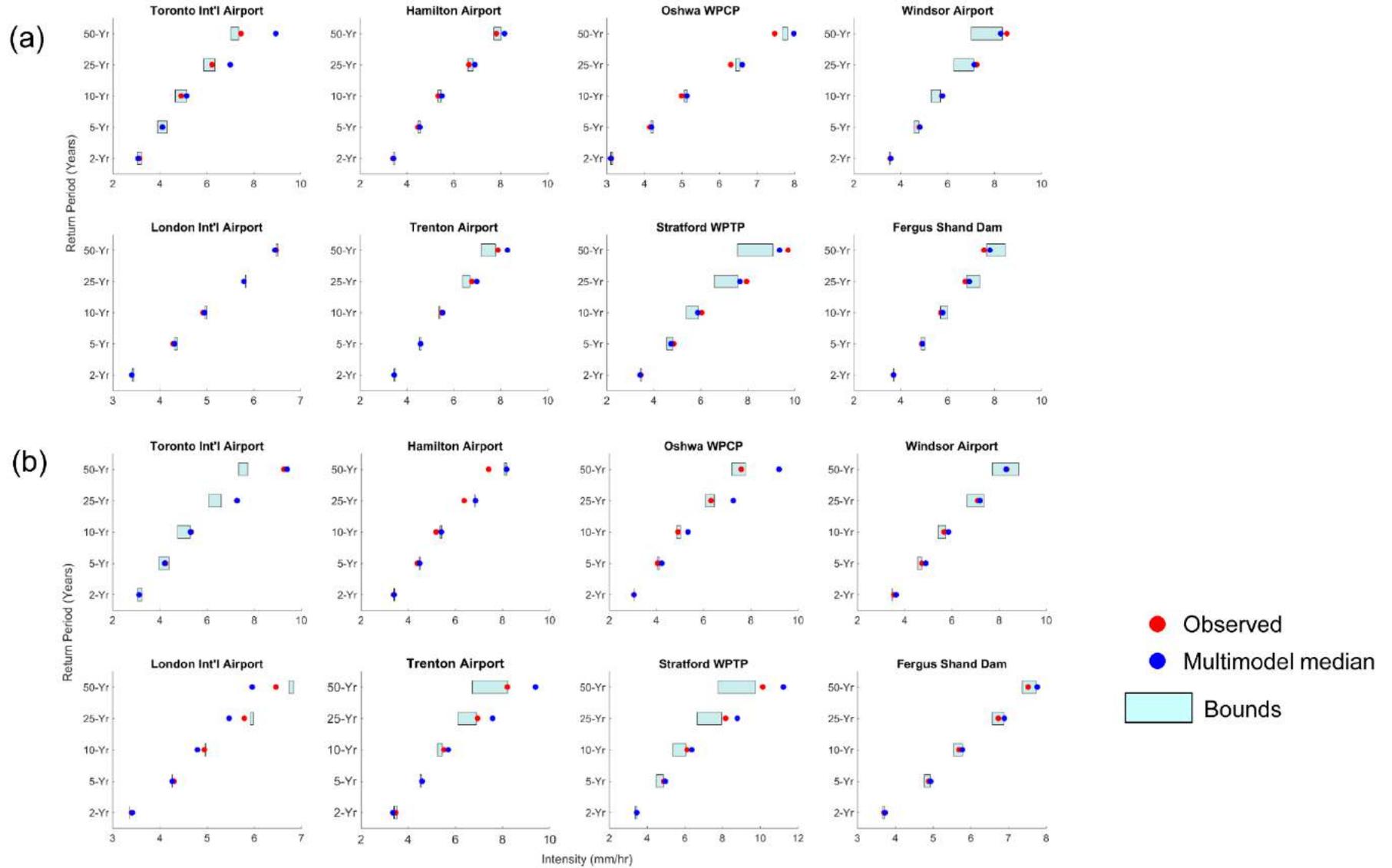

**Figure 15.** Same as Figure 14 but for 24-hour precipitation extreme.

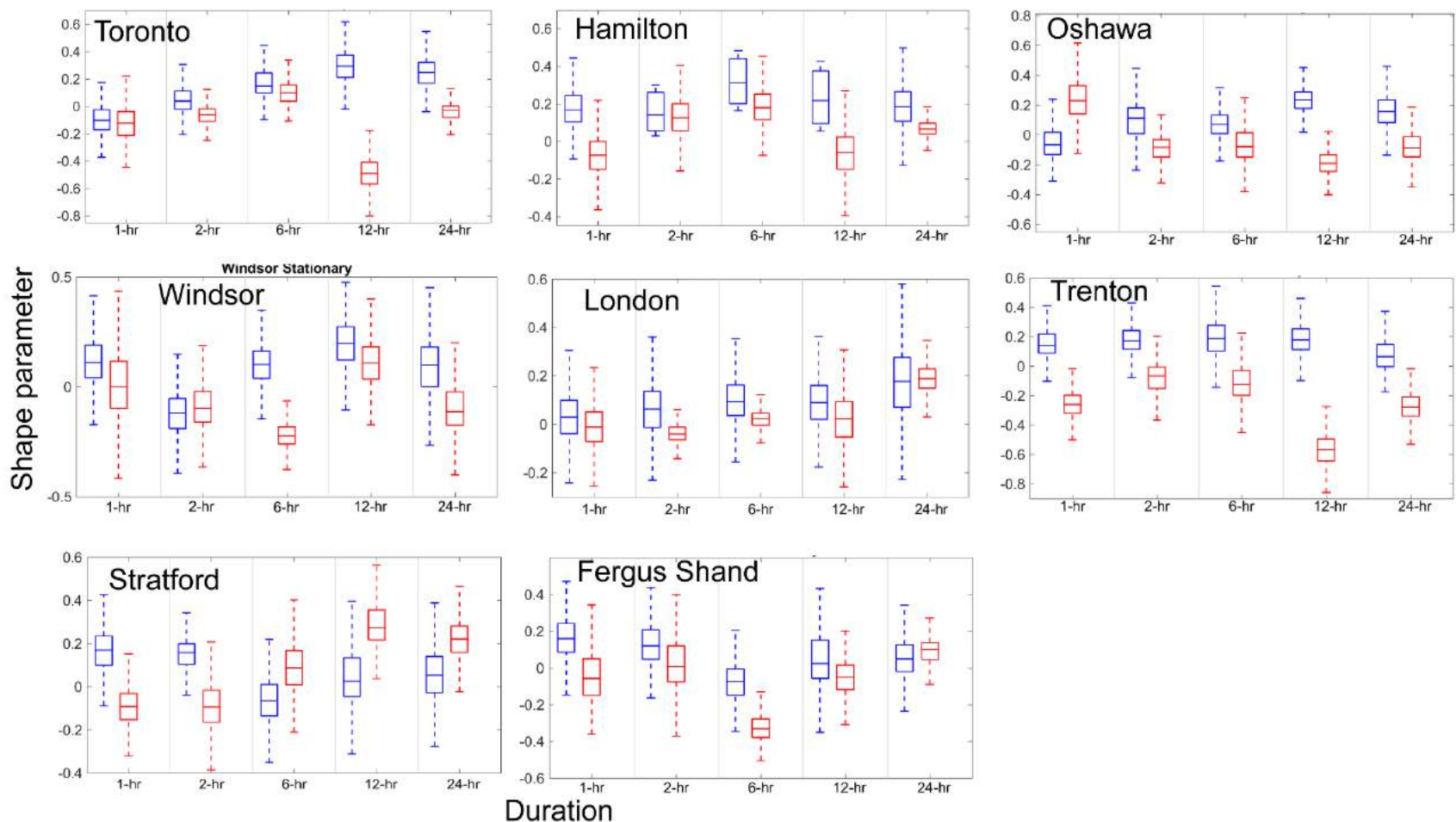

**Figure 16.** The distribution of shape parameters of the simulated GEV distribution at different durations for the future period (2030-2070) assuming stationary GEV models. The box-plot in blue indicates the observed distribution and in red represents multi-model median NA-CORDEX RCM simulated distribution.

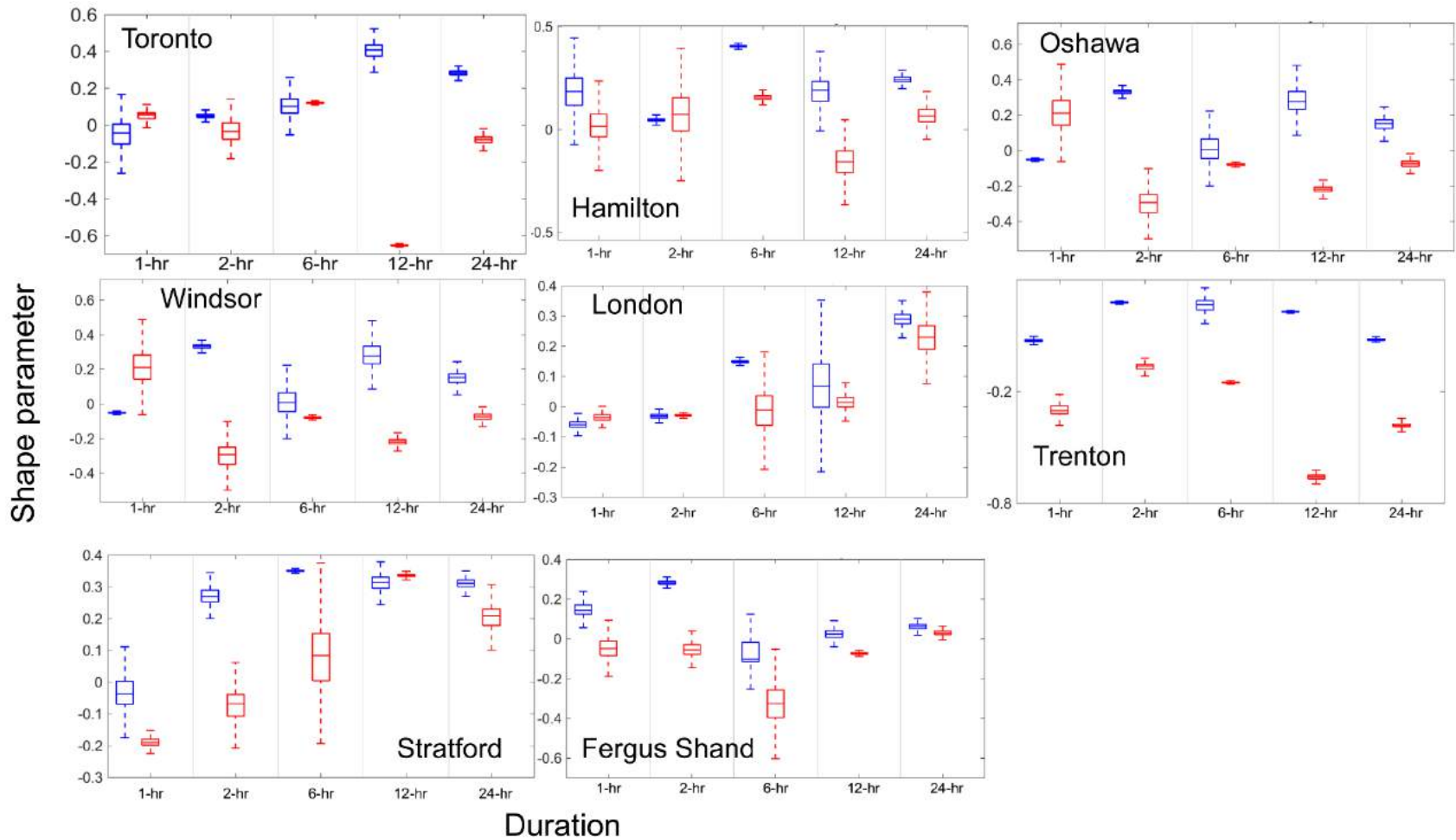

**Figure 17.** Same as Figure 16 but for nonstationary GEV models.

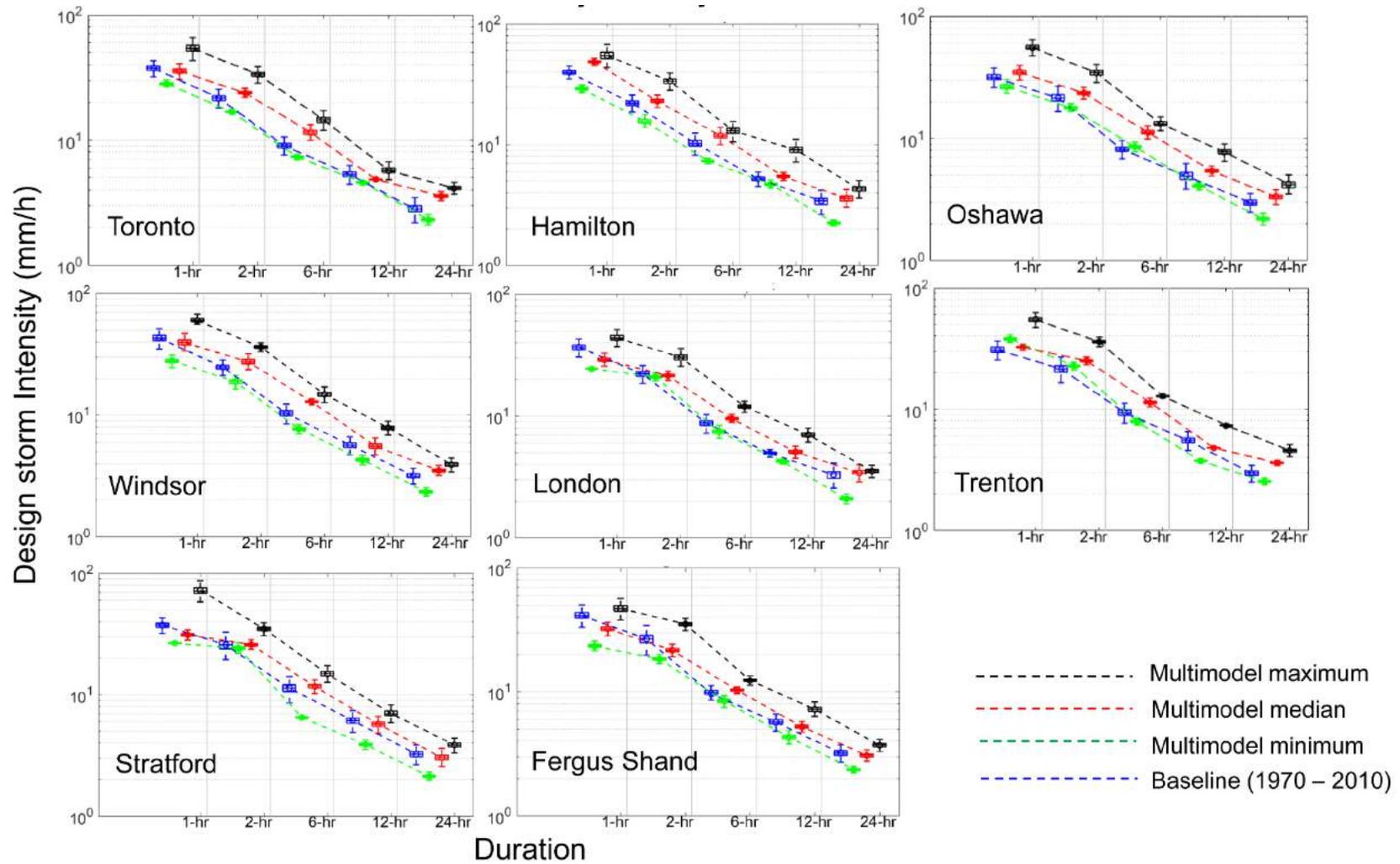

**Figure 18.** Nonstationary present-day (in blue) and Projected (in red; 2030-2070) IDF curves for 10-year return period across eight stations in Southern Ontario. The uncertainty in IDF simulation at different duration is expressed using boxplots. The minimum (in green) and maximum (in red) bounds in IDF curves are shown to express the best and worst plausible scenarios.

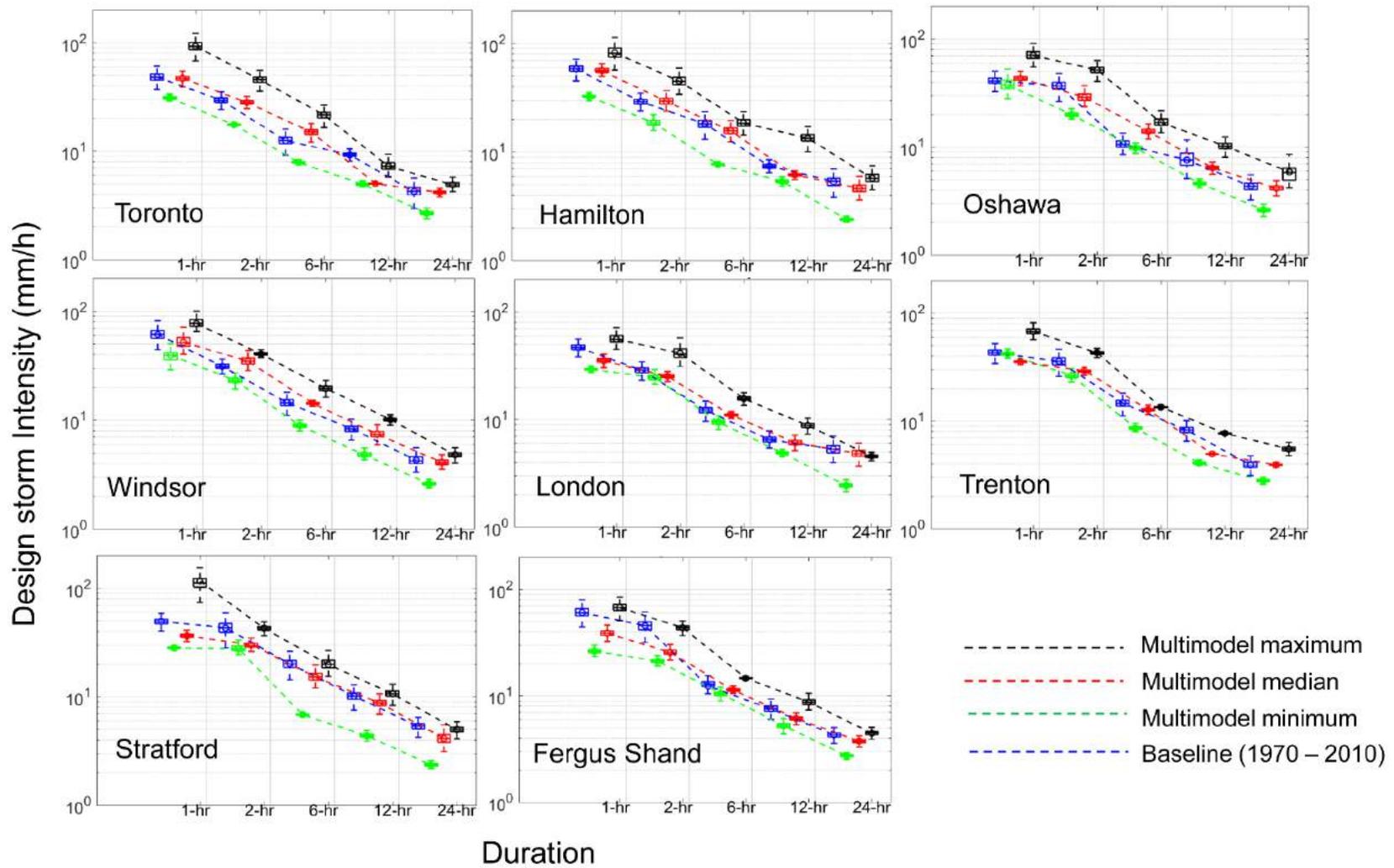

**Figure 19.** Nonstationary present-day and Projected IDF curves for 50-year return period.

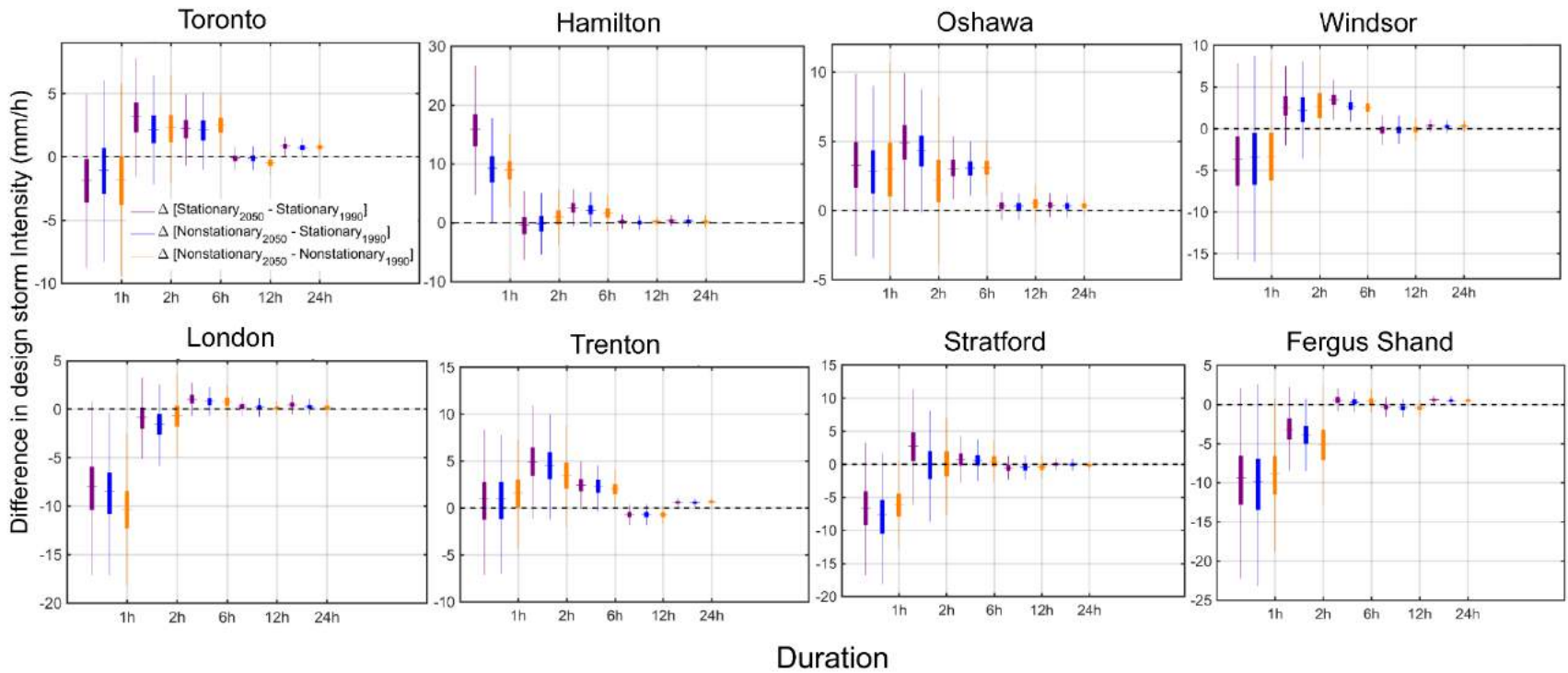

**Figure 20.** Design storm intensity of future scenario versus present day climate for 10-year return period. The variability in the differences in design storm intensity are shown using box-plots representing three different cases: Stationary (2030-2070; or 2050s) versus stationary (1970 – 2010; or 1990s) [in purple]; Nonstationary (2050s) versus stationary (1990s) [in blue]; Nonstationary (2050s) versus nonstationary (1990s) [in orange].

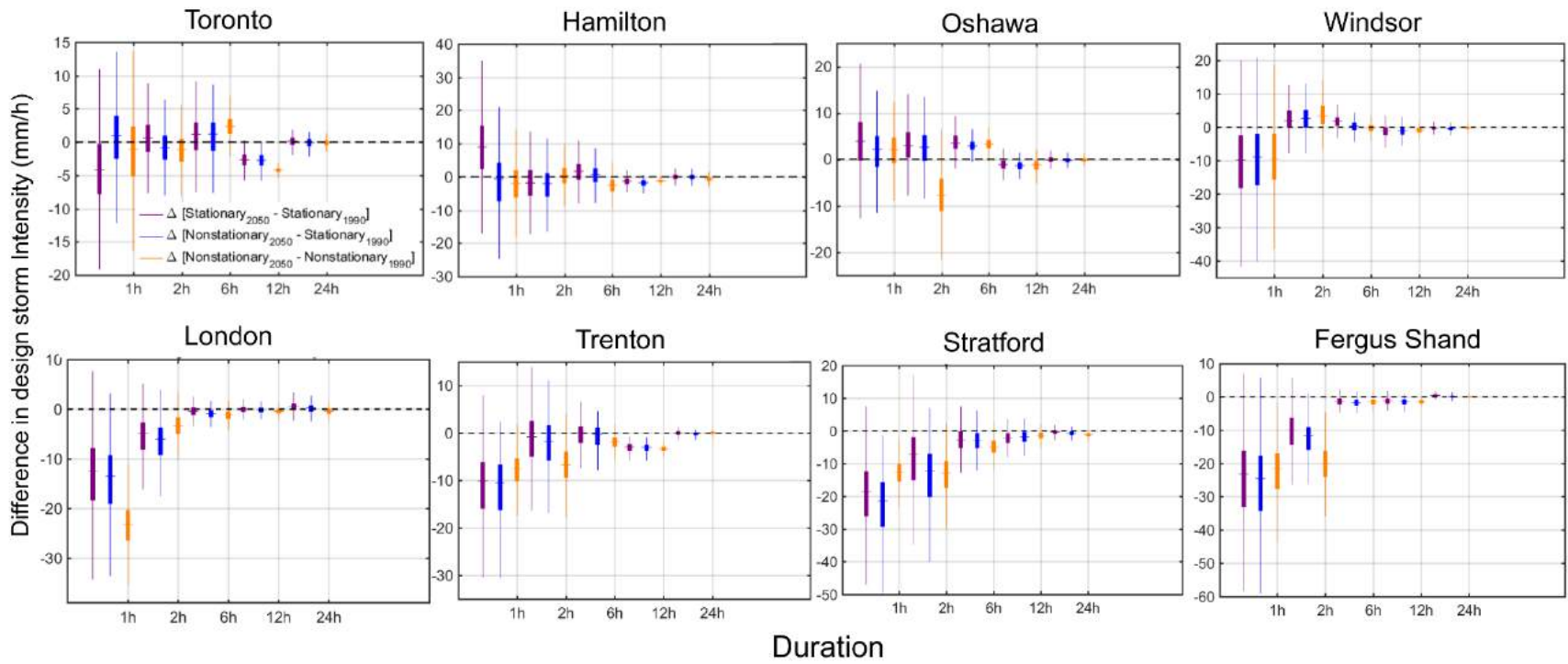

**Figure 21.** Same as Figure 20 but for 50-year return period.

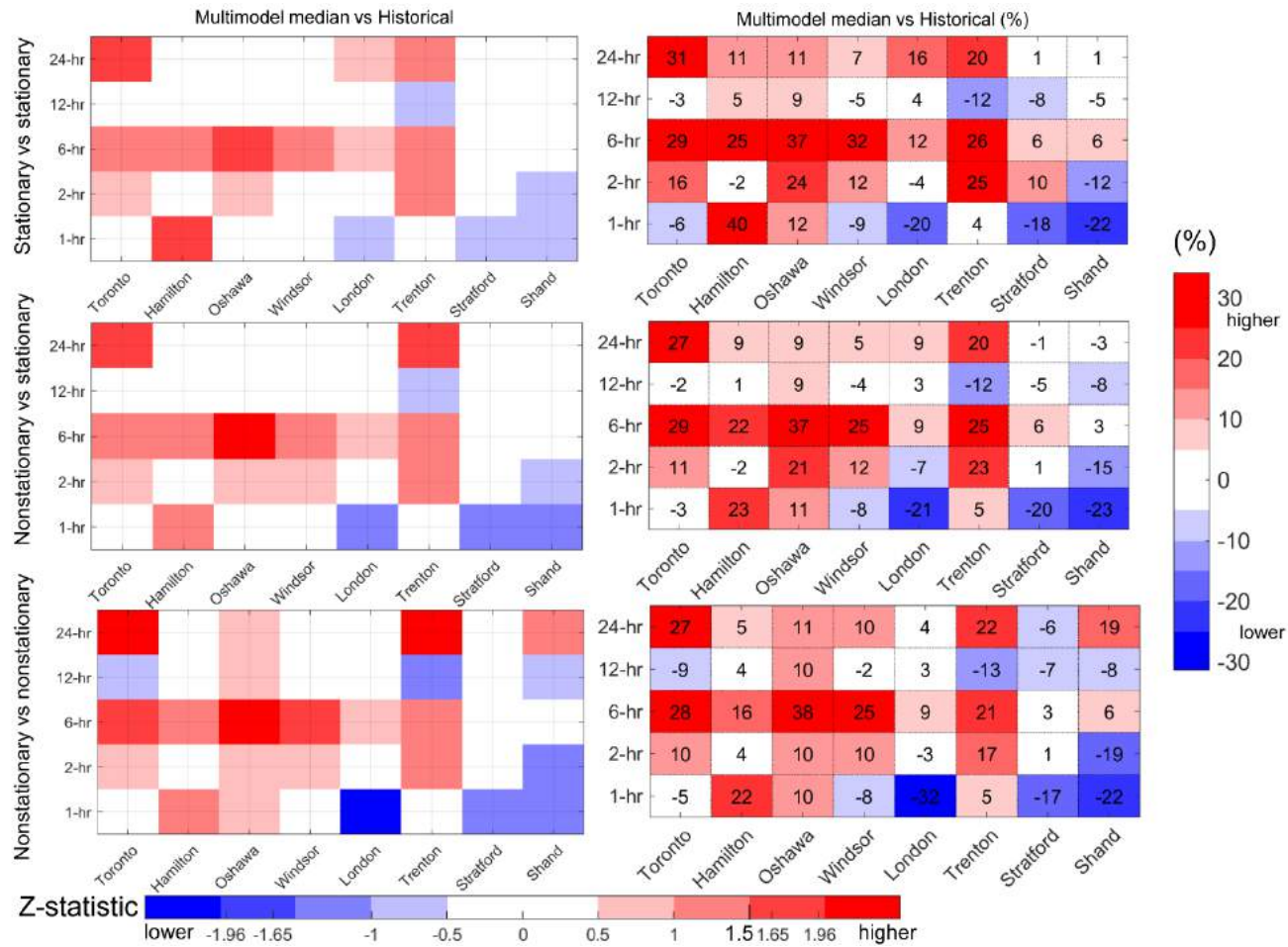

**Figure 22.** Projected changes in rainfall intensity for 10-year return period in three different cases: stationary (2050s) versus stationary (1990s) [*top panel*]; Nonstationary (2050s) versus stationary (1990s) [*middle panel*]; Nonstationary (2050s) versus nonstationary (1990s) [*bottom panel*]. The comparative assessment is performed between projected storm intensity modeled using multi-model median NA-CORDEX RCM ensemble and observed baseline intensity. The shades of the changes express high and low end, with a dark red indicating increase in storm intensity while dark blue show decrease in the intensity. Very small changes [*i.e.,* in and around zero values] are marked with white.

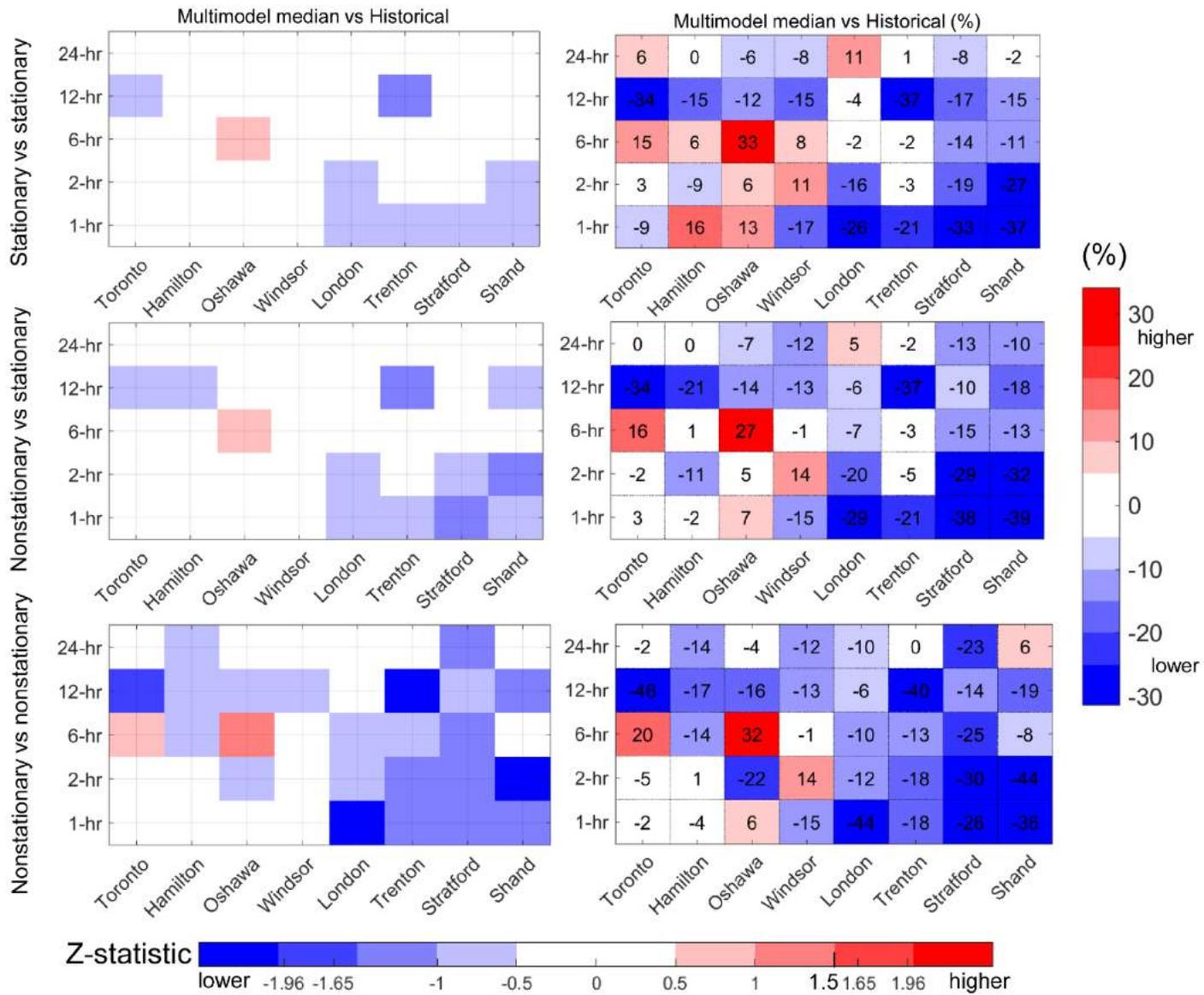

**Figure 23.** Same as in Figure 21 but for 50-year return period.

Supporting Information for

**Assessment of Future Changes in Intensity-Duration-Frequency Curves for Southern Ontario using North American CORDEX Models with Nonstationary Methods**

Poulomi Ganguli[1], Paulin Coulibaly[1]

[1] McMaster Water Resources and Hydrologic Modeling Group, Department of Civil Engineering, McMaster University, 1280 Main Street West, Hamilton ON L8S 4L7, Canada

**SI 1. Temporal Downscaling of Precipitation: Multiplicative Random Cascade-based disaggregation Tool**

The Cascade-based temporal downscaling method in this study is adopted from [*Olsson*, 1995, 1998a]. The method implements a micro-canonical (exact conservation of mass in each cascade branching) approach, in which daily rainfall is disaggregated using a uniformly distributed generator, dependent on rainfall intensity and position of the rain sequence. The method was later developed and tested in different climatic conditions [*Güntner et al.*, 2001; *Rana et al.*, 2013]. In the disaggregation process, each time interval (box) at a given resolution (for example 1 day) is split into two half of the original length (1/2 day). The procedure is continued as a cascade until the desired time resolution is reached, i.e., to ¼ day, then to 1/8 of a day and so on. Each step is termed as a cascade step, with cascade step 0 as the longest time-period with only one box (*i.e.*, a day). The distribution of the volume between two sub-intervals (or smaller boxes) is computed by multiplication with the cascade weights ($W_{i,r}$), $0 \leq W_{i,r} \leq 1$ that assigns $W_{i,r} \bullet R_{i,r}$ to the first half of the period and $\left(1 - W_{i,r}\right) \bullet R_{i,r}$ to the next half. In each branching two possibilities exist: (1) $W_1 = 0$, $W_1 = 1$ (2) $0 < W_1 < 1$. The occurrence of (1) and (2) may be expressed in terms of probabilities, $P_{01} = P_{(1/0)}$ or $P_{(0/1)} = P(W_1 = 0$ or $W_1 = 1)$ and $P_{xx} = P_{(x/x)} = P(0 < W_1 < 1) = 1 - P_{01}$.

Depending on the range of resolution involve, $P_{01}$ either be assumed as resolution independent or parameterized as a scaling law: $Pr_{01}(r) = c_1 r^{c_2}$ where $c_1$ and $c_2$ are constants. The distribution of $W_{i,r}$ is termed as cascade generator, assumed to follow 1-parameter beta distribution [*Olsson*, 2012]. Following [*Olsson*, 1998b], the probabilities *P*, the probability distribution of cascade generator are assumed to be related to (1) position in rainfall sequence, and (2) rainfall volume. The wet boxes, with a rainfall volume $V > 0$, can be characterized by their position in the rainfall series: (1) the starting box, box preceded by a dry box ($V = 0$) and succeeded by a wet box ($V > 0$); (2) the enclosed box, box preceded and succeeded by wet boxes; (3) the ending box, box preceded by a wet box and succeeded by a dry box, and (4) the isolated box, box preceded and succeeded by dry boxes. On the other hand, based on volume dependence, if the volume is large then it is more likely that both halves of the subintervals contribute to nonzero volume than if the volume is small. Following [*Olsson*, 1998b], a partition into three volume classes ($v_c = 1, 2, 3$) was used, separated by percentiles 33rd and 67th of the values at the cascade step. Next, the variation of $P_{(x/x)}$ with volume is parameterised as, $P_{(x/x)} = a + b_m \bullet v_c$, where *a* is the intercept at $v_c = 0$, $b_m$ is the mean slope of linear regression obtained from all cascade steps and $v_c$ is volume class. The theory and implementation issues of MRC-based disaggregation tool are discussed further in [*Olsson*, 1998b, n.d.].

The objectives of employing cascade-based disaggregation tool are as follows: (i) to infill missing values in the station-based AM precipitation time series (ii) to disaggregate NA-CORDEX regional climate model output available at daily time step to an hourly time scale. First, the cascade parameters are calibrated using good quality observed historical dataset available at an hourly resolution. The high-resolution historical (1970 – 2010) data are collected from Toronto Resources Conservation Authority (TRCA). At least 35-years or more available dataset is used to calibrate the model. Table S1 lists the length of the dataset used for model calibration. The parameters are calibrated using 5-cascade steps, which implies, five successive "halving" from one day to generate 45-minute (2700 seconds) data. After calibration, Monte Carlo simulation is performed to gradually fine-grain the data and generate realizations at desired resolution, $R_s$. Since $R_s$ in this cases is not directly achieved by exact resolution doubling from original resolution, $R_l$, the target resolutions are obtained by geometric interpolation of the disaggregated model output at the final time step. Then annual maxima value is obtained from hourly disaggregated time series for each year. The statistic, we use for the validation is by comparing maximum disaggregated rainfall time series reference to observed rainfall. Fig S2 (top panel) show the performance of disaggregation tool for modeling hourly precipitation extremes at Toronto International Airport. The daily AM series is also shown (Fig S2; bottom panel) to compare the magnitude of downscaled rainfall volume relative to the observed daily extreme. The bars in red shows observed series and blue disaggregated AM series. We adjusted occasional overestimation of disaggregated rainfall extreme (for instance, 1980), by a GEV-based bias correction. The bias corrected AM series is shown in black bars. A reasonable agreement is noted between observed and temporally downscaled with bias-corrected precipitation for Toronto Airport. Following a similar procedure, we infilled the missing gaps in the AM series of eight station locations across Southern ON. Fig. S3 shows observed (red) versus disaggregated infilled (blue) hourly AM time series. As observed from the figure, the infilled time series preserve the trend in the data reasonably well, without exhibiting any signature of outliers (*i.e.*, extremely high or low values).

Likewise, to disaggregate the daily NA-CORDEX output to hourly time step, we assume that dynamically downscaled model output is statistically similar to that of station observed precipitation. For this, we compare the distribution of daily precipitation extremes of observed versus NA-CORDEX RCMs (Fig. S4). Fig. S4 presents boxplots of the anomaly of daily maximum precipitation for the observed and regional climate model ensembles. As observed from the figure, the RCMs able to reproduce the observed daily precipitation satisfactorily, although some bias exist. Among individual model performance, in general, CanRCM4 shows maximum variability (or spread) in simulating precipitation extremes than that of the other two RCMs. To disaggregate the daily RCM data to an hourly time step, we use the calibrated model parameters derived in the historical simulation to downscale precipitation in the baseline and future periods.

Then two bias correction schemes are compared (Kernel- and GEV-based) to match station observed precipitation. The performance of bias-corrected RCMs are assessed using various skill metrics as discussed in the main manuscript.

Table S1. Details of available hourly data set for Cascade-based disaggregation model calibration

| Station name | Start Year | End Year | Number of years |
| --- | --- | --- | --- |
| Toronto International Airport | 1961 | 2010 | 50 |
| Hamilton Airport | 1970 | 2002 | 40 |
| Oshawa WPCP | 1970 | 2005 | 36 |
| Windsor Airport | 1960 | 2006 | 47 |
| London International Airport | 1960 | 2001 | 42 |
| Trenton Airport | 1965 | 2000 | 36 |
| Stratford WPTP | 1966 | 2005 | 40 |
| Fergus Shand Dam | 1961 | 2006 | 46 |

Table S2. Skill scores of the Kernel and GEV-based bias correction relative to observations for hourly AM precipitation

| Locations | Kernel-based | | | | | | GEV-based | | | | | |
|---|---|---|---|---|---|---|---|---|---|---|---|---|
| | CanRCM4 | CRCM5 | RegCM4 | MMin | MMed | MMax | CanRCM4 | CRCM5 | RegCM4 | MMin | MMed | MMax |
| Shand | 0.502 | 0.327 | 0.474 | 0.434 | 0.337 | 0.449 | 0.499 | 0.317 | 0.489 | 0.455 | 0.342 | 0.457 |
| Hamilton | 0.420 | 0.631 | 0.453 | 0.398 | 0.416 | 0.641 | 0.424 | 0.626 | 0.449 | 0.429 | 0.405 | 0.636 |
| London | 0.412 | 0.426 | 0.433 | 0.404 | 0.331 | 0.448 | 0.394 | 0.431 | 0.420 | 0.391 | 0.331 | 0.438 |
| Oshawa | 0.394 | 0.413 | 0.612 | 0.405 | 0.535 | 0.505 | 0.399 | 0.426 | 0.609 | 0.409 | 0.541 | 0.504 |
| Stratford | 0.423 | 0.431 | 0.451 | 0.424 | 0.376 | 0.464 | 0.460 | 0.453 | 0.454 | 0.462 | 0.373 | 0.458 |
| Toronto | 0.382 | 0.455 | 0.505 | 0.395 | 0.540 | 0.424 | 0.373 | 0.467 | 0.507 | 0.385 | 0.539 | 0.423 |
| Trenton | 0.544 | 0.440 | 0.436 | 0.544 | 0.492 | 0.383 | 0.542 | 0.441 | 0.441 | 0.541 | 0.481 | 0.392 |
| Windsor | 0.599 | 0.559 | 0.425 | 0.561 | 0.569 | 0.507 | 0.604 | 0.569 | 0.427 | 0.555 | 0.581 | 0.541 |

Table S3. Skill scores of the Kernel and GEV-based bias correction relative to observations for 2-hourly AM precipitation

| Locations | Kernel-based | | | | | | GEV-based | | | | | |
|---|---|---|---|---|---|---|---|---|---|---|---|---|
| | CanRCM4 | CRCM5 | RegCM4 | MMin | MMed | MMax | CanRCM4 | CRCM5 | RegCM4 | MMin | MMed | MMax |
| Shand | 0.555 | 0.407 | 0.493 | 0.435 | 0.405 | 0.570 | 0.552 | 0.404 | 0.489 | 0.435 | 0.398 | 0.561 |
| Hamilton | 0.464 | 0.485 | 0.325 | 0.352 | 0.346 | 0.440 | 0.443 | 0.579 | 0.410 | 0.425 | 0.413 | 0.546 |
| London | 0.411 | 0.391 | 0.488 | 0.438 | 0.360 | 0.411 | 0.403 | 0.398 | 0.489 | 0.432 | 0.359 | 0.416 |
| Oshawa | 0.445 | 0.484 | 0.607 | 0.515 | 0.540 | 0.558 | 0.445 | 0.498 | 0.610 | 0.517 | 0.540 | 0.557 |
| Stratford | 0.391 | 0.530 | 0.390 | 0.402 | 0.488 | 0.394 | 0.457 | 0.531 | 0.364 | 0.397 | 0.479 | 0.256 |
| Toronto | 0.386 | 0.456 | 0.489 | 0.415 | 0.498 | 0.395 | 0.379 | 0.464 | 0.486 | 0.455 | 0.486 | 0.397 |
| Trenton | 0.542 | 0.506 | 0.422 | 0.572 | 0.578 | 0.383 | 0.551 | 0.509 | 0.423 | 0.567 | 0.577 | 0.381 |
| Windsor | 0.592 | 0.557 | 0.490 | 0.564 | 0.607 | 0.540 | 0.596 | 0.561 | 0.493 | 0.564 | 0.605 | 0.543 |

Table S4. Skill scores of the Kernel and GEV-based bias correction relative to observations for 6-hourly AM precipitation

| Locations | Kernel-based | | | | | | GEV-based | | | | | |
|---|---|---|---|---|---|---|---|---|---|---|---|---|
| | CanRCM4 | CRCM5 | RegCM4 | MMin | MMed | MMax | CanRCM4 | CRCM5 | RegCM4 | MMin | MMed | MMax |
| Shand | 0.477 | 0.494 | 0.453 | 0.456 | 0.410 | 0.521 | 0.484 | 0.485 | 0.481 | 0.451 | 0.414 | 0.531 |
| Hamilton | 0.399 | 0.505 | 0.413 | 0.386 | 0.294 | 0.442 | 0.404 | 0.564 | 0.413 | 0.389 | 0.411 | 0.443 |
| London | 0.470 | 0.414 | 0.517 | 0.412 | 0.491 | 0.464 | 0.474 | 0.416 | 0.539 | 0.429 | 0.507 | 0.464 |
| Oshawa | 0.407 | 0.438 | 0.710 | 0.537 | 0.477 | 0.542 | 0.408 | 0.458 | 0.711 | 0.535 | 0.482 | 0.550 |
| Stratford | 0.414 | 0.674 | 0.399 | 0.400 | 0.373 | 0.549 | 0.389 | 0.623 | 0.387 | 0.418 | 0.407 | 0.530 |
| Toronto | 0.433 | 0.456 | 0.447 | 0.326 | 0.350 | 0.513 | 0.439 | 0.485 | 0.490 | 0.359 | 0.382 | 0.509 |
| Trenton | 0.558 | 0.552 | 0.380 | 0.645 | 0.546 | 0.365 | 0.562 | 0.553 | 0.380 | 0.641 | 0.556 | 0.368 |
| Windsor | 0.584 | 0.512 | 0.557 | 0.582 | 0.633 | 0.538 | 0.568 | 0.510 | 0.558 | 0.591 | 0.619 | 0.511 |

Table S5. Skill scores of the Kernel and GEV-based bias correction relative to observations for 12-hourly AM precipitation

| Locations | Kernel-based | | | | | | GEV-based | | | | | |
|---|---|---|---|---|---|---|---|---|---|---|---|---|
| | CanRCM4 | CRCM5 | RegCM4 | MMin | MMed | MMax | CanRCM4 | CRCM5 | RegCM4 | MMin | MMed | MMax |
| Shand | 0.515 | 0.483 | 0.433 | 0.401 | 0.479 | 0.502 | 0.537 | 0.484 | 0.442 | 0.408 | 0.483 | 0.533 |
| Hamilton | 0.382 | 0.525 | 0.459 | 0.442 | 0.430 | 0.439 | 0.382 | 0.522 | 0.462 | 0.442 | 0.433 | 0.433 |
| London | 0.573 | 0.478 | 0.672 | 0.629 | 0.642 | 0.596 | 0.582 | 0.483 | 0.682 | 0.641 | 0.633 | 0.609 |
| Oshawa | 0.436 | 0.363 | 0.535 | 0.443 | 0.395 | 0.494 | 0.423 | 0.415 | 0.606 | 0.530 | 0.478 | 0.487 |
| Stratford | 0.421 | 0.654 | 0.467 | 0.385 | 0.469 | 0.490 | 0.364 | 0.657 | 0.464 | 0.384 | 0.479 | 0.480 |
| Toronto | 0.439 | 0.441 | 0.450 | 0.345 | 0.384 | 0.490 | 0.400 | 0.457 | 0.282 | 0.313 | 0.417 | 0.452 |
| Trenton | 0.582 | 0.516 | 0.373 | 0.633 | 0.443 | 0.468 | 0.584 | 0.503 | 0.379 | 0.592 | 0.449 | 0.469 |
| Windsor | 0.479 | 0.520 | 0.628 | 0.451 | 0.557 | 0.575 | 0.379 | 0.513 | 0.619 | 0.459 | 0.540 | 0.458 |

Table S6. Skill scores of the Kernel and GEV-based bias correction relative to observations for daily AM precipitation

| Locations | Kernel-based | | | | | | GEV-based | | | | | |
|---|---|---|---|---|---|---|---|---|---|---|---|---|
| | CanRCM4 | CRCM5 | RegCM4 | MMin | MMed | MMax | CanRCM4 | CRCM5 | RegCM4 | MMin | MMed | MMax |
| Shand | 0.577 | 0.551 | 0.534 | 0.409 | 0.591 | 0.647 | 0.585 | 0.559 | 0.535 | 0.412 | 0.598 | 0.648 |
| Hamilton | 0.593 | 0.480 | 0.476 | 0.586 | 0.462 | 0.538 | 0.590 | 0.452 | 0.481 | 0.573 | 0.470 | 0.539 |
| London | 0.618 | 0.438 | 0.635 | 0.557 | 0.523 | 0.630 | 0.593 | 0.419 | 0.653 | 0.349 | 0.548 | 0.606 |
| Oshawa | 0.530 | 0.349 | 0.566 | 0.418 | 0.396 | 0.524 | 0.518 | 0.370 | 0.584 | 0.438 | 0.454 | 0.547 |
| Stratford | 0.559 | 0.643 | 0.433 | 0.489 | 0.490 | 0.602 | 0.578 | 0.610 | 0.441 | 0.469 | 0.492 | 0.626 |
| Toronto | 0.526 | 0.471 | 0.472 | 0.371 | 0.493 | 0.511 | 0.507 | 0.462 | 0.470 | 0.410 | 0.477 | 0.498 |
| Trenton | 0.526 | 0.550 | 0.303 | 0.551 | 0.492 | 0.338 | 0.502 | 0.540 | 0.306 | 0.506 | 0.474 | 0.338 |
| Windsor | 0.448 | 0.544 | 0.634 | 0.443 | 0.548 | 0.542 | 0.443 | 0.544 | 0.636 | 0.407 | 0.548 | 0.535 |

Table S7. Mann-Kendall trend statistics and Theil-Sen slope per decade of observed versus MMed CORDEX RCM for hourly AM precipitation

| Stations | Observed | | MMed RCM | |
|---|---|---|---|---|
| | MK Statistics | b | MK Statistics | b |
| Hamilton | -0.048 (0.962) | -0.019 | -0.416 (0.678) | -0.689 |
| London | -1.011 (0.312) | -1.293 | -0.607 (0.544) | -0.823 |
| Oshawa | 0.315 (0.753) | 0.153 | 0.876 (0.381) | 0.940 |
| Shand | 0.146 (0.884) | 0.143 | 0.124 (0.902) | 0.131 |
| Stratford | 0.247 (0.805) | 0.261 | **1.820** (0.069*) | 2.164 |
| Toronto | **-1.775** (0.076*) | -2.567 | 1.135 (0.256) | 1.540 |
| Trenton | 1.011 (0.312) | 1.199 | 1.079 (0.281) | 1.085 |
| Windsor | -0.809 (0.419) | -1.017 | 0.180 (0.857) | 0.287 |

*Value is bracket indicated p-value of the test statistics. Letters in bold indicates statistically significant trend at 10% significance level.*

Table S8. Mann-Kendall trend statistics and Theil-Sen slope per decade of observed versus MMed CORDEX RCM for 2-hourly AM precipitation

| Stations | Observed | | MMed RCM | |
|---|---|---|---|---|
| | MK Statistics | $b$ | MK Statistics | $b$ |
| Hamilton | -0.048 (0.962) | -0.034 | -0.595 (0.552) | -0.753 |
| London | -0.887 (0.375) | -1.109 | -0.966 (0.334) | -1.181 |
| Oshawa | 0.539 (0.590) | 0.578 | 1.292 (0.196) | 1.481 |
| Shand | 0.640 (0.522) | 1.080 | 1.269 (0.204) | 2.129 |
| Stratford | **1.855** (0.064*) | 0.961 | **1.719** (0.086*) | 3.864 |
| Toronto | -1.955 (0.051*) | -2.484 | **2.202** (0.028*) | 2.871 |
| Trenton | 1.178 (0.239) | 1.579 | 0.135 (0.893) | 0.262 |
| Windsor | -0.112 (0.911) | -0.270 | **2.101** (0.036*) | 3.445 |

Table S9. Mann-Kendall trend statistics and Theil-Sen slope per decade of observed versus MMed CORDEX RCM for 6-hourly AM precipitation

| Stations | Observed | | MMed RCM | |
|---|---|---|---|---|
| | MK Statistics | $b$ | MK Statistics | $b$ |
| Hamilton | -0.573 (0.567) | -0.585 | -0.124 (0.902) | -0.114 |
| London | -0.622 (0.534) | -0.652 | 0.887 (0.375) | 1.957 |
| Oshawa | 0.697 (0.486) | 0.988 | 1.143 (0.253) | 1.382 |
| Shand | 0.371 (0.711) | 0.935 | 0.244 (0.808) | 0.207 |
| Stratford | -0.708 (0.479) | -0.927 | 0.708 (0.479) | 1.282 |
| Toronto | -1.393 (0.164) | -2.022 | **1.910** (0.056*) | 2.505 |
| Trenton | -0.622 (0.534) | -0.652 | **2.483** (0.013*) | 2.707 |
| Windsor | 0.607 (0.544) | 1.016 | **1.720** (0.085*) | 2.560 |

Table S10. Mann-Kendall trend statistics and Theil-Sen slope per decade of observed versus MMed CORDEX RCM for 12-hourly AM precipitation

| Stations | Observed | | MMed RCM | |
|---|---|---|---|---|
| | MK Statistics | b | MK Statistics | b |
| Hamilton | -0.584 (0.559) | -1.11 | 0.191 (0.849) | 0.312 |
| London | -0.371 (0.711) | -0.58 | 1.101 (0.271) | 1.851 |
| Oshawa | 1.370 (0.171) | 1.47 | -0.247 (0.805) | -0.540 |
| Shand | -0.687 (0.492) | -0.92 | 0.382 (0.703) | 0.873 |
| Stratford | -1.022 (0.307) | -1.74 | **1.674** (0.094*) | 4.113 |
| Toronto | -0.539 (0.590) | -0.99 | 0.663 (0.507) | 1.154 |
| Trenton | -0.371 (0.711) | -0.58 | 0.748 (0.455) | 1.453 |
| Windsor | 0.247 (0.805) | 0.69 | **2.917** (0.004*) | 2.227 |

Table S11. Mann-Kendall trend statistics and Theil-Sen slope per decade of observed versus MMed CORDEX RCM for daily AM precipitation

| Stations | Observed | | MMed RCM | |
|---|---|---|---|---|
| | MK Statistics | b | MK Statistics | b |
| Hamilton | -1.13 (0.257) | -2.48 | -0.326 (0.745) | -0.515 |
| London | -0.831 (0.406) | -1.48 | 1.359 (0.174) | 1.754 |
| Oshawa | **2.572** (0.010*) | 4.188 | -1.134 (0.257) | -1.746 |
| Shand | -1.574 (0.116) | -1.899 | 0.326 (0.744) | 0.665 |
| Stratford | -1.461 (0.144) | -2.228 | 1.032 (0.302) | 1.965 |
| Toronto | -0.438 (0.661) | -0.638 | -0.775 (0.438) | -1.410 |
| Trenton | **2.741** (0.006*) | 5.018 | 1.606 (0.108) | 2.799 |
| Windsor | 0.899 (0.369) | 1.460 | 0.573 (0.567) | 0.961 |

46  Table S12. Mann-Kendall trend statistics and Theil-Sen slope per decade of historical daily AM temperature anomaly

| Stations | Observed | |
| --- | --- | --- |
| | MK Statistics | b |
| Hamilton | 1.33 (0.18) | 0.254 |
| London | 1.34 (0.18) | 0.308 |
| Oshawa | 0.22 (0.82) | 0.070 |
| Shand | 0.25 (0.80) | 0.00 |
| Stratford | 1.48 (0.14) | 0.463 |
| Toronto | 1.16 (0.25) | 0.370 |
| Trenton | 0.60 (0.55) | 0.142 |
| Windsor | -0.35 (0.73) | -0.035 |

48  Table S13. Estimates of the median shape parameter of observed vs modeled MM-Med RCMs for baseline (1970 – 2010) simulation

| GEV-Type | Sites | Observed | | | | | MM-Med | | | | |
| --- | --- | --- | --- | --- | --- | --- | --- | --- | --- | --- | --- |
| | | 1-hour | 2-hour | 6-hour | 12-hour | 24-hour | 1-hour | 2-hour | 6-hour | 12-hour | 24-hour |
| Stationary | Hamilton | 0.166 | 0.140 | 0.278 | 0.221 | 0.184 | 0.169 | 0.109 | 0.150 | 0.225 | 0.175 |
| | London | 0.029 | 0.060 | 0.128 | 0.077 | 0.177 | 0.053 | 0.047 | 0.123 | 0.060 | 0.037 |
| | Oshawa | -0.066 | 0.113 | 0.078 | 0.235 | 0.156 | -0.046 | 0.122 | 0.186 | 0.268 | 0.211 |
| | Shand | 0.159 | 0.121 | -0.073 | 0.024 | 0.049 | 0.172 | 0.262 | 0.152 | 0.053 | 0.055 |
| | Stratford | 0.170 | 0.158 | 0.276 | 0.337 | 0.052 | 0.074 | 0.159 | 0.407 | 0.275 | 0.287 |
| | Toronto | -0.102 | 0.039 | 0.149 | 0.294 | 0.247 | -0.079 | 0.026 | 0.036 | 0.419 | 0.158 |
| | Trenton | 0.142 | 0.172 | 0.188 | 0.178 | 0.064 | 0.146 | 0.130 | 0.183 | 0.238 | 0.151 |
| | Windsor | 0.111 | -0.118 | 0.102 | 0.198 | 0.100 | 0.165 | -0.068 | -0.032 | 0.144 | 0.190 |
| Nonstationary | Hamilton | 0.184 | 0.046 | 0.404 | 0.190 | 0.243 | 0.222 | 0.075 | 0.216 | 0.260 | 0.152 |
| | London | -0.060 | -0.031 | 0.150 | 0.068 | 0.291 | 0.127 | 0.105 | 0.224 | -0.022 | 0.048 |
| | Oshawa | -0.052 | 0.333 | 0.008 | 0.276 | 0.152 | -0.361 | 0.157 | 0.148 | 0.376 | 0.247 |
| | Shand | 0.144 | 0.283 | -0.102 | 0.024 | 0.061 | 0.155 | 0.182 | 0.202 | 0.054 | 0.073 |
| | Stratford | -0.037 | 0.270 | 0.351 | 0.314 | 0.311 | 0.157 | 0.196 | 0.499 | 0.369 | 0.308 |
| | Toronto | -0.044 | 0.049 | 0.101 | 0.408 | 0.281 | -0.180 | 0.063 | 0.066 | 0.435 | 0.146 |
| | Trenton | 0.074 | 0.279 | 0.268 | 0.230 | 0.080 | 0.114 | 0.248 | 0.166 | 0.333 | 0.118 |
| | Windsor | 0.116 | -0.101 | 0.109 | 0.193 | 0.072 | 0.196 | -0.055 | -0.016 | 0.132 | 0.228 |

Table S14. GEV shape parameters for stationary and nonstationary models of observed versus MM-Med RCM simulation during baseline period for Toronto International Airport

| Time Slice | Data | Model | Location parameter $\mu_0$ | Location parameter $\mu_1$ | Scale parameter $\sigma$ | Shape parameter $\xi$ |
|---|---|---|---|---|---|---|
| 1-hr | Observation | $GEV_{sta}$ | 21.38 (19.07, 23.14) | | 7.59 (6.32, 9.1) | -0.10 (-0.24, 0.08) |
| | | $GEV_{Nonsta}$ | 27.28 (23.51, 31.26) | -0.27 (-0.41, -0.14) | 7.46 (6.38, 8.97) | -0.04 (-0.20, 0.10) |
| | MM-Med CORDEX | $GEV_{sta}$ | 21.04 (18.85, 22.87) | | 7.73 (6.64, 9.88) | -0.07 (-0.22, 0.10) |
| | | $GEV_{Nonsta}$ | 19.51 (15.03, 23.54) | 0.10 (-0.07, 0.29) | 8.0 (6.76, 9.89) | -0.18 (-0.26, -0.12) |
| 2-hr | Observation | $GEV_{sta}$ | 11.82 (10.75, 12.96) | | 4.23 (3.46, 5.26) | 0.03 (-0.08, 0.25) |
| | | $GEV_{Nonsta}$ | 13.65 (11.53, 15.71) | -0.08 (-0.17, -0.006) | 4.11 (3.39, 5.05) | 0.05 (0.03, 0.07) |
| | MM-Med CORDEX | $GEV_{sta}$ | 11.83 (10.63, 12.92) | | 4.39 (3.70, 5.05) | -1.53 e-5 (-0.14, 0.20) |
| | | $GEV_{Nonsta}$ | 9.47 (7.18, 11.66) | 0.12 (0.02, 0.21) | 4.08 (3.39, 4.93) | 0.04 (0.003, 0.096) |
| 6-hr | Observation | $GEV_{sta}$ | 5.08 (4.6, 5.43) | | 1.55 (1.25, 1.89) | 0.18 (0.005, 0.43) |
| | | $GEV_{Nonsta}$ | 5.58 (5.19, 6.01) | -0.02 (-0.04, -0.002) | 1.55 (1.29, 1.92) | 0.10 (0.008, 0.21) |
| | MM-Med CORDEX | $GEV_{sta}$ | 3.95 (3.57, 4.34) | | 1.48 (1.43, 1.53) | 0.036 (-0.12, 0.27) |
| | | $GEV_{Nonsta}$ | 3.20 (2.54, 3.93) | 0.033 (0.0002, 0.061) | 1.37 (1.12, 1.68) | 0.06 (-0.01, 0.14) |
| 12-hr | Observation | $GEV_{sta}$ | 2.88 (2.71, 3.06) | | 0.65 (0.52, 0.83) | 0.28 (0.11, 0.54) |
| | | $GEV_{Nonsta}$ | 2.87 (2.52, 3.25) | 0.0001 (-0.015, 0.014) | 0.67 (0.52, 0.87) | 0.40 (0.32, 0.47) |
| | MM-Med CORDEX | $GEV_{sta}$ | 2.82 (2.68, 3.04) | | 0.62 (0.46, 0.82) | 0.42 (0.26, 0.66) |
| | | $GEV_{Nonsta}$ | 2.90 (2.73, 3.04) | -0.001 (-0.009, 0.006) | 0.64 (0.49, 0.81) | 0.43 (0.25, 0.67) |
| 24-hr | Observation | $GEV_{sta}$ | 1.68 (1.59, 1.79) | | 0.37 (0.31, 0.46) | 0.25 (0.08, 0.43) |
| | | $GEV_{Nonsta}$ | 1.81 (1.63, 1.98) | -0.006 (-0.013, 0.001) | 0.36 (0.28, 0.46) | 0.28 (0.26, 0.31) |
| | MM-Med CORDEX | $GEV_{sta}$ | 1.47 (1.39, 1.52) | | 0.40 (0.35, 0.49) | 0.16 (0.005, 0.34) |
| | | $GEV_{Nonsta}$ | 1.58 (1.49, 1.67) | -0.005 (-0.012, 0.001) | 0.42 (0.35, 0.49) | 0.15 (0.003, 0.35) |

\* GEV parameters are estimated using Bayesian Inference. The reported values are median of DE-MC sampled parameters. Bracketed values indicate $5^{th}$ and $95^{th}$ percentile of simulated GEV parameters obtained from DE-MC simulations

Table S15. GEV shape parameters for stationary and nonstationary models of observed versus MM-Med RCM simulation during baseline period for Oshawa WPCP

| Time Slice | Data | Model | Location parameter | | Scale parameter | Shape parameter |
|---|---|---|---|---|---|---|
| | | | $\mu_0$ | $\mu_1$ | $\sigma$ | $\xi$ |
| 1-hr | Observation | GEV$_{sta}$ | 17.39 (15.27, 19.03) | | 6.69 (5.89, 7.93) | -0.06 (-0.23, 0.22) |
| | | GEV$_{Nonsta}$ | 17.06 (13.61, 20.07) | 0.02 (-0.10, 0.15) | 6.66 (5.58, 8.26) | -0.05 (-0.06, -0.045) |
| | MM-Med CORDEX | GEV$_{sta}$ | 17.44 (15.61, 19.41) | | 6.60 (5.58, 7.98) | -0.046 (-0.24, 0.12) |
| | | GEV$_{Nonsta}$ | 14.46 (10.24, 18.06) | 0.17 (0.05, 0.32) | 8.66 (7.80, 9.98) | -0.36 (-0.41, -0.29) |
| 2-hr | Observation | GEV$_{sta}$ | 10.70 (9.54, 11.69) | | 3.43 (2.71, 4.55) | 0.11 (-0.08, 0.33) |
| | | GEV$_{Nonsta}$ | 10.02 (8.67, 11.25) | 0.01 (-0.03, 0.06) | 3.32 (2.67, 4.32) | 0.33 (0.30, 0.35) |
| | MM-Med CORDEX | GEV$_{sta}$ | 10.75 (9.91, 11.76) | | 3.23 (2.58, 4.22) | 0.12 (-0.07, 0.26) |
| | | GEV$_{Nonsta}$ | 9.07 (7.54, 10.64) | 0.07 (0.01, 0.14) | 3.11 (2.54, 3.93) | 0.15 (0.003, 0.27) |
| 6-hr | Observation | GEV$_{sta}$ | 4.81 (4.55, 5.18) | | 1.37 (1.11, 1.69) | 0.078 (-0.06, 0.19) |
| | | GEV$_{Nonsta}$ | 4.33 (3.93, 4.75) | 0.025 (0.009, 0.04) | 1.45 (1.22, 1.77) | 0.007 (-0.11, 0.15) |
| | MM-Med CORDEX | GEV$_{sta}$ | 3.53 (3.25, 3.89) | | 1.08 (0.88, 1.37) | 0.19 (0.006, 0.33) |
| | | GEV$_{Nonsta}$ | 3.06 (2.36, 3.68) | 0.022 (-0.005, 0.05) | 1.10 (0.89, 1.39) | 0.15 (0.11, 0.19) |
| 12-hr | Observation | GEV$_{sta}$ | 2.82 (2.62, 3.01) | | 0.77 (0.64, 0.93) | 0.23 (0.07, 0.43) |
| | | GEV$_{Nonsta}$ | 2.62 (2.33, 2.83) | 0.007 (-0.002, 0.02) | 0.69 (0.58, 0.85) | 0.28 (0.16, 0.41) |
| | MM-Med CORDEX | GEV$_{sta}$ | 2.81 (2.58, 3.05) | | 0.75 (0.59, 0.94) | 0.27 (0.08, 0.51) |
| | | GEV$_{Nonsta}$ | 2.72 (2.35, 3.08) | 0.003 (-0.011, 0.016) | 0.72 (0.57, 0.94) | 0.37 (0.36, 0.38) |
| 24-hr | Observation | GEV$_{sta}$ | 1.64 (1.48, 1.77) | | 0.52 (0.41, 0.64) | 0.15 (0.006, 0.28) |
| | | GEV$_{Nonsta}$ | 1.54 (1.36, 1.71) | 0.005 (-0.002, 0.011) | 0.50 (0.43, 0.62) | 0.15 (0.07, 0.20) |
| | MM-Med CORDEX | GEV$_{sta}$ | 1.43 (1.34, 1.50) | | 0.37 (0.31, 0.46) | 0.21 (0.06, 0.38) |
| | | GEV$_{Nonsta}$ | 1.46 (1.28, 1.64) | -0.001 (-0.007, 0.004) | 0.38 (0.32, 0.45) | 0.25 (0.10, 0.39) |

Table S16. GEV shape parameters for stationary and nonstationary models of observed versus MM-Med RCM simulation during baseline period for Wndsor Airport

| Time Slice | Data | Model | Location parameter | | Scale parameter | Shape parameter |
|---|---|---|---|---|---|---|
| | | | $\mu_0$ | $\mu_1$ | $\sigma$ | $\xi$ |
| 1-hr | Observation | GEV$_{sta}$ | 23.08 (21.18, 25.32) | | 7.73 (6.38, 9.60) | 0.11 (-0.06, 0.34) |
| | | GEV$_{Nonsta}$ | 26.35 (22.69, 30.5) | -0.15 (-0.32, -0.03) | 7.67 (6.47, 9.60) | 0.11 (0.01, 0.22) |
| | MM-Med CORDEX | GEV$_{sta}$ | 22.48 (20.54, 24.82) | | 8.10 (6.6, 10.07) | 0.16 (-0.02, 0.31) |
| | | GEV$_{Nonsta}$ | 21.12 (17.91, 24.2) | 0.06 (-0.05, 0.20) | 7.67 (6.14, 9.61) | 0.20 (0.16, 0.23) |
| 2-hr | Observation | GEV$_{sta}$ | 14.38 (13.07, 15.83) | | 5.15 (4.33, 6.11) | -0.11 (-0.29, 0.042) |
| | | GEV$_{Nonsta}$ | 15.15 (12.54, 17.62) | -0.04 (-0.13, 0.05) | 5.23 (4.46, 6.27) | -0.10 (-0.13, -0.07) |
| | MM-Med CORDEX | GEV$_{sta}$ | 14.32 (13.05, 15.60) | | 5.12 (4.31, 6.37) | -0.07 (-0.23, 0.13) |
| | | GEV$_{Nonsta}$ | 10.39 (7.51, 13.16) | 0.19 (0.074, 0.30) | 4.72 (4.06, 5.66) | -0.05 (-0.09, -0.017) |
| 6-hr | Observation | GEV$_{sta}$ | 5.82 (5.35, 6.30) | | 1.78 (1.46, 2.28) | 0.10 (-0.068, 0.29) |
| | | GEV$_{Nonsta}$ | 5.72 (4.81, 6.54) | 0.004 (-0.034, 0.044) | 1.78 (1.46, 2.19) | 0.10 (0.06, 0.15) |
| | MM-Med CORDEX | GEV$_{sta}$ | 4.73 (4.26, 5.12) | | 1.71 (1.46, 2.04) | -0.03 (-0.16, 0.17) |
| | | GEV$_{Nonsta}$ | 4.06 (3.09, 4.84) | 0.035 (-0.004, 0.077) | 1.65 (1.42, 1.98) | -0.016 (-0.04, 0.004) |
| 12-hr | Observation | GEV$_{sta}$ | 3.20 (2.96, 3.45) | | 0.90 (0.75, 1.13) | 0.19 (0.0008, 0.37) |
| | | GEV$_{Nonsta}$ | 3.19 (2.75, 3.58) | 0.0009 (-0.016, 0.018) | 0.88 (0.71, 1.08) | 0.19 (0.17, 0.21) |
| | MM-Med CORDEX | GEV$_{sta}$ | 3.22 (2.94, 3.48) | | 0.96 (0.76, 1.22) | 0.14 (-0.02, 0.30) |
| | | GEV$_{Nonsta}$ | 2.99 (2.61, 3.33) | 0.013 (-0.005, 0.03) | 0.98 (0.80, 1.20) | 0.13 (-0.007, 0.32) |
| 24-hr | Observation | GEV$_{sta}$ | 1.90 (1.77, 2.08) | | 0.57 (0.46, 0.68) | 0.10 (-0.12, 0.33) |
| | | GEV$_{Nonsta}$ | 1.69 (1.63, 1.75) | 0.008 (0.004, 0.014) | 0.53 (0.43, 0.69) | 0.07 (-0.13, 0.31) |
| | MM-Med CORDEX | GEV$_{sta}$ | 1.60 (1.48, 1.74) | | 0.42 (0.33, 0.57) | 0.19 (0.02, 0.45) |
| | | GEV$_{Nonsta}$ | 1.41 (1.23, 1.63) | 0.009 (0.007, 0.016) | 0.40 (0.32, 0.52) | 0.23 (0.12, 0.32) |

Table S17. The AICc values of stationary and nonstationary GEV models at 1- and 6-hour duration during baseline (1970-2010) period

| Duration (hour) | Stations | Observation | | MM-Med NA-CORDEX RCMs | |
|---|---|---|---|---|---|
| | | $GEV_{sta}$ | $GEV_{Nonsta}$ | $GEV_{sta}$ | $GEV_{Nonsta}$ |
| 1- | Hamilton | **-273.68** | -273.67 | **-290.91** | -285.85 |
| | London | **-252.00** | -251.15 | **-286.09** | -283.10 |
| | Oshawa | -268.21 | **-272.28** | **-244.83** | -238.16 |
| | Shand | -302.63 | **-305.21** | **-275.15** | -273.87 |
| | Stratford | **-266.32** | -256.17 | -260.26 | **-277.12** |
| | Toronto | **-275.29** | -270.06 | **-275.08** | -269.83 |
| | Trenton | -276.34 | **-278.01** | -298.49 | **-295.16** |
| | Windsor | -268.32 | **-268.59** | -285.05 | **-284.94** |
| 6- | Hamilton | -285.46 | **-295.95** | -245.22 | **-246.73** |
| | London | **-306.55** | -298.91 | -241.78 | **-245.62** |
| | Oshawa | **-288.58** | -277.32 | **-300.48** | -288.05 |
| | Shand | **-296.63** | -278.46 | -277.43 | **-278.89** |
| | Stratford | -262.79 | **-269.36** | -284.68 | **-292.52** |
| | Toronto | **-284.60** | -277.39 | -284.48 | **-295.38** |
| | Trenton | -252.40 | **-264.43** | -283.42 | **-289.46** |
| | Windsor | -287.92 | **-299.88** | -289.94 | **-289.89** |

*The AICc values are calculated between design storm estimate from GEV model and empirical distribution, fitted with Gringorten's plotting position formula. The expression for the performance statistics, corrected AIC is given as, $AIC_c(m) = AIC + 2m(m+1)/(n-m-1)$, in which n is the number of observations, m denotes the number of fitted model parameters. The best distribution is marked as bold. In this case, the best distribution is selected from the minimum AICc statistics.*

Table S18. Estimates of the median shape parameter of observed baseline (1970 – 2010) vs MM-Med RCM modeled projected (2030 – 2070) scenario

| GEV-Type | Sites | Observed | | | | | MM-Med | | | | |
|---|---|---|---|---|---|---|---|---|---|---|---|
| | | 1-hour | 2-hour | 6-hour | 12-hour | 24-hour | 1-hour | 2-hour | 6-hour | 12-hour | 24-hour |
| Stationary | Hamilton | 0.166 | 0.140 | 0.278 | 0.221 | 0.184 | -0.075 | 0.125 | 0.180 | -0.059 | 0.044 |
| | London | 0.029 | 0.060 | 0.128 | 0.077 | 0.177 | -0.012 | -0.040 | 0.024 | 0.021 | 0.188 |
| | Oshawa | -0.066 | 0.113 | 0.078 | 0.235 | 0.156 | 0.230 | -0.083 | -0.078 | -0.191 | -0.087 |
| | Shand | 0.159 | 0.121 | -0.073 | 0.024 | 0.049 | -0.058 | 0.007 | -0.331 | -0.052 | 0.102 |
| | Stratford | 0.170 | 0.158 | 0.276 | 0.337 | 0.052 | -0.092 | -0.094 | 0.087 | 0.273 | 0.221 |
| | Toronto | -0.102 | 0.039 | 0.149 | 0.294 | 0.247 | -0.123 | -0.063 | 0.098 | -0.490 | -0.032 |
| | Trenton | 0.142 | 0.172 | 0.188 | 0.178 | 0.064 | -0.260 | -0.067 | -0.123 | -0.567 | -0.278 |
| | Windsor | 0.111 | -0.118 | 0.102 | 0.198 | 0.100 | 0.022 | -0.097 | -0.225 | 0.108 | -0.112 |
| Nonstationary | Hamilton | 0.184 | 0.046 | 0.404 | 0.190 | 0.243 | 0.017 | 0.073 | 0.157 | -0.157 | 0.067 |
| | London | -0.060 | -0.031 | 0.150 | 0.068 | 0.291 | -0.036 | -0.029 | -0.012 | 0.014 | 0.230 |
| | Oshawa | -0.052 | 0.333 | 0.008 | 0.276 | 0.152 | 0.210 | -0.294 | -0.079 | -0.220 | -0.075 |
| | Shand | 0.144 | 0.283 | -0.102 | 0.024 | 0.061 | -0.049 | -0.056 | -0.326 | -0.073 | 0.030 |
| | Stratford | -0.037 | 0.270 | 0.351 | 0.314 | 0.311 | -0.190 | -0.068 | 0.084 | 0.335 | 0.209 |
| | Toronto | -0.044 | 0.049 | 0.101 | 0.408 | 0.281 | 0.056 | -0.036 | 0.121 | -0.653 | -0.079 |
| | Trenton | 0.074 | 0.279 | 0.268 | 0.230 | 0.080 | -0.303 | -0.063 | -0.152 | -0.659 | -0.381 |
| | Windsor | 0.116 | -0.101 | 0.109 | 0.193 | 0.072 | 0.021 | -0.012 | -0.255 | 0.131 | -0.112 |

Table S19. The AICc values of stationary and nonstationary GEV models at 1- and 6-hour duration during projected (2030-2070) period

| Duration (hour) | Stations | GEV$_{sta}$ | GEV$_{Nonsta}$ |
|---|---|---|---|
| 1- | Hamilton | **-278.51** | -189.54 |
|  | London | **-277.65** | -268.92 |
|  | Oshawa | **-290.30** | -281.95 |
|  | Fergus Shand | **-288.61** | -279.86 |
|  | Stratford | **-242.03** | -240.96 |
|  | Toronto | **-271.06** | -263.22 |
|  | Trenton | -263.46 | **-284.98** |
|  | Windsor | -220.39 | **-221.36** |
| 6- | Hamilton | -277.22 | **-280.78** |
|  | London | **-281.53** | -252.72 |
|  | Oshawa | -245.97 | **-280.53** |
|  | Fergus Shand | **-270.07** | -250.75 |
|  | Stratford | -308.54 | **-311.28** |
|  | Toronto | -289.57 | **-291.09** |
|  | Trenton | **-259.90** | -253.57 |
|  | Windsor | **-299.06** | -224.51 |

*The AICc values are calculated between design storm estimate from GEV model and empirical distribution, fitted with Gringorten's plotting position formula.*

## List of Supplementary Figure Captions

**Figure S1.** Flow chart of the comparison procedure.

**Figure S2.** Selected station locations in Southern Ontario. The Southern Ontario (41° - 44°N, 84° - 76°W) is the southernmost region of Canada and is situated on a southwest-northeast transect, bounded by lakes Huron, Erie, and Ontario. The eight locations on the map are (*from southwest to northeast corner*): Windsor Airport, London International Airport, Stratford Wastewater Treatment Plant (WWTP), Fergus Shand Dam, Hamilton Airport, Toronto International Airport, Oshawa Water Pollution Control Plant (WPCP) and Trenton Airport. Topography map is obtained from 90-m digital elevation model (DEM; SRTM-90m, [*Jarvis et al.*, 2008] indicates shallow slope with elevation of maximum 670 m above mean sea level.

**Figure S3.** Observed versus temporally downscaled hourly AM series (*top panel*). The daily AM series (*bottom panel*).

**Figure S4.** Observed versus infilled hourly AM series.

**Figure S5.** Distribution of annual maximum precipitation anomaly in observations and NA-CORDEX simulated RCMs during baseline (1970-2010) period.

**Figure S6.** PDF of observed versus multi-model median 2-hour AM precipitation in eight station locations. The minimum and the maximum bounds are shown using dotted blue and red lines respectively.

**Figure S7.** Same as Figure S6 but for 6-hour precipitation extreme.

**Figure S8.** Same as Figure S7 but for 12-hour precipitation extreme.

**Figure S9.** Present-day (in blue) versus Projected (in red; 2030-2070) IDF curves for 10-year return period across eight stations in Southern Ontario using stationary GEV models. The uncertainty in IDF simulation at different duration is expressed using boxplots. The minimum (in green) and maximum (in red) bounds in IDF curves are shown to express the best and worst plausible scenarios.

**Figure S10.** Same as Figure S7 but for 50-year return period.

**Figure S11.** Projected changes in rainfall intensity for 25-year return period in three different cases: stationary (2050s) versus stationary (1990s) [*top panel*]; Nonstationary (2050s) versus stationary (1990s) [*middle panel*]; Nonstationary (2050s) versus nonstationary (1990s) [*bottom panel*]. The comparative assessment is performed between projected storm intensity modeled using multi-model median NA-CORDEX RCM ensemble and observed baseline intensity. The shades of the changes express high and low end, with dark red indicating increase in storm intensity while dark blue show decrease in the intensity. Very small changes [*i.e.,* in and around zero values] are marked with white.

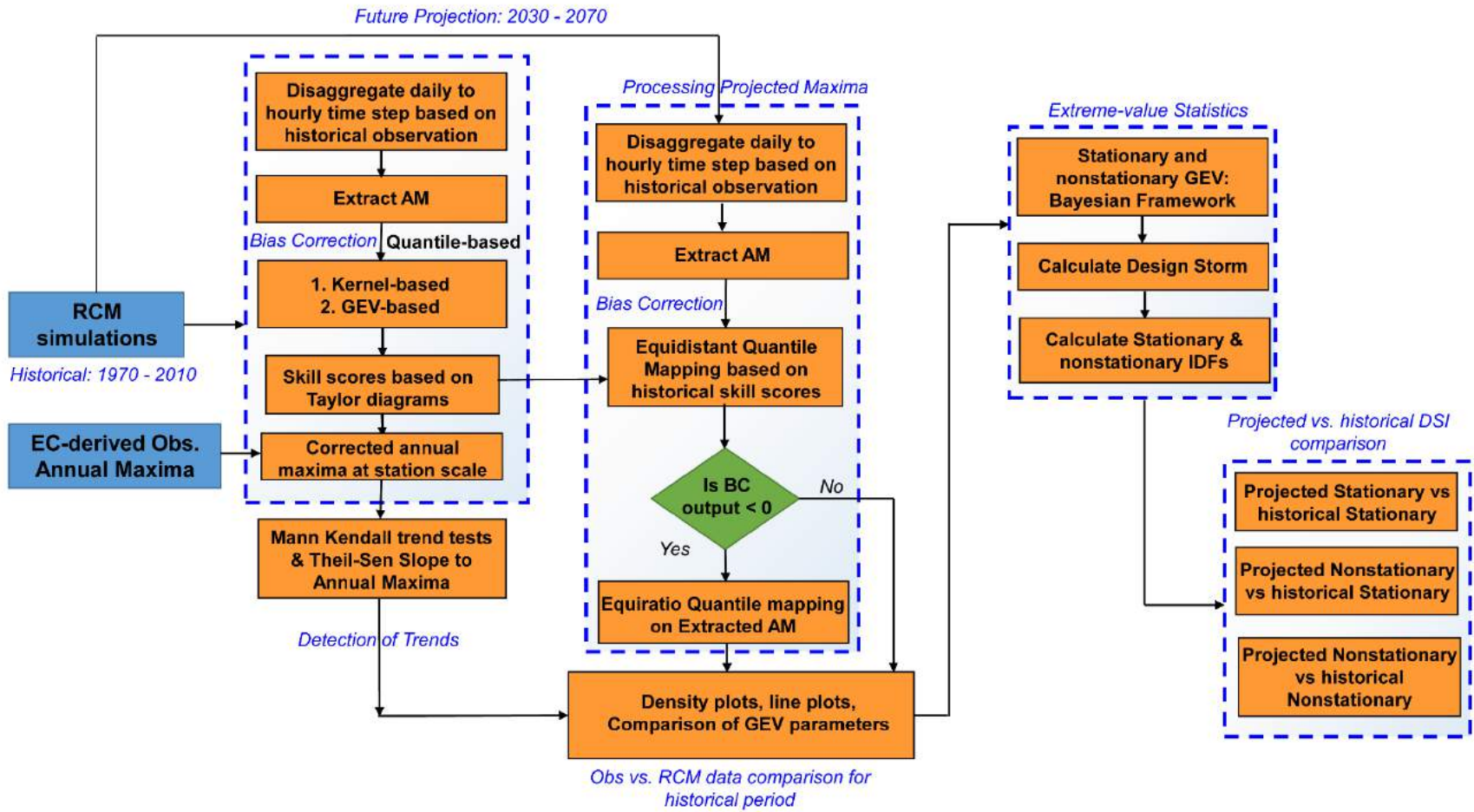

**Figure S1.** Flow chart of the comparison procedure

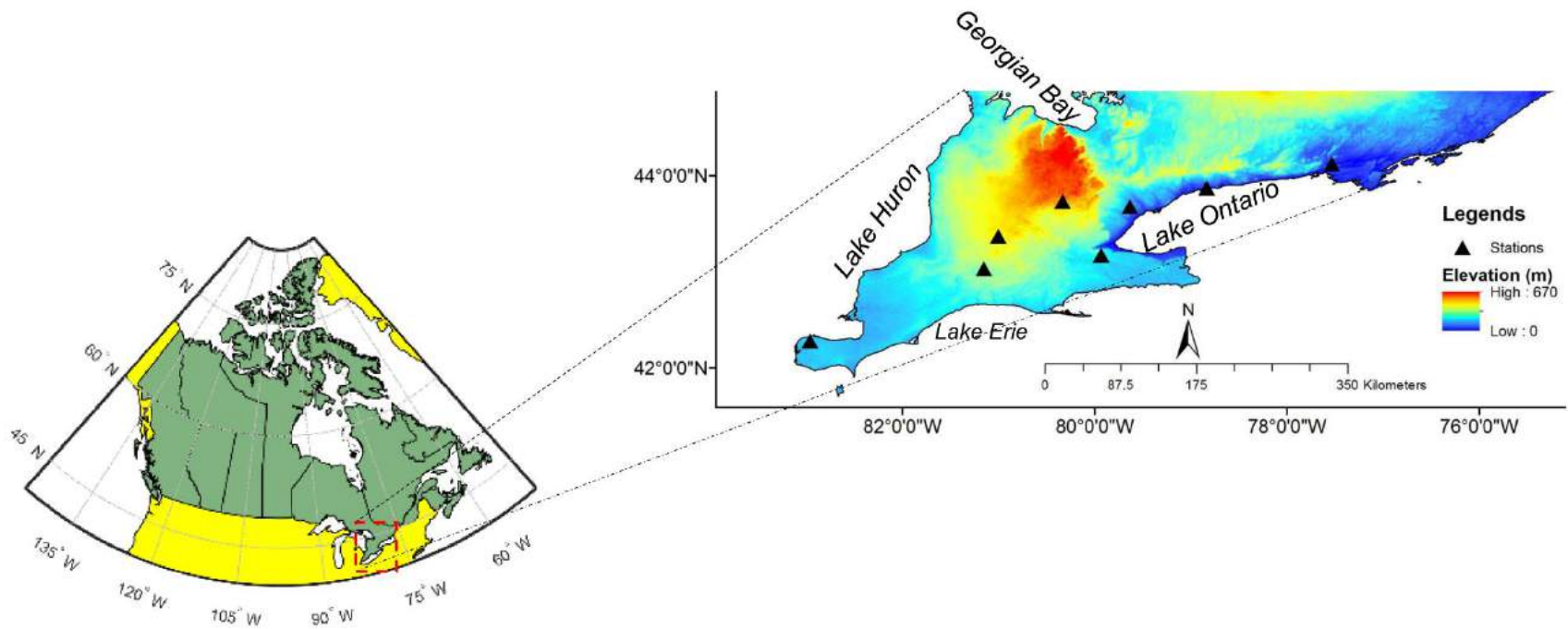

**Figure S2.** Selected station locations in Southern Ontario. The Southern Ontario (41° - 44°N, 84° - 76°W) is the southernmost region of Canada and is situated on a southwest-northeast transect, bounded by lakes Huron, Erie, and Ontario. The eight locations on the map are (*from southwest to northeast corner*): Windsor Airport, London International Airport, Stratford Wastewater Treatment Plant (WWTP), Fergus Shand Dam, Hamilton Airport, Toronto International Airport, Oshawa Water Pollution Control Plant (WPCP) and Trenton Airport. Topography map is obtained from 90-m digital elevation model (DEM; SRTM-90m, [*Jarvis et al.*, 2008] indicates shallow slope with maximum elevation of 670 m above mean sea level.

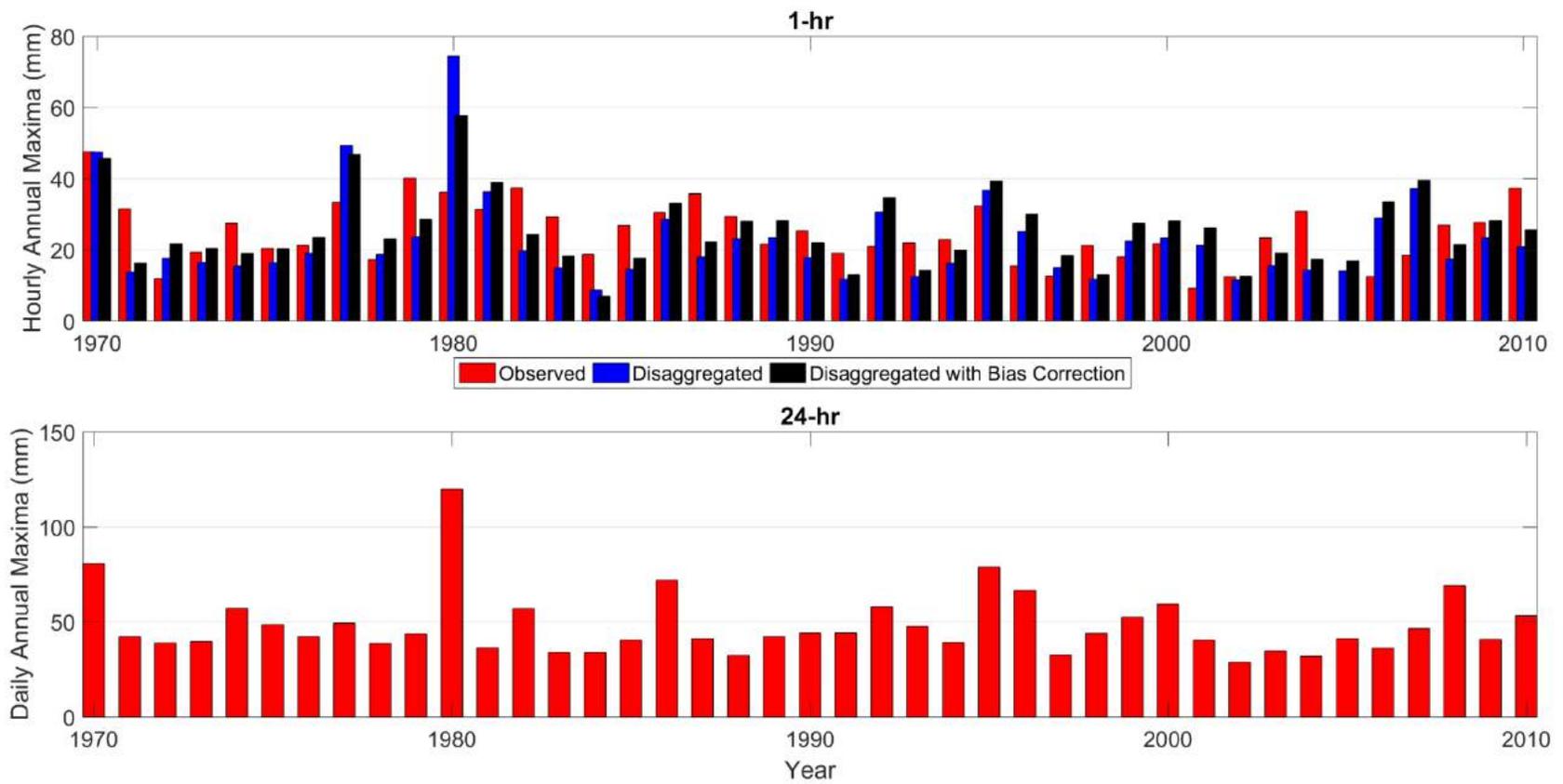

**Figure S3.** Observed versus temporally downscaled hourly AM series (*top panel*). The daily AM series (*bottom panel*).

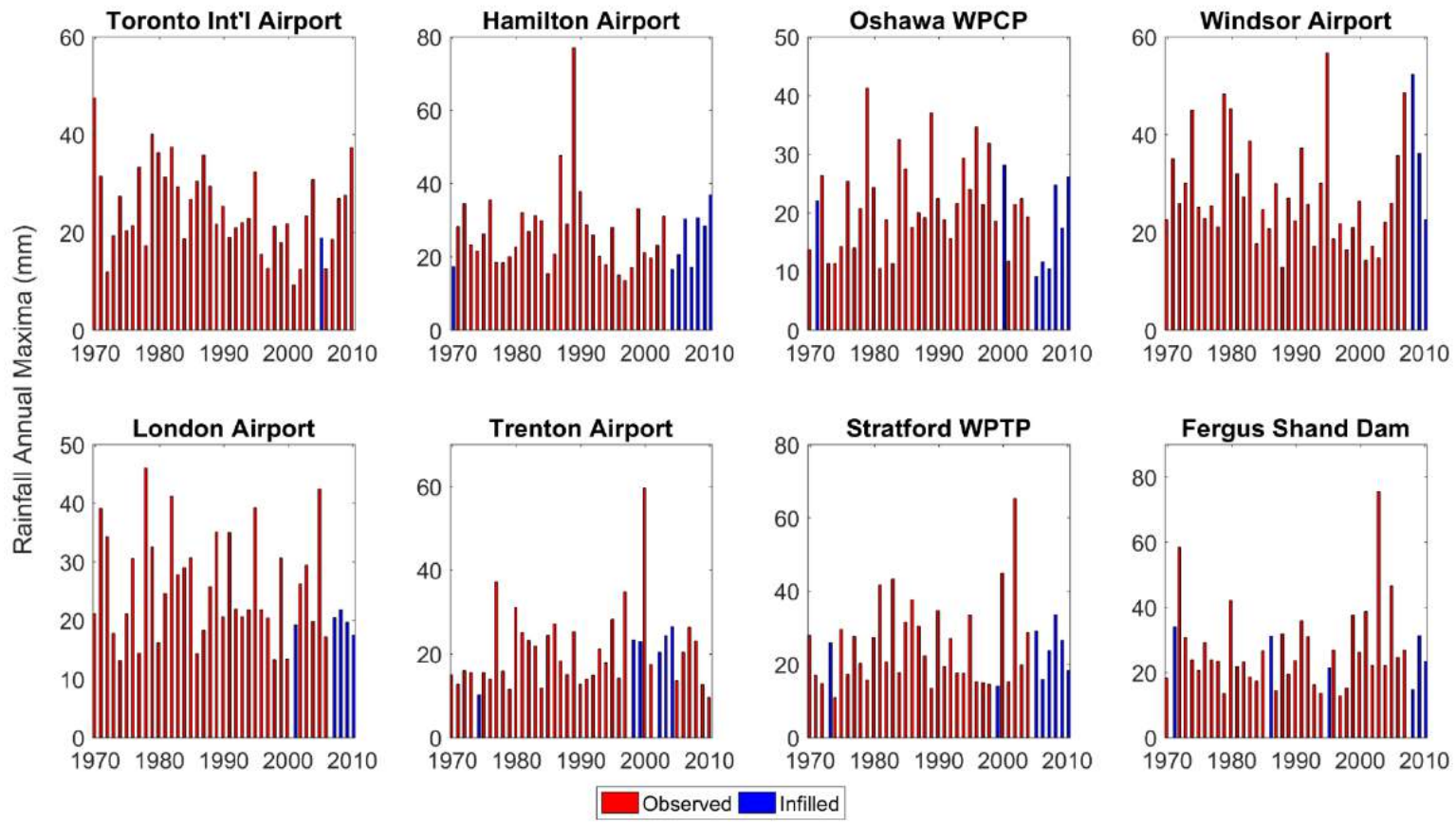

**Figure S4.** Observed versus infilled hourly AM series

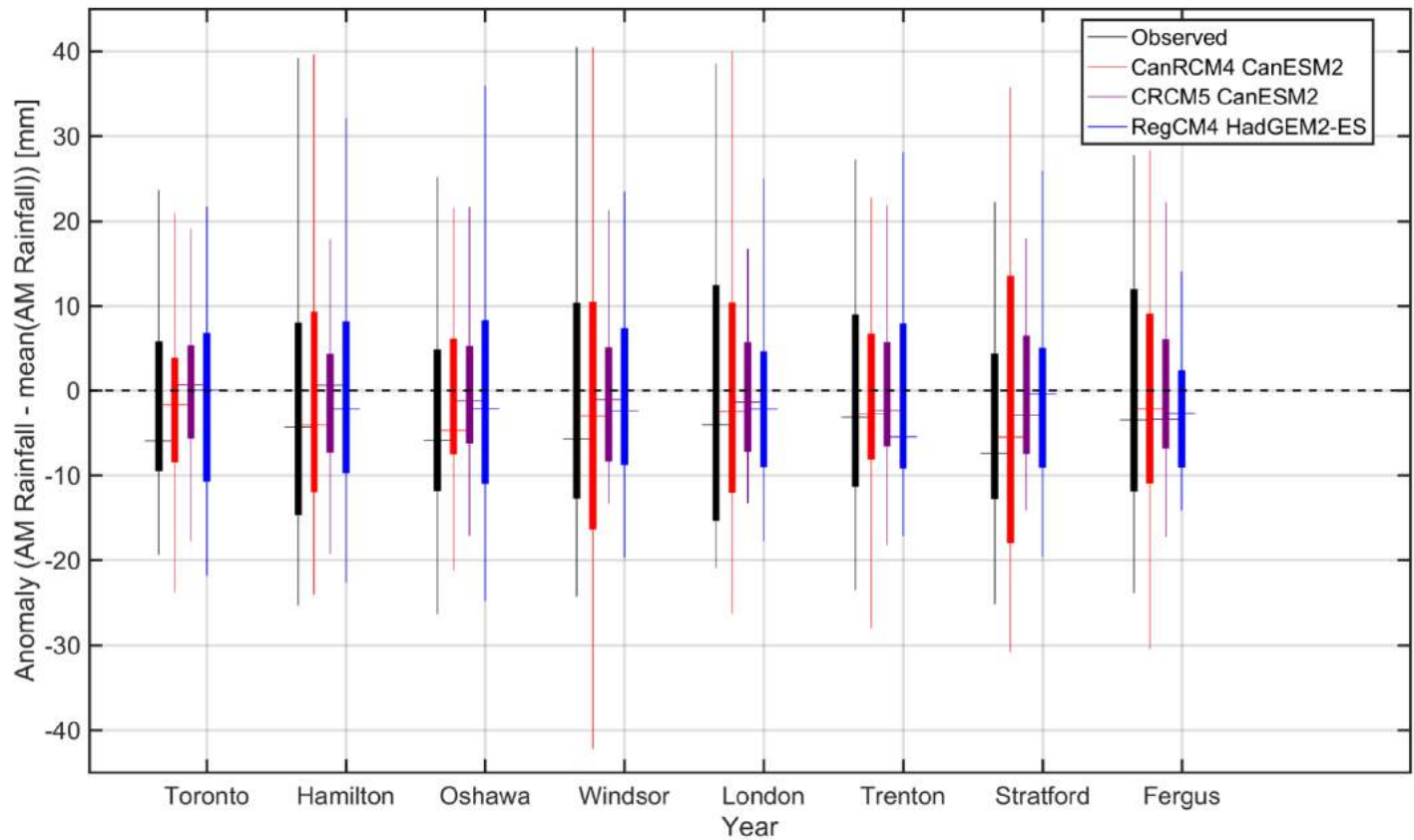

**Figure S5.** Distribution of annual maximum precipitation anomaly in observations and NA-CORDEX simulated RCMs during baseline (1970-2010) period.

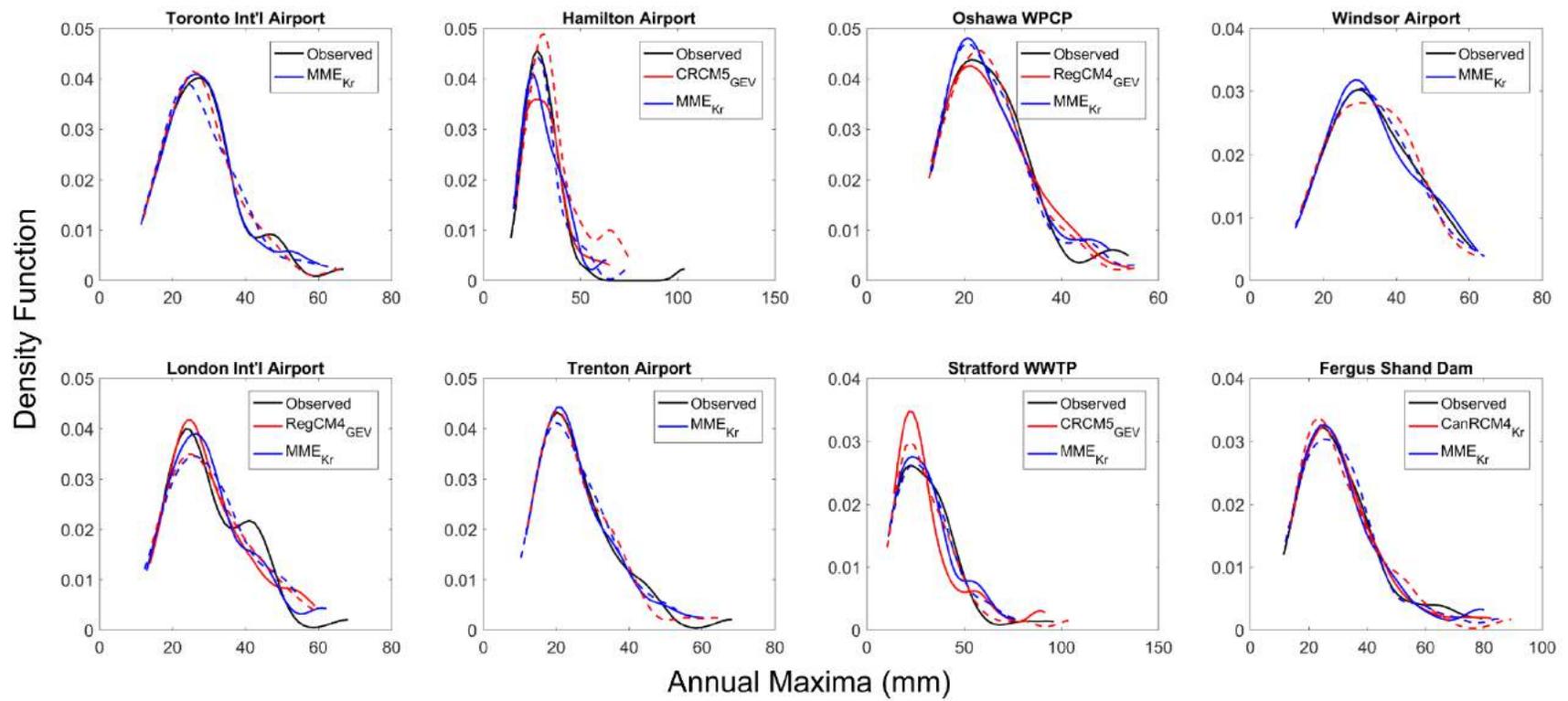

**Figure S6.** PDF of observed versus multi-model median 2-hour AM precipitation in eight station locations. The minimum and the maximum bounds are shown using dotted blue and red lines respectively.

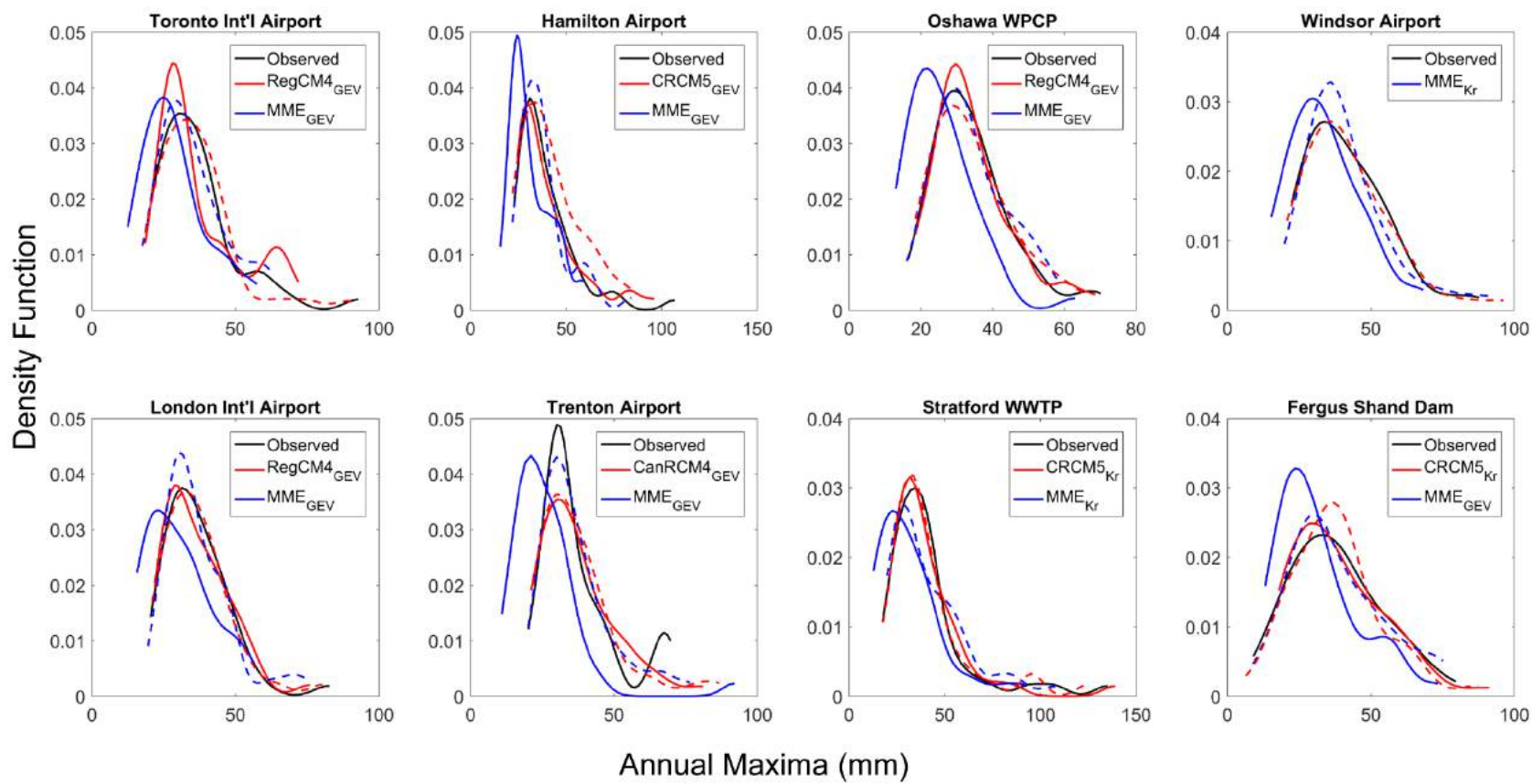

**Figure S7.** Same as Figure S6 but for 6-hour precipitation extreme.

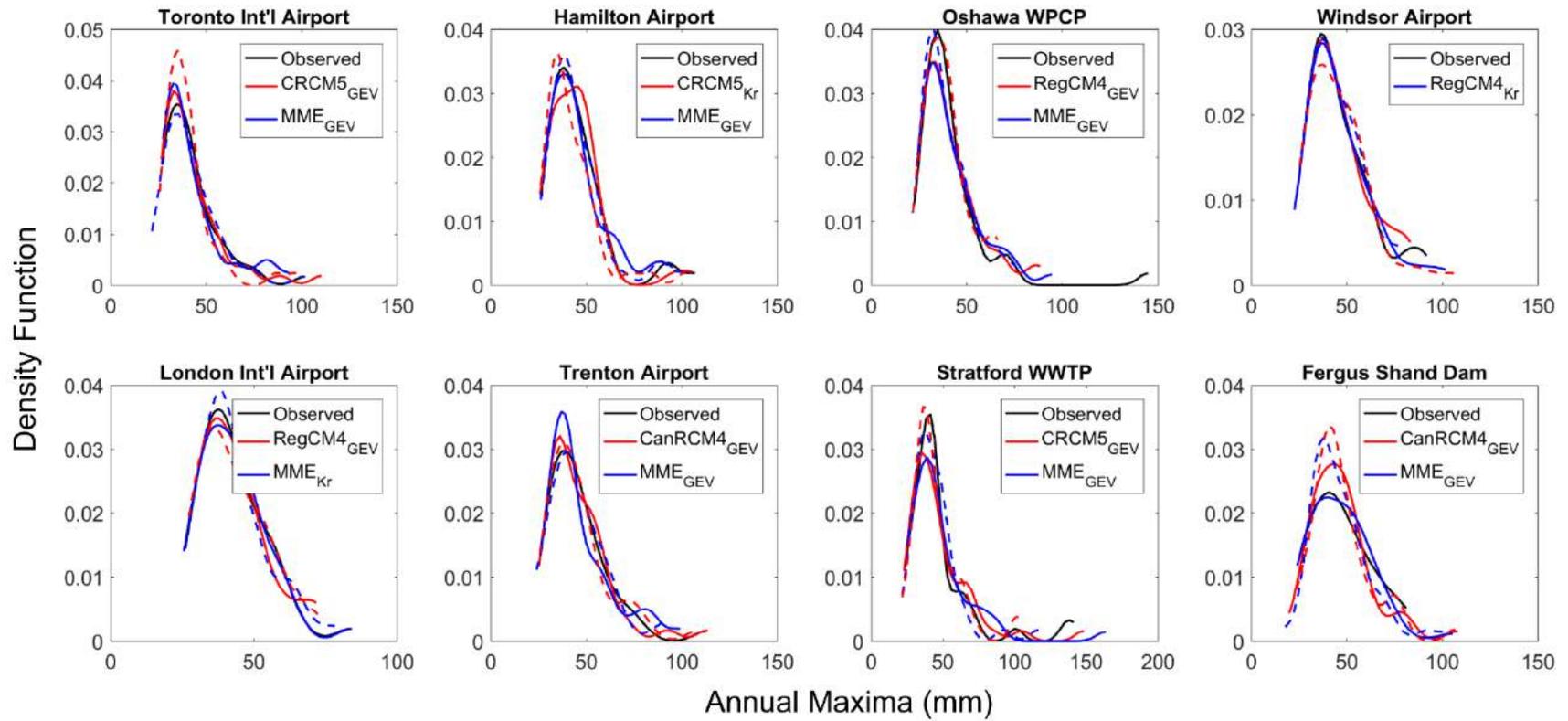

**Figure S8.** Same as Figure S7 but for 12-hour precipitation extreme.

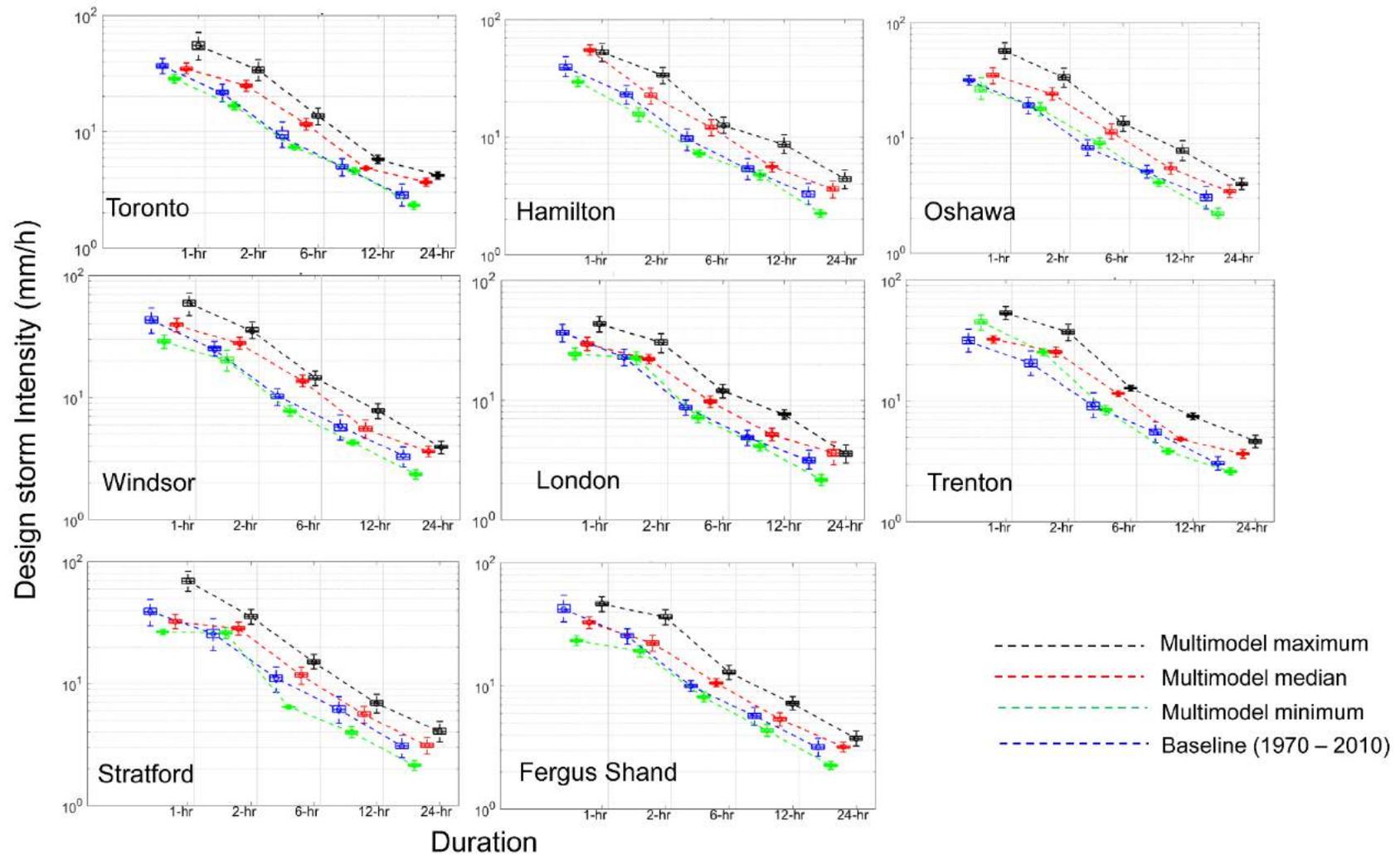

**Figure S9.** Present-day (in blue) versus Projected (in red; 2030-2070) IDF curves for 10-year return period across eight stations in Southern Ontario using stationary GEV models. The uncertainty in IDF simulation at different duration is expressed using boxplots. The minimum (in green) and maximum (in red) bounds in IDF curves are shown to express the best and worst plausible scenarios.

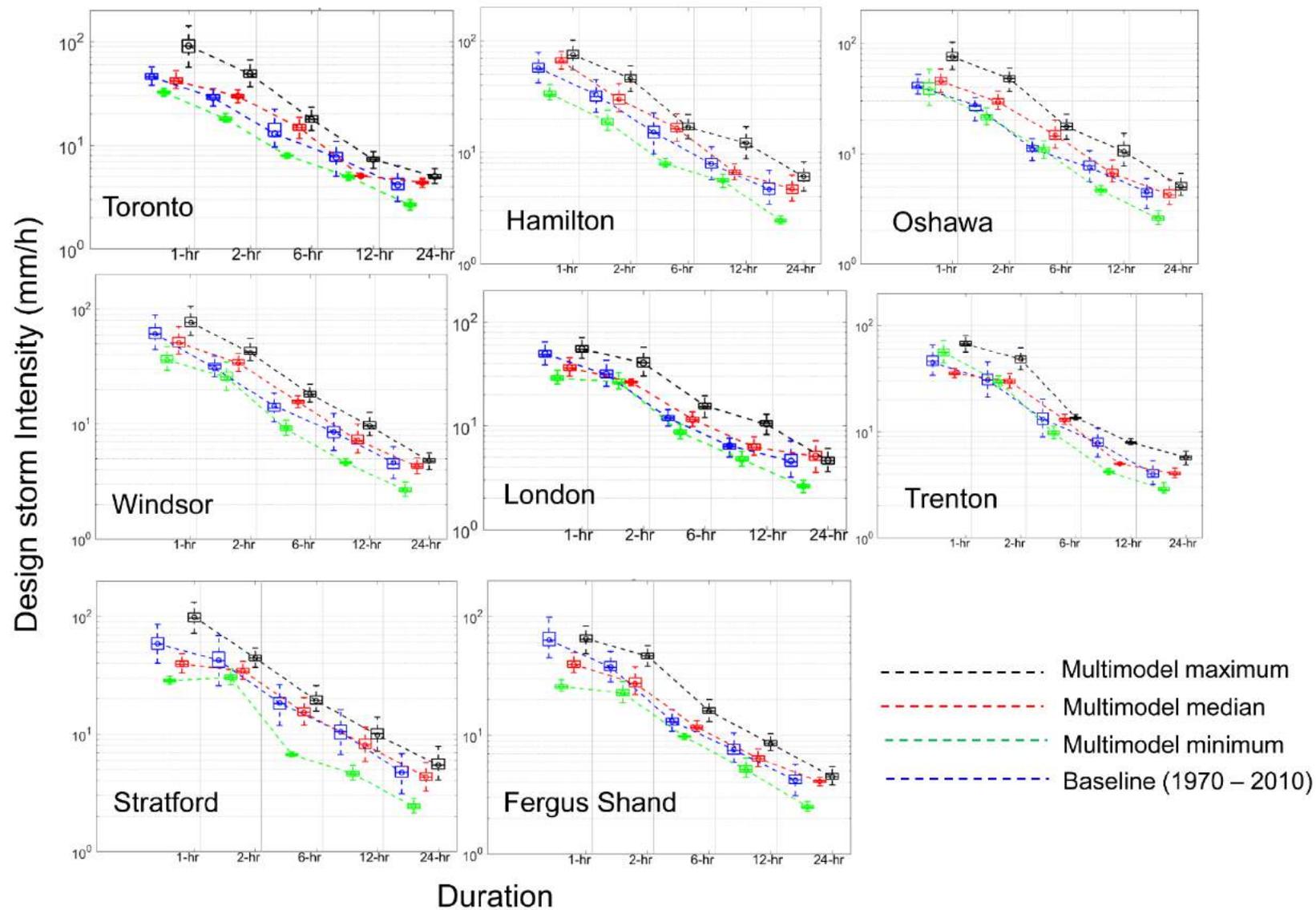

**Figure S10.** Same as Figure S7 but for 50-year return period.

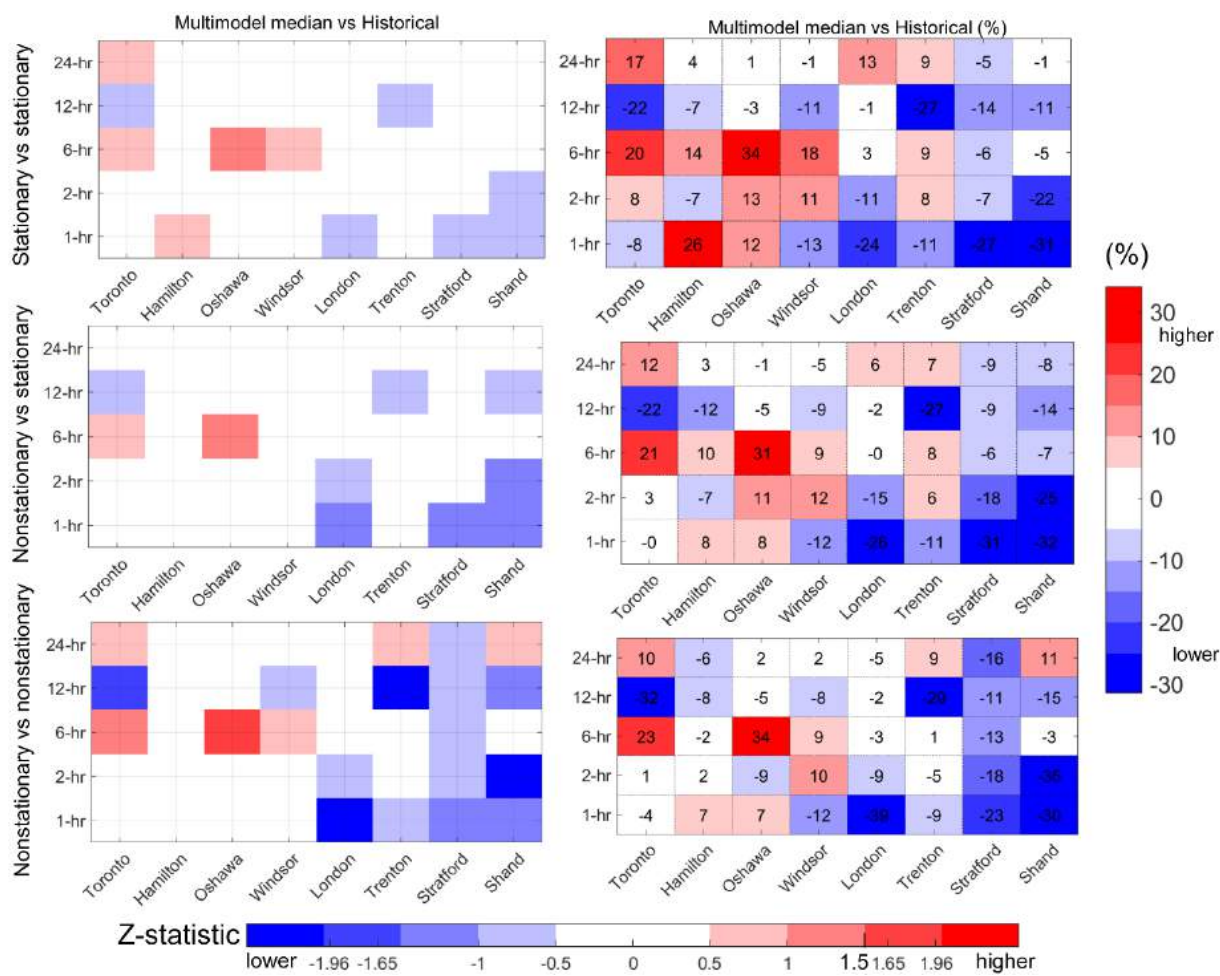

**Figure S11.** Projected changes in rainfall intensity for 25-year return period in three different cases: stationary (2050s) versus stationary (1990s) [*top panel*]; Nonstationary (2050s) versus stationary (1990s) [*middle panel*]; Nonstationary (2050s) versus nonstationary (1990s) [*bottom panel*]. The comparative assessment is performed between projected storm intensity modeled using multi-model median NA-CORDEX RCM ensemble and observed baseline intensity. The shades of the changes express high and low end, with dark red indicating increase in storm intensity while dark blue show decrease in the intensity. Very small changes [*i.e.,* in and around zero values] are marked with white.